\documentclass{aa}
\pdfoutput=1

\usepackage{graphicx}
\usepackage{txfonts}
\usepackage{pictex}
\usepackage{latexsym}
\usepackage{graphics}
\usepackage{afterpage}
\usepackage[]{hyperref}

\usepackage{units}

\begin{document} 

   \title{A comparison of compact, presumably young with extended, evolved  radio active galactic nuclei}

   \author{Helmut Meusinger\inst{1} 
          and
           Mukul Mhaskey\inst{1}
          }

   \institute{Th\"uringer Landessternwarte, 07778 Tautenburg, Germany 
                 }

  \date{Received XXXXXX; accepted XXXXXX}

 \abstract{
 The triggering and evolution of active galactic nuclei (AGNs) and the interaction of the AGN with its host galaxy is an important topic in extragalactic astrophysics. Radio sources with peaked spectra (peaked spectrum sources, PSS) and compact symmetric objects (CSO) are powerful, compact, and presumably young AGNs and therefore particularly suitable to study aspects of the AGN-host connection.
 }
 {
We use a statistical approach to investigate  properties of a PSS-CSO sample that are related to host galaxies and could potentially shed light on the link between host galaxies and AGNs. The main goal is to compare the PSS-CSO sample with a matching comparison sample of extended sources (ECS) to see if the two have significant differences.
 }
 {
We analysed composite spectra, diagnostic line diagrams,  multi-band spectral energy distributions (MBSEDs), star formation (SF) indicators, morphologies, 
and cluster environments for a sample of 121 PSSs and CSOs for which spectra are available from the Sloan Digital Sky Survey (SDSS). 
The statistical results were compared with those of the ECS sample, where we  generally considered the two subsamples of quasi-stellar objects (QSOs) and radio galaxies separately.
The analysis is based on a large set of archival data in the spectral range from the ultraviolet to mid-infrared. 
 }
 {
We find significant differences between the PSS-CSO and the ECS sample.  In particular, we find that the ECS sample has a higher proportion of passive galaxies with a lower star formation activity. This applies to both sub-samples (QSOs or radio galaxies) as well as to the entire sample. The star formation rates of the PSS-CSO host galaxies with available data are typically in the range $\sim 0$  to 5\,$\mathcal{M}_\odot$\,yr$^{-1}$, and the stellar masses are in the  range $3\cdot10^{11}$ to $10^{12} \mathcal{M}_\odot$. Secondly, in agreement with previous results, we find a remarkably high proportion of PSS-CSO host galaxies with merger signatures.  The merger fraction of the PSS-CSO sample is $0.61 \pm 0.07$,  which is significantly higher than that of the comparison sample  ($0.15 \pm 0.06$).  We suggest that this difference can be explained by assuming that the majority  of the PSSs and CSOs cannot evolve to extended radio sources and are therefore not represented in our comparison sample. 
}
{}


  \titlerunning{Presumed young radio galaxies}
  \authorrunning{H. Meusinger and M. Mhaskey}

   \maketitle

%
\section{Introduction}\label{Intro}
%

Radio-loud active galactic nuclei (AGNs) are typically associated with powerful bipolar radio jets  \citep{Tadhunter_2016, Hardcastle_2020, Saikia_2022}. The jets transport energy and momentum from the accreting supermassive black hole (SMBH) into the surrounding interstellar medium of the host galaxy and the intra-cluster medium of a galaxy cluster.  This so-called jet (or kinematic) feedback is thought to have an impact on the fate of the ambient gas and thus on the star formation (SF) history of galaxies \citep{McNamara_2007, Fabian_2012, Morganti_2017, Venturi_2021}.

The projected size of the radio jets is usually larger than that of typical host galaxies and can reach a total extent of up to several megaparsecs.  On the other side, there are also compact radio galaxies that are just as energetic as the most luminous extended ones, but much smaller and confined within their host galaxies.  The  gigahertz peaked spectrum (GPS) sources, compact steep spectrum (CSS) sources, and the high-frequency peakers (HFPs), collectively referred to as  peaked spectrum sources (PSSs), belong to this class of radio AGNs.  In the milliarcsecond resolution radio images taken with the very long baseline interferometry technique, these sources show a similar morphology to typical radio AGNs,  albeit with a linear size extending only to a few kiloparsecs \citep{Wilkinson_1994, Stanghellini_1997, Orienti_2006}. The flat-spectrum radio core is surrounded by symmetric steep-spectrum lobes on both sides, similar to the large-scale morphology of extended radio galaxies.  The radio spectra of PSSs deviate from the canonical straight power law and are characterised by a concave (inverted) shape with steep slopes on either side of the peak.  The  peak frequencies (in the rest frame) are around 1 GHz for GPS,  below 500 MHz for CSS, and above 5 GHz for HFP sources.  Synchrotron self-absorption by relativistic electrons inside the radio-emitting source \citep{Slish_1963} or free-free absorption by ionised  plasma either inside or outside is responsible for the inverted shape of the radio spectrum of the PSSs  \citep{Kellermann_1966}. 
A comprehensive review on the PSSs was given by \citet{Odea_1998} and \citet{Odea_2021}.

The origin and evolution of the PSSs are still a matter of debate.  Kinematic ages derived from the motion of the hot spots indicate a young age (< 5000 years) for the radio activity \citep{Owsianik_1998, Polatidis_2003, Gugliucci_2005}.  This is also supported by the spectral age estimates from modelling the high-frequency spectral break \citep{Murgia_1999, Orienti_2010}.  Statistical studies, however, indicate an overabundance of PSSs over the extended radio galaxies \citep{Readhead_1996, Odea_1997, An_2012}. \citet{Slob_2022} assumed that the PSSs selected at low frequencies (< 144 MHz) represent  young radio galaxies.  Their conclusion is based on the analysis of the luminosity function and radio source counts of PSSs, both of which are found to be scaled down by a significant factor. In an alternative evolution scenario, the PSSs are considered as `old and frustrated' sources that are confined to the host galaxies due to an over-dense interstellar medium \citep{vanBreugel_1984, Bicknell_1997, Dicken_2012}. The high detection rate of \ion{H}{i} absorption amongst PSSs over extended radio galaxies may indicate that a high proportion of PSSs are embedded in a rich interstellar medium \citep{Morganti_2018}.   Furthermore, for some  PSSs, the intermittent nature of AGN activity may lead to their truncated evolution \citep{Reynolds_1997, Kunert_2006}.

Concerning the present work, we briefly summarise the following findings on the host galaxies of compact radio sources from \citet{Odea_2021}:
The typical hosts appear to be large elliptical galaxies with magnitudes around the Schechter luminosity, but there are also exceptions with significant disk components.  The optical spectrum is dominated by an old stellar population, but often current or ongoing SF is also indicated, where the star formation rates (SFRs) are typically a few to tens $\mathcal{M}_\odot$\,yr$^{-1}$. The optical emission line spectra  are very similar to those of the extended radio sources, both types can be divided into high-excitation radio galaxies (HERGs) and low-excitation radio galaxies (LERGs). Some well-studied hosts of compact radio sources are classified as galaxy mergers and a few systematic studies indicate that mergers and interactions are common.  Their environment  appears similar to that of powerful large radio sources and they are not predominantly in rich galaxy clusters.

A critical source of information for a systematic investigation of the host galaxies are optical spectra. \citet{Liao_2020} compiled a sample of 126 compact radio galaxies with spectroscopy available from the Sloan Digital Sky Survey \citep[SDSS;][]{York_2000}.  Besides GPS, CSS, and HFP sources, this sample also includes Compact Symmetric Objects (CSOs), which have similar properties as GPS radio galaxies.   \citet{Liao_2020} derived the SMBH masses and the Eddington ratios and found that they cover broad ranges with $\log {\mathcal M}_{\rm SMBH}/\mathcal{M}_\odot \sim 7$ to 10 and $\log R_{\rm Edd} \sim -4.9$ to 0.4. \citet{Nascimento_2022} investigated the optical and mid-infrared (MIR) properties of a sample of compact radio sources with SDSS spectra consisting of 58 CSS or GPS sources and 14 Megahertz-Peaked-Spectrum (MPS) sources at $z \le 1$ from various radio-selection catalogues publicly available in the literature.  Based on stellar population synthesis,  they concluded that their sample is dominated by intermediate to old stellar populations in a wide range of morphological galaxy types. The SFRs cover the range from zero to $\sim 20 \mathcal{M}_\odot$\,yr$^{-1}$, for most sources the SFR is $\la 5 \mathcal{M}_\odot$\,yr$^{-1}$.   No strong correlation between the optical and radio properties of these sources was found. Recently, \citet{Gordon_2023} investigated the relationship between SF and radio source size in a sample of CSS sources and concluded that the bulk of SF ceased several hundred Myr before the jet was triggered, where the broad range of SFRs is possibly a consequence of episodic SF that can result from galaxy-galaxy interactions. Using  complete  samples  of  CSOs,  \citet{Kiehlmann_2023}  argued that  their relative numbers and the distributions of redshifts and sizes show conclusively that most CSOs belong to a distinct population of jetted AGNs that do not evolve into larger radio sources and should be described as `short-lived', rather than  `young'.  \citet{Duggal_2023} carried out a systematic search for UV signatures from AGN feedback in CSS radio galaxies based on new HST images.   They claim that their observations  are in line with recent SF activity `likely triggered by current or an earlier episode  of  radio  emission,  or  by  a  confined  radio  source  that  has  frustrated  growth  due  to  a  dense environment'.

The present work compares compact and presumably young radio galaxies from \citet{Liao_2020} with old, extended radio galaxies. Special attention is paid to the properties of the host galaxies. Because the  \citet{Liao_2020} sample is heterogeneous,  \citet{Odea_2021} warned that it may not match well with other samples in the literature. For this reason,  we have created a  matched `comparison sample' of extended radio galaxies and QSOs which we assume represent advanced stages of radio source evolution. We look for significant differences between  the \citet{Liao_2020}  sample and the comparison sample, where we focus on three main aspects: (i) spectral energy distributions (SEDs), in particular, composite spectra and multi-band SEDs (MBSEDs) from the far ultraviolet (UV) to MIR and their implications for the stellar population of the host galaxies, (ii) SF activity, and (iii) morphological indicators of gravitational interactions and merging.   Our study is based entirely on existing archival data. It complements the study of \citet{Nascimento_2022} in which they compare the different types of compact sources and look for correlations between optical/MIR and radio properties,  while our work focuses on the comparison of the sample of compact radio sources with the comparison sample of extended sources. Of course, their sample of 72 compact radio sources at $z \le 1$ with spectra in SDSS Data Release (DR) 12 \citep{Alam_2015} strongly overlaps with the sample of our study.  The low-redshift ($z \le 1$) subsample of the \citet{Liao_2020} sample contains 94 sources, the total sample covers a larger redshift range up 3.6. We used the optical spectra from the SDSS DR16  \citep{Ahumada_2020}.

This paper is structured as follows. 
In Sect.\,\ref{sect:Samples}, we describe our sample of compact radio sources and explain in detail how the comparison sample of extended radio sources was constructed.
In Sect.\,\ref{sect:line_ratios}, we discuss three different diagnostic diagrams and compare
 the distributions of the spectral types from the two samples.
In  Sect.\,\ref{sect:SED}, we use the SDSS composite spectra, the MBSEDs, and the MIR data to discuss the spectral energy distributions. 
The SFRs and stellar masses are discussed in Sect.\,\ref{sect:SFR}.
In Sect.\,\ref{sect:morph}, we determine the merger fractions in the two samples and discuss possible reasons for the differences.
The summary is presented in Sect.\,\ref{sect:conclusion}.
Additional data are provided in the Appendices.

We assume Lambda Cold Dark Matter ($\Lambda$CDM) cosmology with $H_0 = 73$ km s$^{-1}$ Mpc$^ {-1}$, $\Omega_{\rm \Lambda} =0.73$, and $\Omega_{\rm M}=0.27$.

%
\section{Samples}\label{sect:Samples}
%


\subsection{PSS-CSO sample}\label{sect:PSS_sample}


\citet{Liao_2020}  collected all the radio samples of PSSs and CSOs available in the literature and created a large sample of 250 CSS, 148 GPS, 116 HFP sources, and 80 CSOs. By cross-identification with the SDSS, a sample of 147 sources with spectroscopic counterparts was selected from this, which they called the `optical sample'.   After removing blazar-type sources,  which contaminate this sample, their `final optical sample' consists of 126 objects. \citet{Liao_2020} and  \citet{Liao_2021} have described the sample in detail and listed several properties of the individual objects. The largest angular size (LAS) and the largest linear size  (LLS) of the radio structure are given there for 109 sources (87\%), all sources larger than 0.6 kpc are classified as CSS sources.  With one exception,  LLS = 56\,kpc for J090933.49+425346.5, all listed sizes are smaller than 40\,kpc.

\begin{figure}[htbp]
\begin{center}
\includegraphics[viewport= 0 0 360 356,width=4.4cm,angle=0]{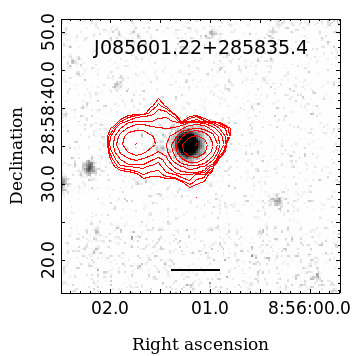}       
\includegraphics[viewport= 0 0 360 356,width=4.4cm,angle=0]{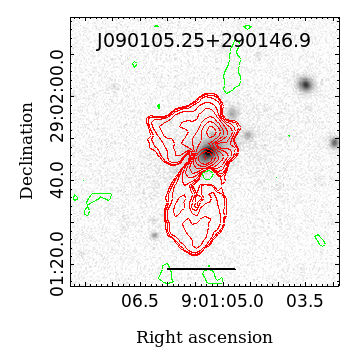}       
\includegraphics[viewport= 0 0 360 356,width=4.4cm,angle=0]{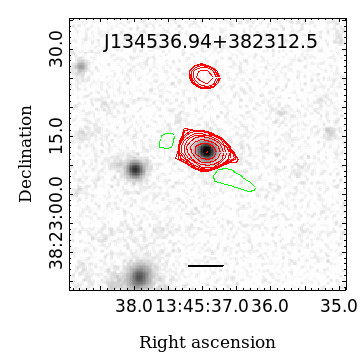}       
\end{center}
\caption{
VLASS contours of the three sources from the \protect\citet{Liao_2020} sample with VLASS sizes larger than 50\,kpc. The ten red contour lines are logarithmically spaced between  $\sim 5\sigma$ of the local noise and the maximum flux of the source. The green dashed lines are negative contours at $-5\sigma$ to $-3\sigma$. The background images are from the SDSS i band.  The vertical bar at the bottom indicates a length of 50 kpc at the redshift of the source. 
}
\label{fig:SDSS_VLASS_PSS}
\end{figure}

In the present work, this PSS-CSO sample was slightly reduced for two reasons. Firstly, as will be explained in Sect.\,\ref{sect:comp_sample} below,  the comparison sample is selected among extended radio AGNs with LLS $ >  50$\,kpc in the  Very Large Array Sky Survey \citep[VLASS;][]{Lacy_2020}. 
For consistency, we therefore also checked the extents of the PSSs and CSOs in the VLASS images.  For eleven objects that were found to be extended or possibly double we measured the FR type and size (Table\,\ref{tab:PSS_sizes}). All remaining objects do not appear to have been resolved in our visual inspection of the VLASS images. Depending on the FR type \citep{FR_1974}, the LAS was identified with the distance between the hot spots for FR\,II and with the extent of the 5$\sigma$ contour for FR\,I and uncertain classifications, where $\sigma$ is the r.m.s. of the local background. The three sources with LLS > 50\,kpc are shown in Fig.\,\ref{fig:SDSS_VLASS_PSS}. Traditionally the size limit has been put at 20 kpc for PSSs and CSOs  \citep{Odea_2021}. The PSS-CSO sample contains 4 sources with sizes larger than 50 kpc and seven sources with LLS between 20 and 50\,kpc. For the present study, we have thus decided to use a strict size limit of 50 kpc to distinguish compact from extended sources.   Therefore, we removed SDSS\,J090933.49+425346.5 and the three sources shown in Fig. 1.  The case of J134536.94+382312.5 is not entirely clear. We cannot be sure that the two radio sources seen in the image are physically connected, but we cannot rule it out either. We decided to remove this object from the sample as a precaution.

Secondly, we removed  the CSS source  \object{4C\,+39.29}, which was assigned to \object{SDSS\,J101714.23+390121.1} at $z = 0.211$ by \citet{Liao_2020}.   We argue  that the host galaxy of \object{4C\,+39.29} is \object{SDSS\,J101714.10+390123.8} at $z = 0.536$, for which no SDSS spectrum is available. This interesting source is discussed in more detail in  Appendix\,\ref{sect:SDSSJ1017}.

We refer to the remaining sample of 121 sources as the  `PSS-CSO sample'.  The visual inspection of the SDSS spectra of all sources revealed incorrect redshifts from the SDSS spectroscopic pipeline for \object{SDSS\,J014109.16+135328.3}  ($ z = 0.621$ instead of 0.236 given by SDSS) and \object{SDSS\,J080133.55+141442.8} ($ z = 1.196$ instead of 0.246).  We note that, in both cases, the correct redshift is given by \citet{Liao_2020}. 
For the source \object{SDSS\,J085408.44+021316.1} we prefer the value $z = 0.400$ from SDSS to the value 0.459 given by \citet{Liao_2020}.  
The PSS-CSO table containing the properties derived here is published at CDS, an excerpt from the table (the first six lines) is given in Table\,\ref{tab:PSS}. 
With the exception of four sources outside the FIRST footprint area, the radio luminosity  $L_{\rm 1.4GHz}$ was calculated from the integrated 1.4 GHz flux density provided by FIRST \citep{Helfand_2015}.  For the remaining four sources, we took the flux densities from the 
NED\footnote{https://ned.ipac.caltech.edu/}. For all other data we refer to \citet{Liao_2020}.

\begin{table}[htbp]
\caption{Sizes of sources from the \protect\citet{Liao_2020}  sample found to be extended in the VLASS. }
 \begin{tabular}{lcrr} 
\hline\hline 
\noalign{\smallskip}
Name                           & FR type & LAS & LLS \\ 
                                     &              &  (")   & (kpc)\\
\hline        
\noalign{\smallskip}      
J080442.23+301237.0   &    II   &    4.4   &    37   \\
J080447.96+101523.7   &    II   &    4.0   &    33   \\
J085601.22+285835.4   &    II   &    7.6   &    60   \\
J090105.25+290146.9   &    I    &   37.4   &   116   \\
J101251.77+403903.4   &    II   &    5.9   &    23   \\
J115919.97+464545.1   &    II   &    6.0   &    20   \\
J131718.64+392528.1   &    II   &    8.3   &    33   \\
J134536.94+382312.5   &    II:  &   13.0  &    107   \\ 
J144712.76+404744.9   &    II   &    3.1   &    27   \\
J154754.12+351842.2   &     :   &    6.6   &    40   \\
J161823.57+363201.7   &   II:  &     4.7  &     33   \\ 
\hline                                    
\end{tabular}   
\tablefoot{
The angular sizes are as measured from the VLASS images.
A colon in the second column indicates uncertain classification.
}
\label{tab:PSS_sizes}                    
\end{table}


\subsection{Selection of the comparison sample}\label{sect:comp_sample}


\begin{figure*}[htbp]
\begin{center}
\includegraphics[width=5.8cm,angle=0]{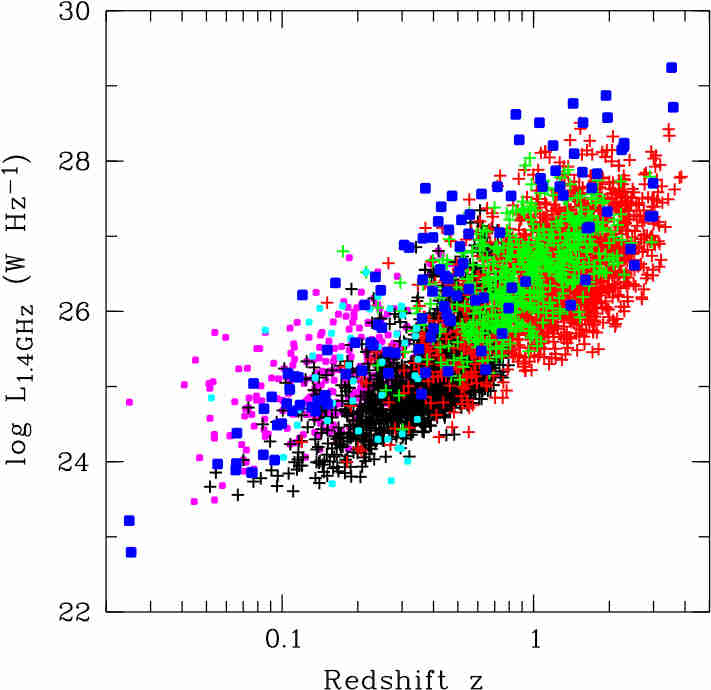}  \
\includegraphics[width=5.8cm,angle=0]{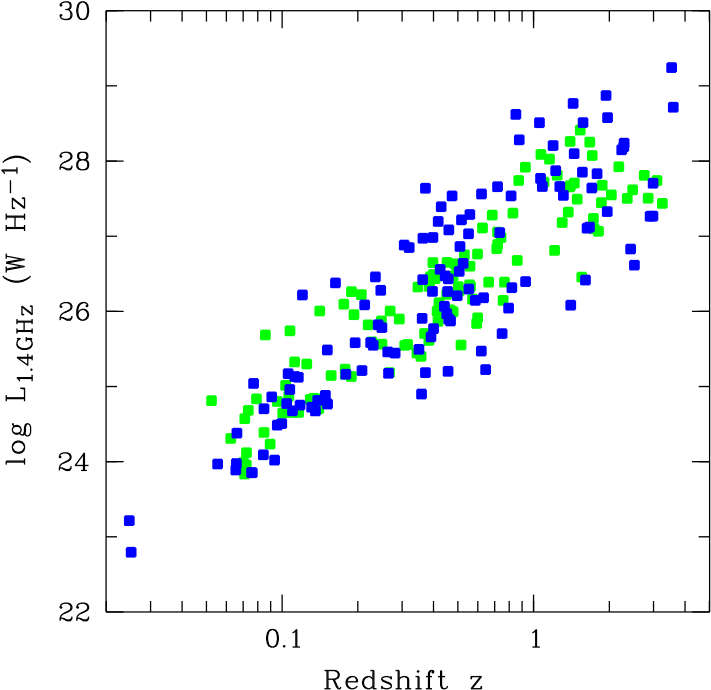} \ 
\includegraphics[width=5.6cm,angle=0]{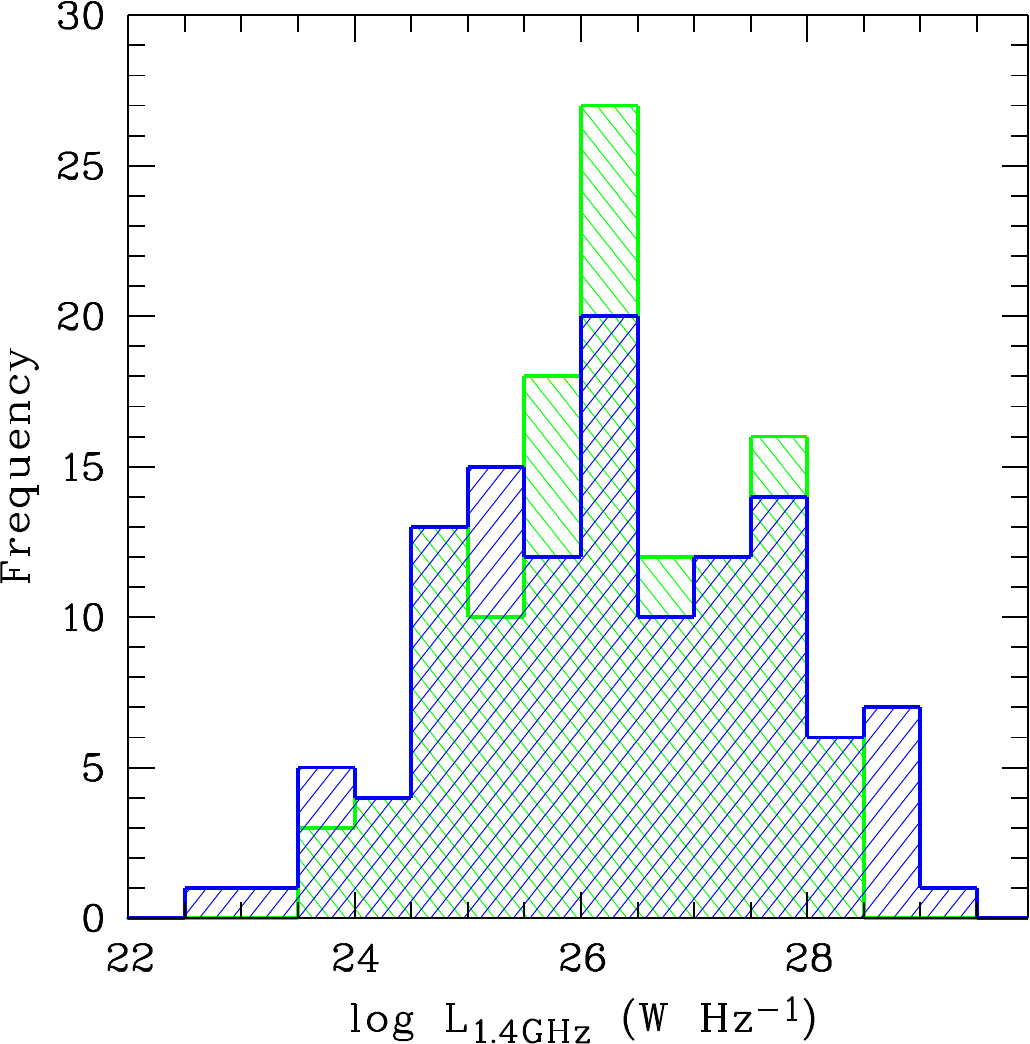} \\
\includegraphics[width=5.8cm,angle=0]{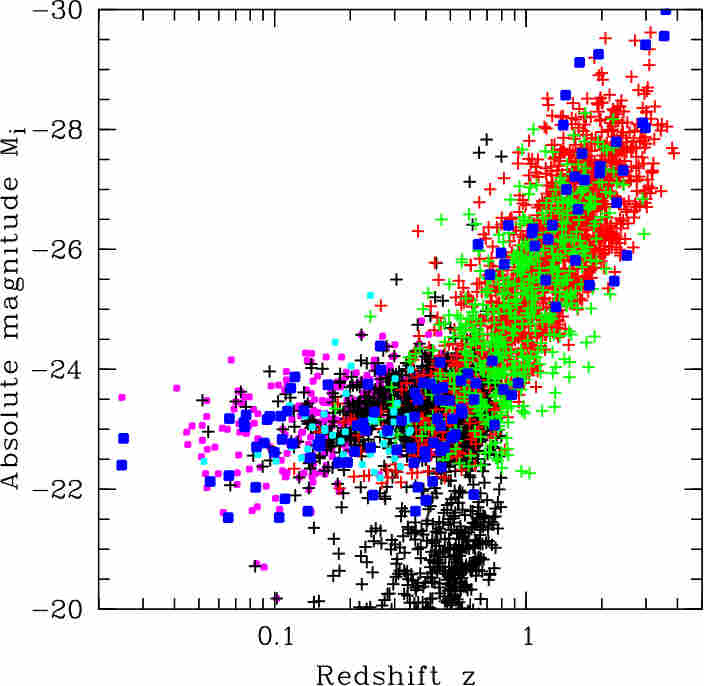} \
\includegraphics[width=5.8cm,angle=0]{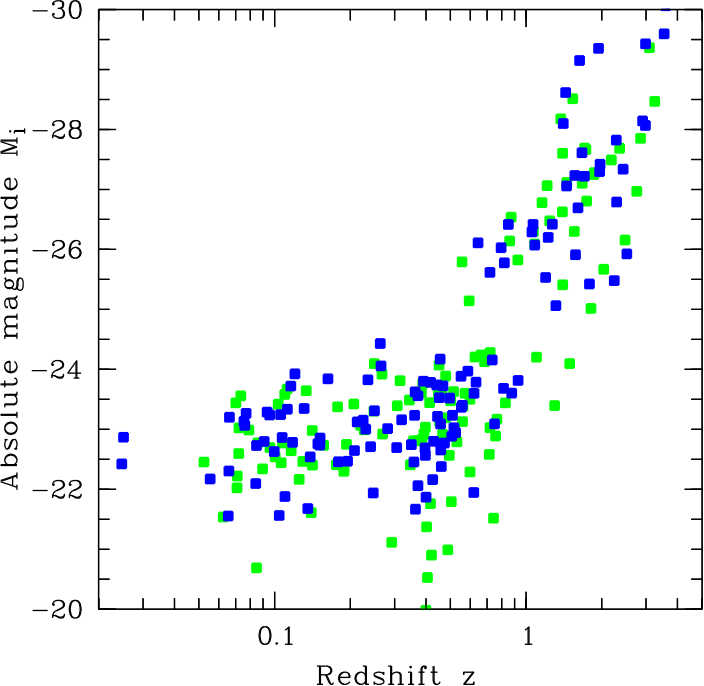} \ 
\includegraphics[width=5.6cm,angle=0]{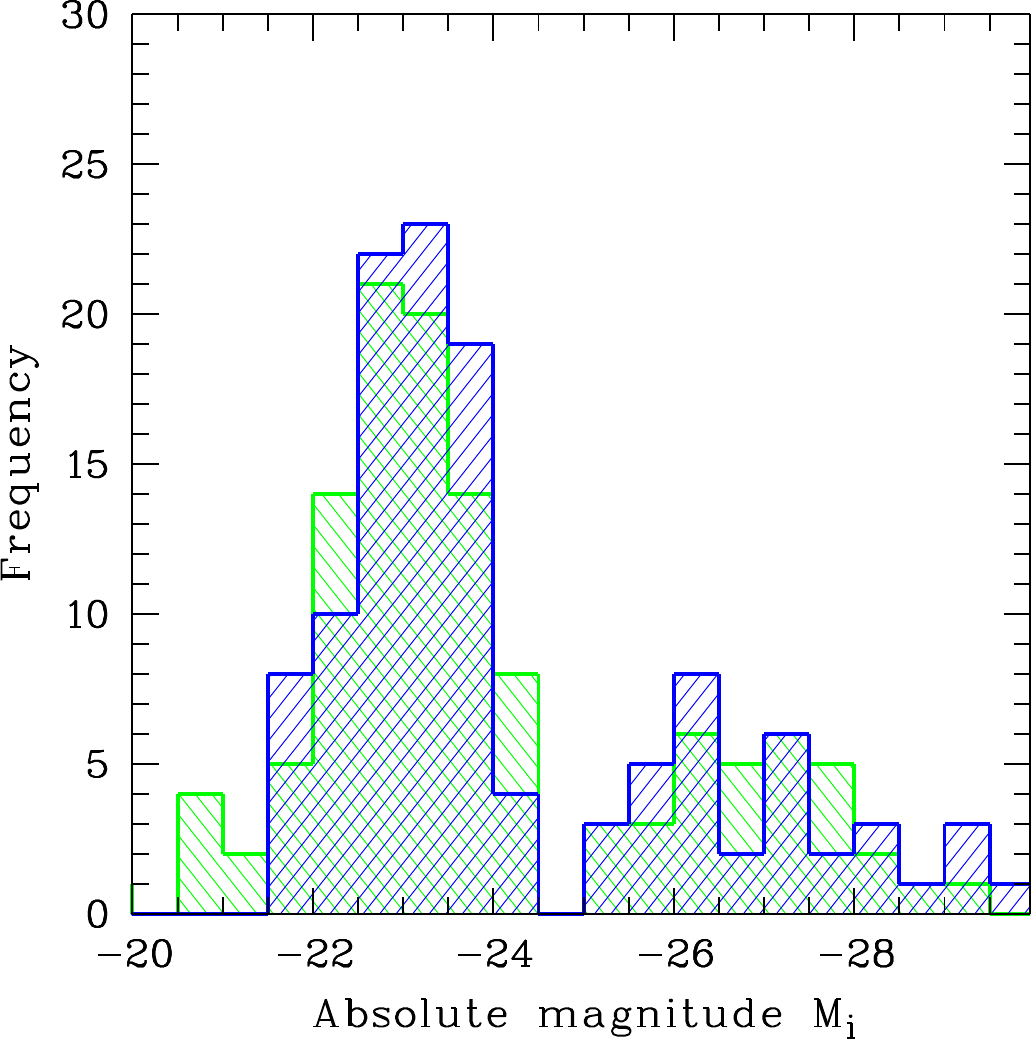} \\
\includegraphics[width=5.8cm,angle=0]{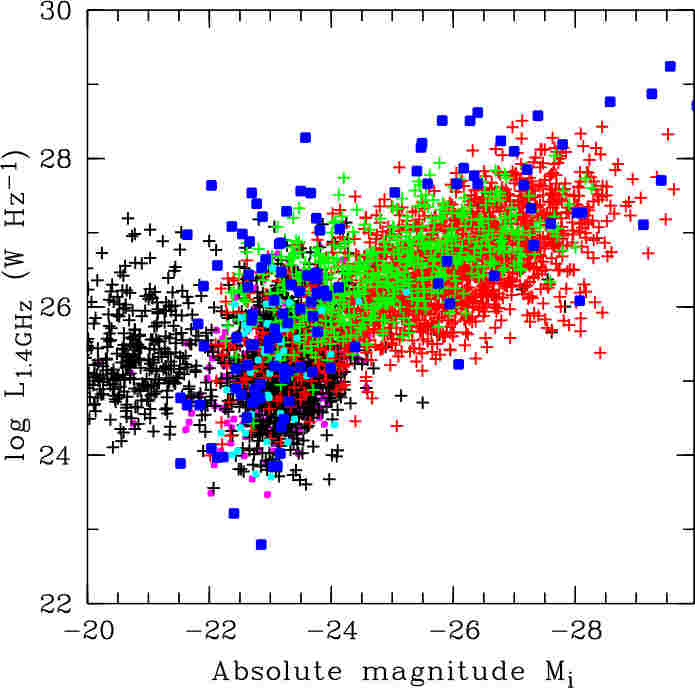}  \
\includegraphics[width=5.8cm,angle=0]{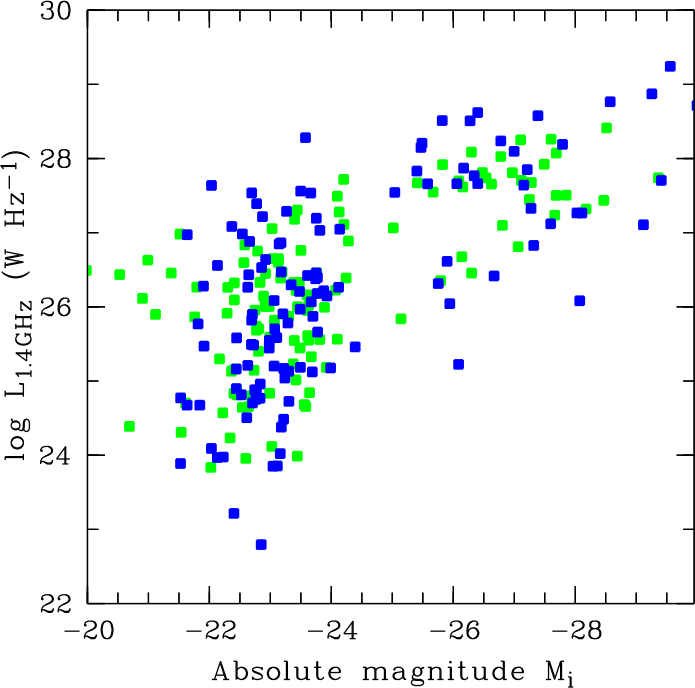} \ 
\includegraphics[viewport= 0 -8 490 500,width=5.6cm,angle=0]{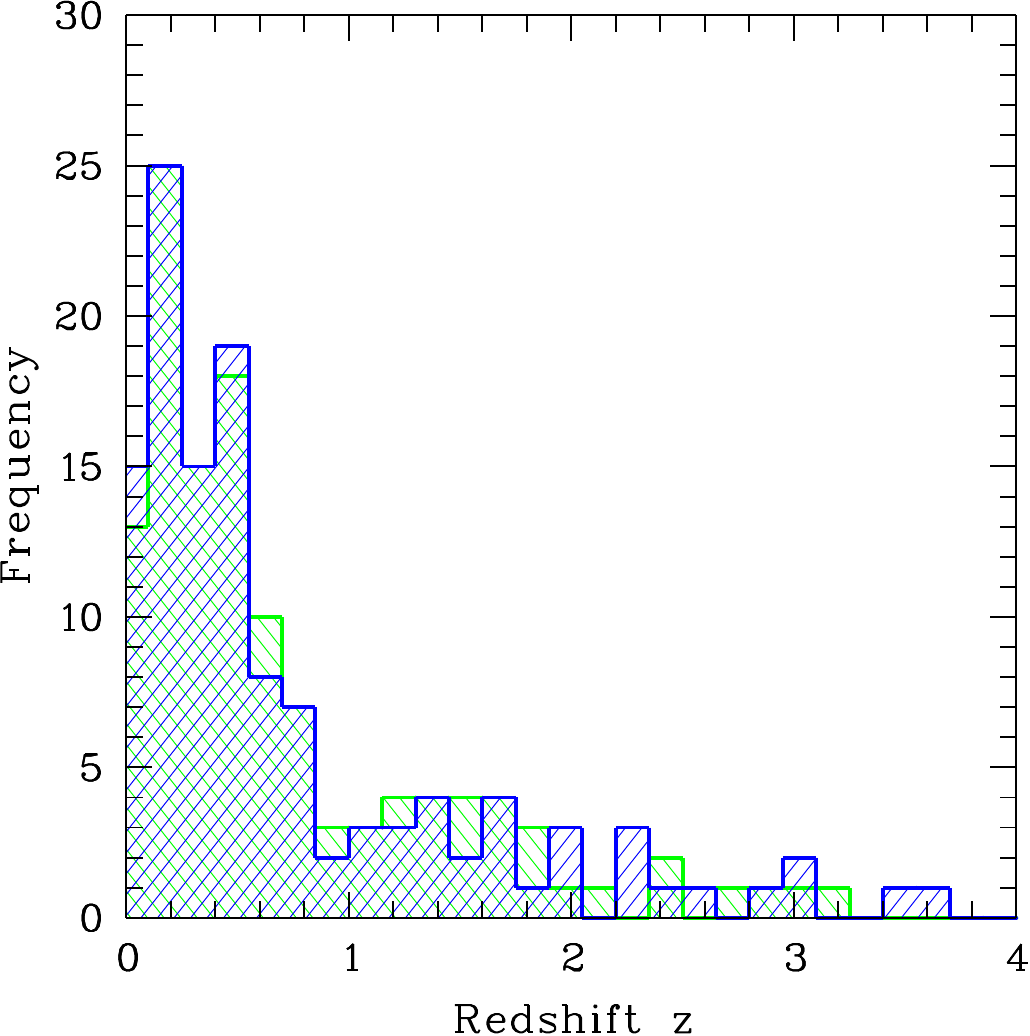} \\
\end{center}
\caption{
Comparison of the PSSs and CSOs with extended radio galaxies and quasars in the  $z - L_{\rm 1.4GHz} - M_{\rm i}$ parameter space.
Left and middle column:  Projections of the distributions of the PSS-CSO sample  (blue filled squares) and the extended comparison galaxies and quasars.  The panels on the left-hand side show the different contributions to the basic sample from which the comparison sample is drawn 
(small symbols: 
red plus signs - \protect\citet{Shen_2011}, 
black plus signs - \protect\citet{Garon_2019},
cyan filled squares - \protect\citet{Rafter_2011},
green plus signs - \protect\citet{Kuzmicz_2021}, 
magenta filled squares - \protect\citet{Koziel_2011}). 
In the middle column, the final ECS sample (green) is compared with the PSS-CSO sample (blue).
Right column: histogram distributions of the PSSs and CSOs (blue) and the ECSs (green).
}
\label{fig:LogLr_z}
\end{figure*}

As mentioned above, the \citet{Liao_2020} sample was inhomogeneously selected and cannot be considered statistically complete and representative of the entire population.  In particular, their optical sample is biased towards low redshifts and, as a consequence, towards low luminosities. For the present study, it is thus of key importance to select a control sample in a consistent way. The construction of such a `matched' comparison sample consists of two steps. First, a basic sample must be defined from which the control sample is to be selected. Especially in the frame of the evolutionary scenario,  it is meaningful to compare the PSS-CSO sample with their evolved counterparts.   Because the physical size of a radio AGN is expected to grow with time,   extended radio galaxies can be considered later stages of evolution than compact sources, if a  similar environment is assumed.  Therefore, a possible way is to select the comparison galaxies from a basic sample of extended radio galaxies and QSOs with sizes larger than the upper limit  for  the PSS-CSO sources.   For the present study we use a size limit of 50\,kpc (Sect\,\ref{sect:PSS_sample}).

The main difficulty in creating such a basic sample of extended radio galaxies is the fact that only a small percentage of radio sources identified in SDSS  show extended radio emission \citep{Ivezic_2002} and not all of them have SDSS spectra.  As a result of an extensive literature search, we found several suitable data collections based on radio data from  the Faint Images of the Radio Sky at Twenty-Centimeters \citep[FIRST;][]{Becker_1995}.

\citet{Garon_2019} extensively discussed a  sample of 4304 extended radio galaxies from Radio Galaxy Zoo (RGZ). Their selection was restricted `to sources with at least a 0.65 consensus in the radio component classification from RGZ to remove sources with ambiguous structure, as well as non-physical associations between radio lobes and coincident galaxies'. The host galaxies were identified via cross-matching with the  images from the Wide-Field Infrared Survey Explorer \citep[WISE;][]{Wright_2010}, redshift data were taken from SDSS. Here, we extracted the subsample of the 1743 sources with spectroscopic redshifts.  \citet{Kuzmicz_2021} presented a sample of 174 newly discovered giant radio quasars  (GRQs; their Table 1), which are defined as having a projected linear size greater than 700 kpc, and 367 smaller, lobe-dominated radio quasars found during their search for GRQs (their Table 2).  This search was based on the cross-match of spectroscopically identified quasars and quasar candidates from SDSS with the NRAO Very Large Array Sky Survey \citep[NVSS;][]{Condon_1998} and the FIRST survey.  Another list of extended radio galaxies with SDSS spectroscopy was provided by \citet{Koziel_2011}, who created a sample of 401 FR II radio galaxies (no quasars) that have optical counterparts with SDSS spectra.  Furthermore, we included the sample of 63 broad-line AGNs from the SDSS with extended emission in FIRST from \citet{Rafter_2011}.

All these data cover the redshift range $z>1$ only insufficiently. Therefore,  we added a subsample of quasars from the `Catalog of Quasar Properties from SDSS DR7' \citep{Shen_2011}.  To include radio properties, \citet{Shen_2011} matched the SDSS DR7 quasar catalogue with the FIRST catalogue with a matching radius of $30\arcsec$  (corresponding to $\sim 250$\,kpc at $z = 2$) and flagged those quasars which have multiple FIRST sources within $30\arcsec$  as lobe-dominated ({\tt FIRST} = 2).  We extracted a list of 2105 lobe-dominated quasars with $z < 4$ from that catalogue.  Unlike the other data sources described above, the \citet{Shen_2011} catalogue does not contain information of the LAS of the FIRST source. In the first step, we identified the LAS with the largest distance between two radio components from the FIRST catalogue within an aperture of 500 kpc diameter centred on the respective target, where only sources with a probability of being a side lobe less than 0.1 were considered. The final sizes were measured more precisely on the VLASS images when selecting the comparison sample (see below).

The combination of the above-mentioned five compilations results in a total of 3930 sources covering a wide range of radio power and size, with the vast majority having LLS $> 50$\,kpc in the literature.  As discussed above, we use a size limit of 50\,kpc and define the sample of 3875 FIRST-SDSS sources with LLS $> 50$\,kpc  as the basic sample for selecting our comparison sample.   The left column of Fig.\,\ref{fig:LogLr_z} illustrates the distributions of $z$, radio luminosity $L_{\rm 1.4GHz}$ at 1.4 GHz (rest frame), and absolute i band magnitude $M_{\rm i}$.   The latter two were $k$-corrected following, for example, \citet{Kennefick_2008},  where we assumed  $F_\nu \propto \nu^{\,\alpha}$ with a mean spectral index $\alpha_{\rm o} = -0.5$ in the optical and $\alpha_{\rm r} = -0.7$ in the radio.  This sample of extended radio galaxies with SDSS spectroscopy populates a region of parameter space that is similar but not identical to that of the PSS-CSO sample.  In particular, it extends towards fainter 1.4 GHz luminosities, illustrating the well-known fact that PSSs and CSOs are among the most luminous radio AGNs. We note that all radio sources from this sample are more luminous than the threshold $L_{\rm 1.4GHz} =  10^{23}$\,W\,Hz$^{-1}$ above which  the radio luminosity distribution of radio galaxies is dominated by AGNs \citep{Sadler_2002}.

In the second step, the comparison sample was selected. \citet{Liao_2020} explicitly  excluded relativistically beamed sources from their sample.
To obtain a comparison sample that is also free from such sources,  we have reduced the basic sample only to objects with steep radio spectra.
In addition, we also checked the radio morphology and excluded compact, core-dominated sources in a later step (see below).   We used the SPECFIND V3.0 catalogue \citep{Stein_2021} and found that 978 sources from our basic sample have steep radio spectra with  $\alpha_{\rm r} <  -0.5$  measured at two or more frequencies  $\nu > 1$\,GHz. This is the parent sample from which the comparison sample was finally drawn. The comparison sample was selected so that the distributions of the main physical parameters are as similar as possible to those of the PSS-CSO sample. Here, we consider the redshift $z$,  the radio luminosity $L_{\rm 1.4GHz}$, the i band absolute magnitude $M_{\rm i}$, and the SDSS spectral class  as the most relevant properties. In particular, this means that for a PSS-CSO with a host-dominated spectrum, the host galaxy of the corresponding source in the comparison sample has a similar mass.  In practice, the apparent SDSS i band magnitude and the radio loudness $R_i$  are used after $z$ is fixed.\footnote{$R_i$ is the decadic logarithm  of the ratio of the 1.4\,GHz flux to the flux in the SDSS i band without including the $k$-corrections \citep{Ivezic_2002}.}  That is, we selected for each PSS-CSO $k$ with $(z_k, i_k, R_{i,k})$  a counterpart from the basic sample of extended sources with  $(z, i, R_i)$ within   $(z_k\pm \Delta z, i_k\pm \Delta i, R_{i,k}\pm \Delta R_i)$, where the interval widths $\Delta z, \Delta i$, and $\Delta R_i$ were chosen as small as possible. The resulting comparison sample thus has the same size as the PSS-CSO sample, that is, it consists of 121 sources. We henceforth refer to these sources as `extended comparison sources' (ECSs).

\begin{figure}[htbp]
\vspace{0.3cm}
\includegraphics[width=4.2cm,angle=0]{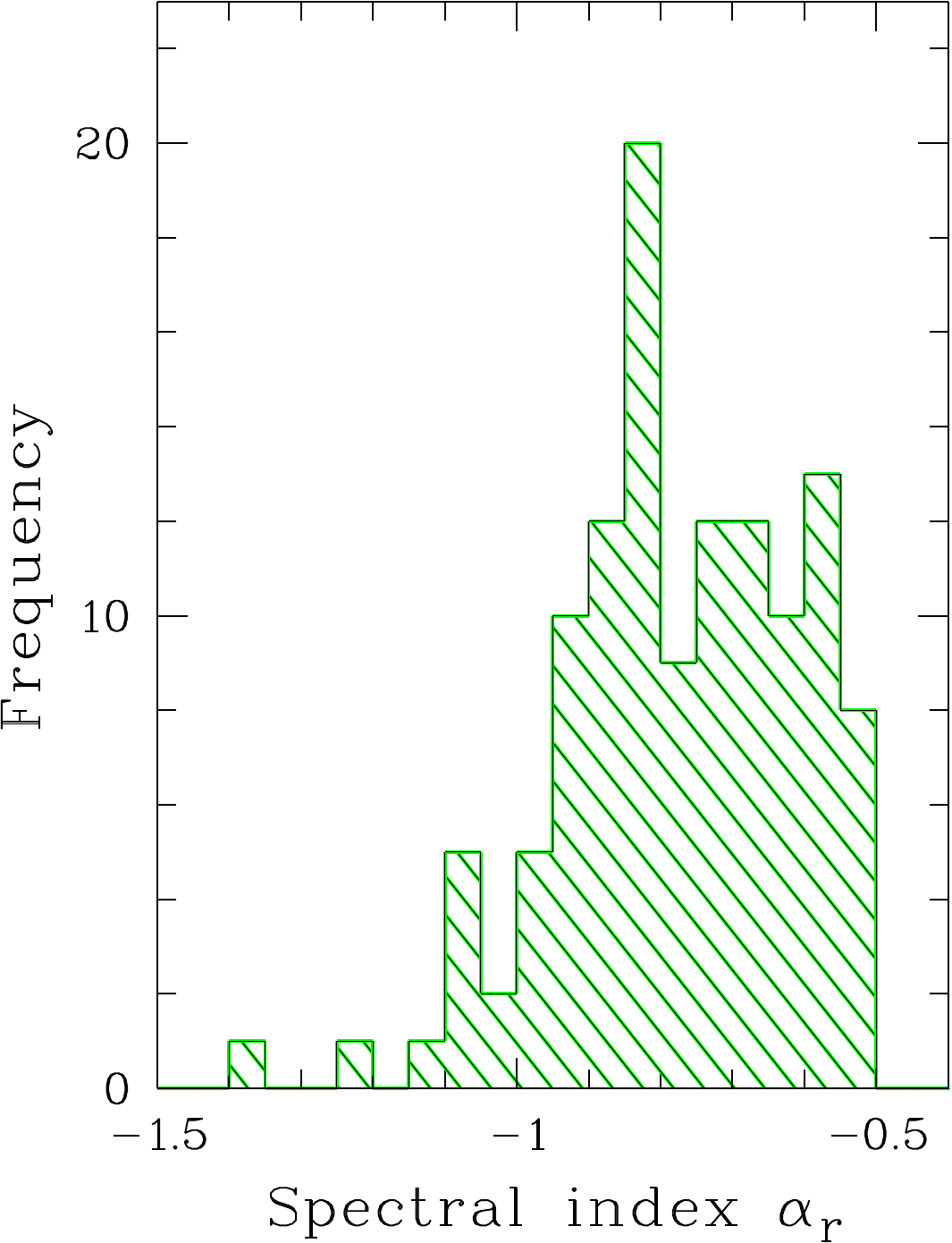} 
\hspace{0.2cm}
\includegraphics[width=4.2cm,angle=0]{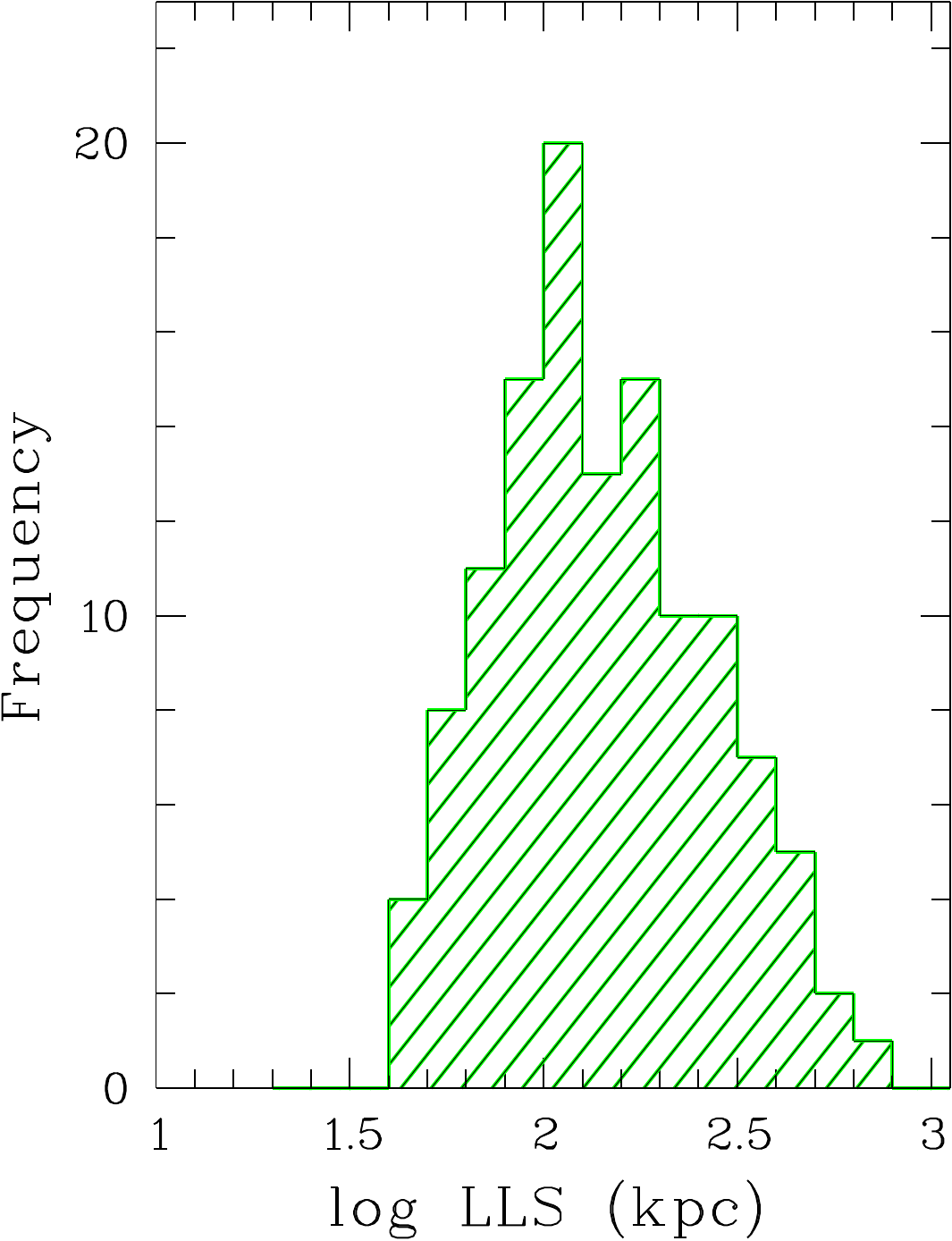}
\caption{
Frequency distribution of the spectral slope $\alpha_{\rm r}$ (left) and the LLS (right) for the ECS sample.
}
\label{fig:slope_distribution}
\end{figure}

Finally, we used  the VLASS-SDSS overlays to determine the FR type  and the LAS of the selected ECSs to ensure that they are indeed larger than 50\,kpc. 
The LAS was measured in the same way as for the PSSs and CSOs (Sect.\,\ref{sect:PSS_sample}). Compact, core-dominated and unclear or suspicious radio sources were removed from the selection and replaced by other sources from the parent sample in an iterative process. In the final sample, 112 ECSs were classified as FR\,II and 9 as FR\,I,  where for both FR types the classification is uncertain due to more complex structures for $\sim$10\% of the sources.  A mosaic of the VLASS-SDSS overlays  is shown in Figure\,\ref{fig:SDSS_VLASS_all}.   Figure\,\ref{fig:slope_distribution} shows the histogram distributions of the radio slope $\alpha_{\rm r}$ (left) and the LLS (right).   The full table of the ECS sample is published in electronic form at CDS, an excerpt is shown in Table\,\ref{tab:ECS}.

\begin{figure}[htbp] 
\includegraphics[viewport= 20 0 590 8200,width=6.2cm,angle=270]{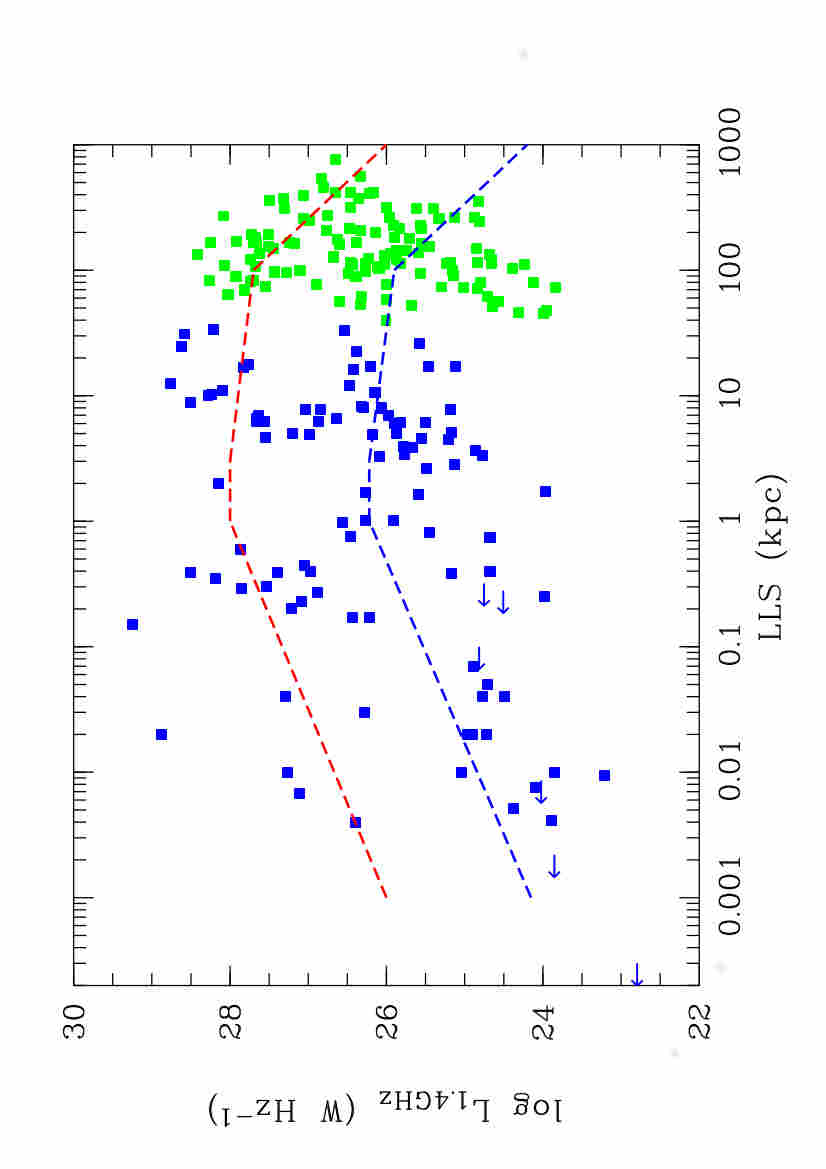}   
\caption{
Power - linear size diagram for the PSS-CSO (blue) and the ECS (green) samples.  Left arrows indicate upper limits for the LLS.  For the PSS-CSO sample, the sizes are taken from \protect\citet{Liao_2020}. The dashed lines depict evolutionary tracks from \citet{An_2012} for sources of high radio-power (red) and low radio-power (blue).
}
\label{fig:PD_diagram}
\end{figure}
                                                                                                                                                                                
In the middle column of Fig.\,\ref{fig:LogLr_z}, the ECS positions  in the parameter space $z - M_{\rm i} - \log\,L_{\rm 1.4\,GHz}$ are compared with those of the PSSs and CSOs. The histogram distributions of $\log\,L_{\rm 1.4GHz}, M_{\rm i}$, and $z$ are shown in the right column. It is obvious that both samples populate almost the same area of the parameter space.  A small but notable difference is the lack of ECSs with $\log\,L_{\rm 1.4GHz} > 28.4$.  On the other hand, adopting a threshold $R_i = 1$ to distinguish radio-quiet from radio-loud sources, 100\% of the ECS and 98\% of the PSS-CSO sample are radio-loud.   The two sources J103719.33+433515.3 ($R_i = 0.61$) and J105731.17+405646.1 ($R_i  = 0.01$), which have the lowest redshifts ($z = 0.025$) and the lowest 1.4 GHz luminosity, are not radio-loud according to our criterion.  As can be seen in Fig.\,\ref{fig:LogLr_z}, the parameter space around these two sources is almost empty. This means that the selection intervals $\Delta z, \Delta i$,  and  $\Delta R_i$ had to be extended considerably. Both selected ECS counterparts are radio-loud. We performed the two-sided two-sample Kolmogorov-Smirnov (KS) test with the null hypothesis ${\rm H}^0$ that the histogram distributions  in the panels of the right column of Fig.\,\ref{fig:LogLr_z} are the same for the two samples against the alternative hypothesis ${\rm H}^{\rm A}$ that they are different. ${\rm H}^0$ is discarded in favour of ${\rm H}^{\rm A}$ if the test statistic exceeds a critical value that depends on the significance level $\alpha$. The test result says that there is no reason to discard ${\rm H}^0$  for all three quantities at a significance level $\alpha = 0.05$  (i.e. the probability of mistakenly not rejecting   ${\rm H}^0$  is $< 0.05$).   We also note here that the proportion of the SDSS spectral class {\tt GALAXY} is very similar for the two samples: $43\pm 5$\,\% for the ECSs and  $45\pm 5$\,\%  for the PSS-CSO sample.

Fig.\,\ref{fig:PD_diagram} shows the PSS-CSO and the ECS samples in the power-size diagram.  The LLSs of the PSSs and CSOs are from \citet{Liao_2020}.  The size distributions of the two samples do not overlap. We over-plotted two evolutionary pathways  from \citet{An_2012} based on parametrised models.  It is obvious that the positions of the selected ECSs are not inconsistent with the assumption that they represent later evolutionary stages of sources similar to those from the PSS-CSO sample.

%
\section{Diagnostic emission line diagrams}\label{sect:line_ratios}
%

\begin{figure}[htbp] 
\begin{center}
\includegraphics[viewport= 15  15 570 780,width=6.15cm,angle=270]{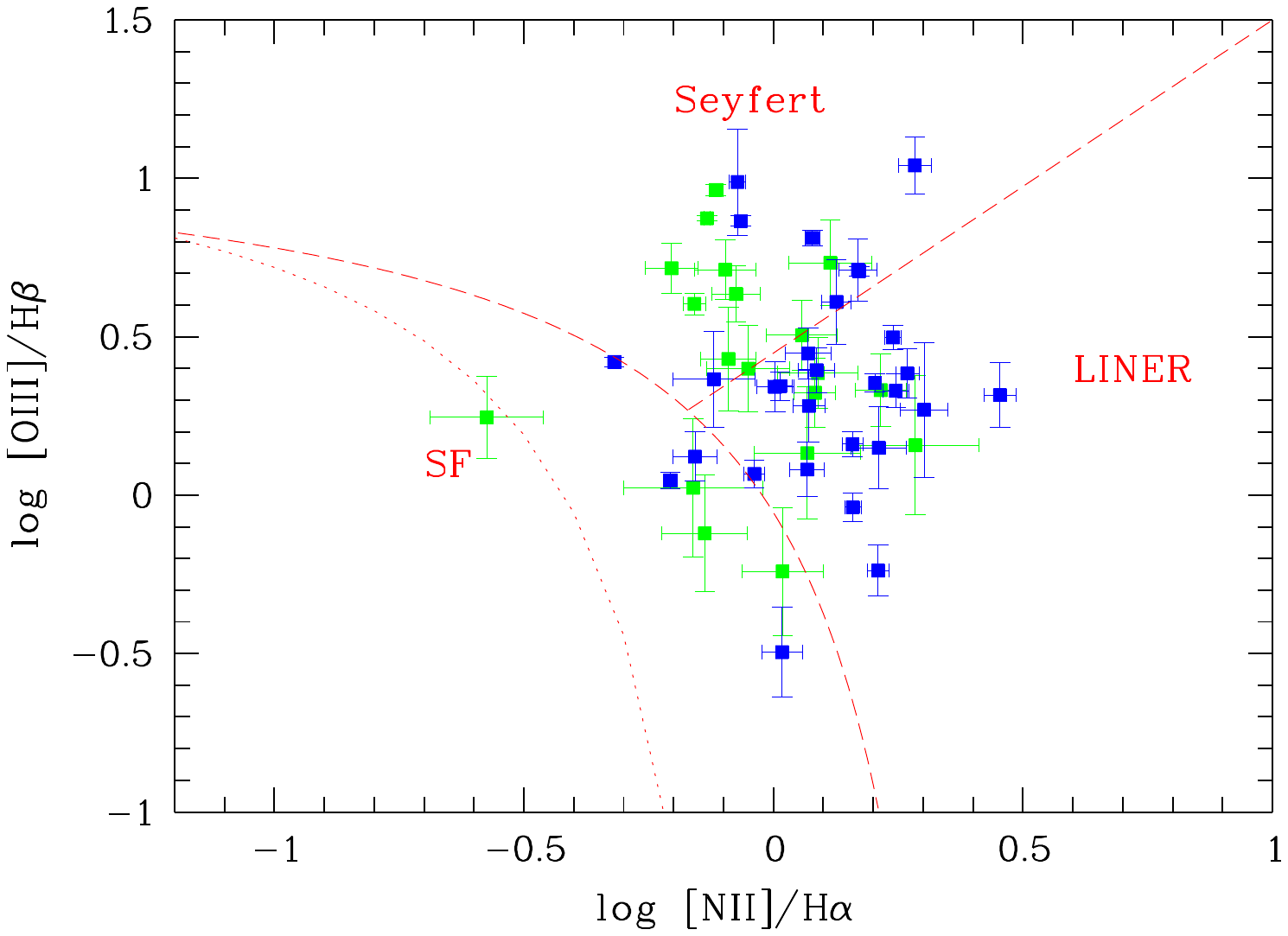} \\
\includegraphics[viewport= 15  15 570 780,width=6.05cm,angle=270]{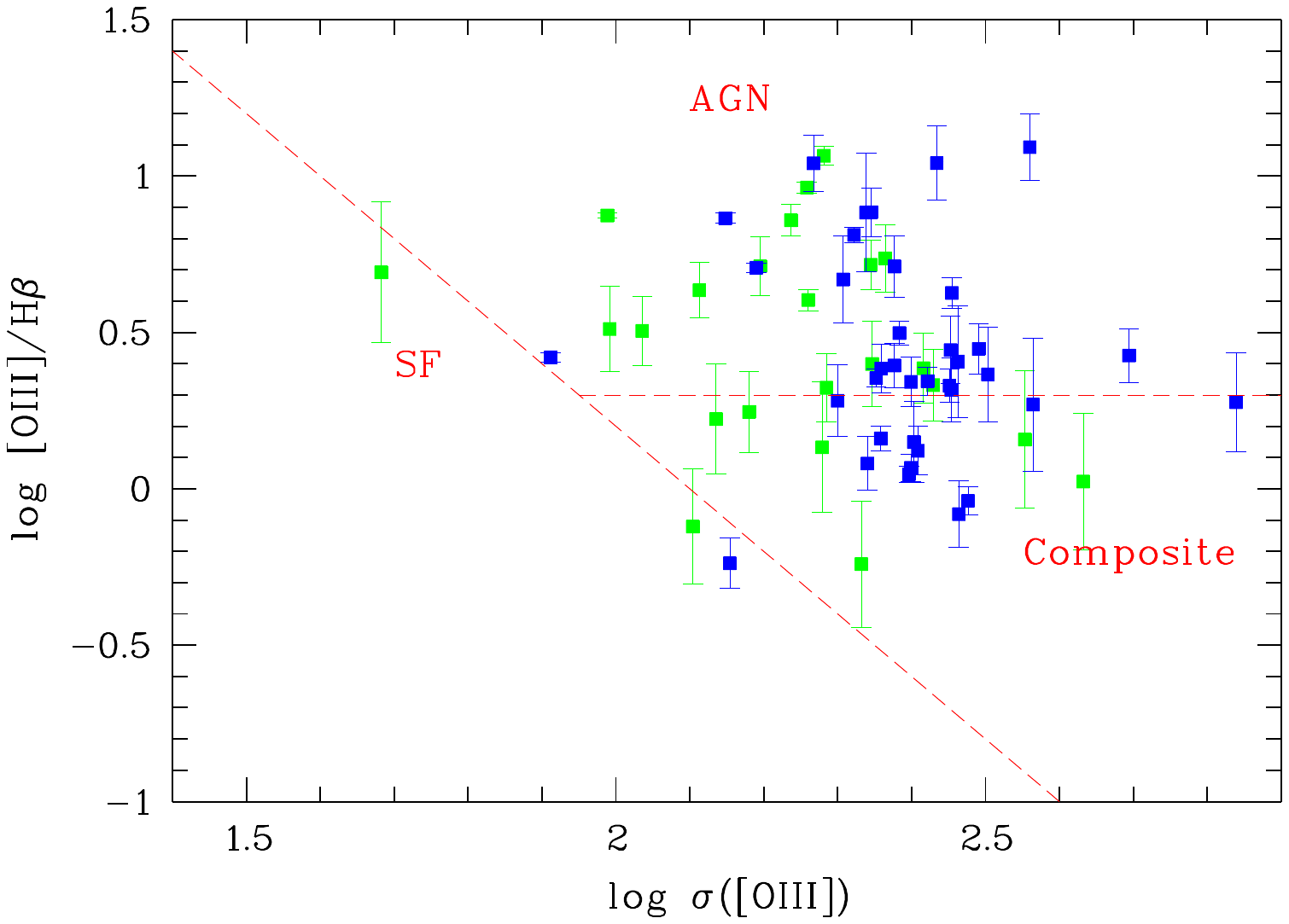} \\
\includegraphics[viewport= 15  15  570 780,width=6.05cm,angle=270]{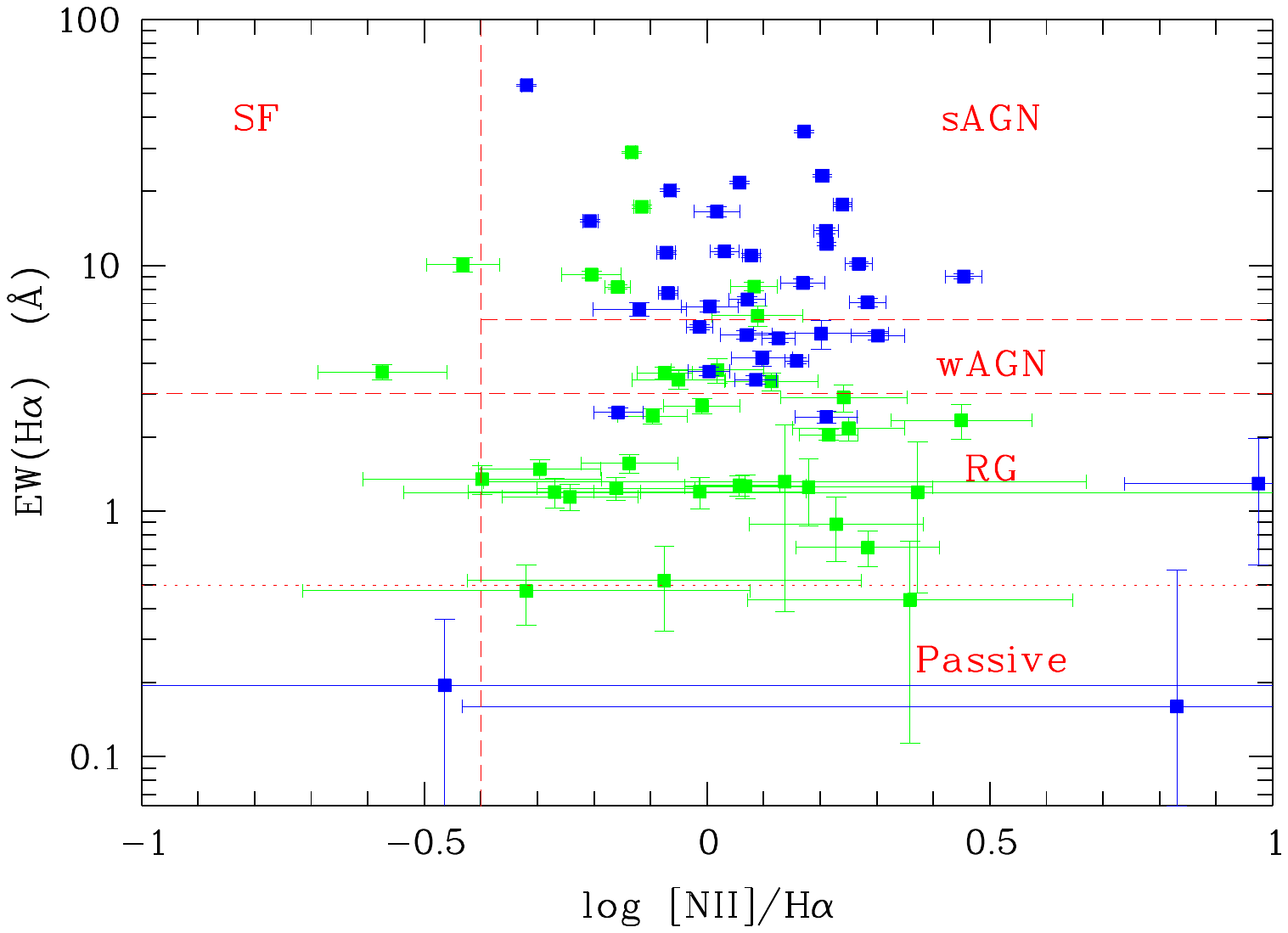}
\end{center}
\caption{
Diagnostic diagrams  for the galaxies  (SDSS spectral class = {\tt GALAXY}) from the PSS-CSO (blue) and the ECS  (green) samples with  $\mbox{S/N} \ge 3$ for used lines.   
Top:   [\ion{N}{ii}] based BPT diagram for $z < 0.35$ with the upper limit for SF galaxies from  \protect\citet{Kewley_2006} (dashed),  the lower limit for AGNs from  \protect\citet{Kauffmann_2003b} (dotted), and  the separation between AGNs and LINERs from \protect\citet{Schawinski_2007} (dashed). 
Middle: KEx diagram with demarcation lines from \protect\citet{Zhang_2018}.
Bottom: WHAN diagram  with demarcation lines from \protect\citet{CidFernandes_2011}.
}
\label{fig:diagn_diagrams}
\end{figure}

\begin{figure}[htbp] 
\begin{center}
\includegraphics[viewport= 0 0 822 580,width=8.6cm,angle=0]{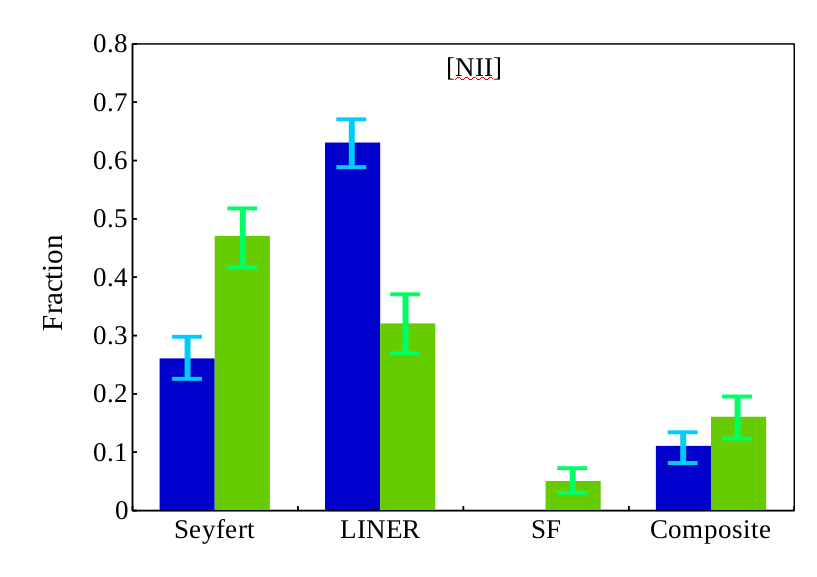}  \\
\includegraphics[viewport= 0 0 822 580,width=8.6cm,angle=0]{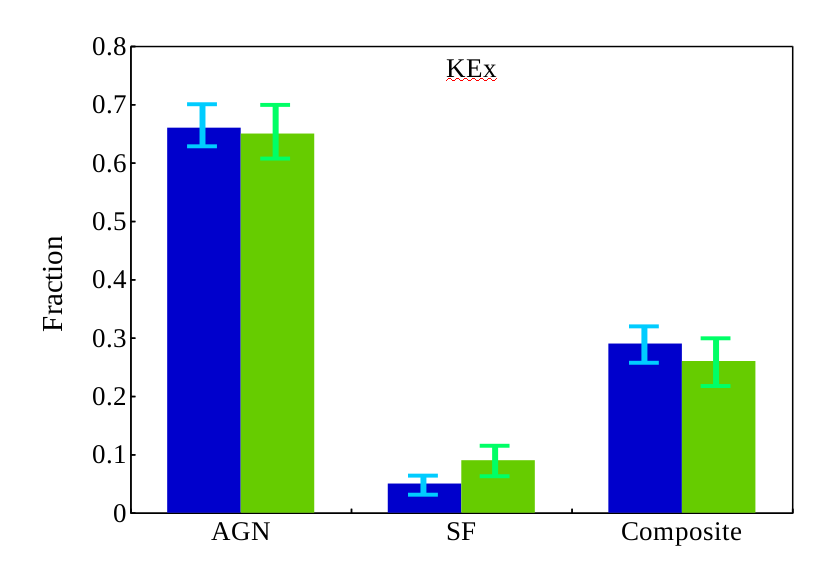} \\
\includegraphics[viewport= 0 0 822 580,width=8.6cm,angle=0]{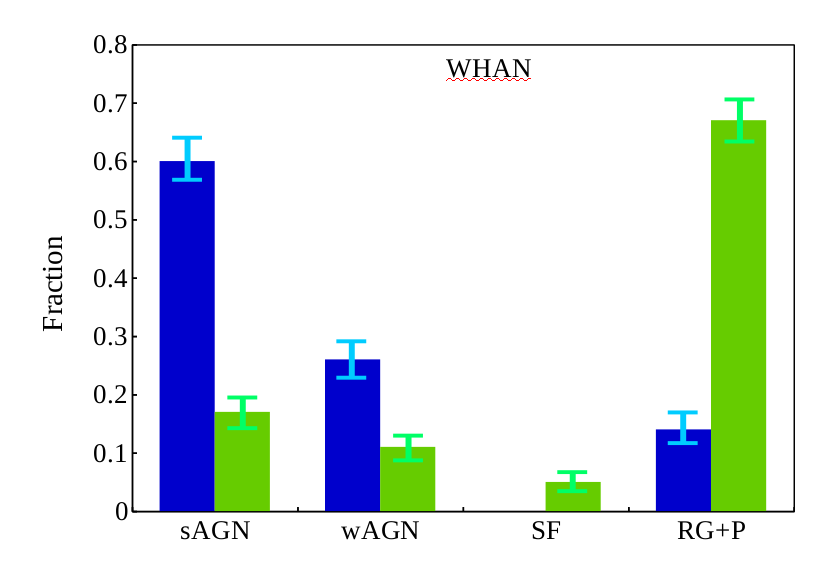}  
\end{center}
\caption{
Distributions of the types from the diagnostic diagrams in Fig.\,\ref{fig:diagn_diagrams} for the galaxies from the PSS-CSO  (blue) and the ECS (green) samples: [\ion{N}{ii}]-based BPT diagram (top),  KEx diagram (middle), and WHAN diagram (bottom).  The vertical bars indicate the standard errors of the proportions.
}
\label{fig:distribution_types}
\end{figure}

\begin{table}[htbp]
\caption{Properties (mean values with standard errors of the mean) 
of galaxies (SDSS spectral class = {\tt GALAXY}) from the PSS-CSO and the ECS samples. 
}
\begin{tabular}{lrr} 
\hline\hline 
\noalign{\smallskip}
                                                                                               & PSS-CSO sample            & ECS sample \ \ \  \\
\hline        
\noalign{\smallskip}
$\log\,[\ion{O}{iii}]$/H$\beta$\tablefootmark{\ a}            & $0.32\pm0.06$ (29)  & $0.41\pm0.08$ (19)  \\  
$\log\,[\ion{N}{ii}]$/H$\alpha$\tablefootmark{\ a}            & $0.10\pm0.03$ (29)   & $-0.05\pm0.04$ (32)   \\  
$N_{\rm HERG}/N_{\rm LERG}$\tablefootmark{\ a}                & $0.17\pm0.12$ (27)   & $0.50\pm0.35$ (19)   \\ 
\hline        
\noalign{\smallskip}
EW($[\ion{O}{ii}]$) ($\AA$)\tablefootmark{\ b}               &    $-27.4\pm2.7$ (51)   & $-12.1\pm2.1$ (29)    \\   
EW(H$\beta$)  ($\AA$)\tablefootmark{\ b}                      &   $-2.9\pm0.4 $  (35)  &  $-2.6\pm0.7$ (23)   \\
EW($[\ion{O}{iii}]$) ($\AA$) \tablefootmark{\ b}              &   $-8.9\pm1.3$ (49)  & $-11.1\pm0.4$ (29)   \\  
$\log\,[\ion{O}{iii}]$/H$\beta$\tablefootmark{\ b}            &    $0.41\pm0.06$ (38)  &  $0.5\pm0.1$ (23)   \\  
$\sigma_{\rm [OIII]}\,(\mbox{km\,s}^{-1})$\tablefootmark{\ b} &   $259\pm16$  (50)      & $208\pm15$  (38)   \\  
D4000n\tablefootmark{\ b}                                     &   $1.70\pm0.04$ (41)    &  $1.88\pm0.02$ (41)   \\  
\hline        
\noalign{\smallskip}
log\,${\mathcal M_\ast / \mathcal M_\odot}$\tablefootmark{\ c}& $11.28\pm0.06$ (19)  &  $11.23\pm0.06$ (27)   \\
log\,SFR ($\mathcal M_\odot$/yr)\tablefootmark{\ c}           &  $-1.64\pm0.45$ (19) &  $-2.08\pm0.36$ (27)   \\
log\,sSFR (Gyr$^{-1}$)\tablefootmark{\ c}                     &  $-3.92\pm0.49$ (19) &  $-4.32\pm0.38$ (27)   \\
\hline        
\noalign{\smallskip}
merger fraction $f_{\rm m}$\tablefootmark{\ d}                & $0.61\pm0.07$  (43)    & $0.15\pm0.06$  (41)     \\
\hline                                    
\end{tabular}             
\tablefoot{
The upper index in the first column refers to the selection criteria for the corresponding subsamples: 
\tablefoottext{a}{$z < 0.35$, S/N $>3$;}
\tablefoottext{b}{all galaxies with  S/N $>3$;}
\tablefoottext{c}{data from \protect\citet{Chang_2015} with {\tt flag} = 1;}
\tablefoottext{d}{low-$z$ galaxy sample ($z < 0.3$).}
The number of objects used for the statistics is given in brackets.
}
\label{tab:gal_subsamples}                    
\end{table}

The BPT diagrams of diagnostic line ratios \citep{Baldwin_1981},  in particular  the  [\ion{O}{iii}]\,5007\AA/H$\beta$  versus  [\ion{N}{ii}]\,6585\AA/H$\alpha$ diagram,  are commonly used to classify emission line galaxies. (Hereafter, we refer to these lines as [\ion{O}{iii}] and [\ion{N}{ii}].)  For line data from SDSS spectra, the diagram  applies to sources at $z \la 0.35$. The top panel of Fig.\,\ref{fig:diagn_diagrams} shows the [\ion{N}{ii}]-based BPT diagram for the galaxies (SDSS spectral class {\tt GALAXY}) with  $z < 0.35$ and a signal-to-noise ratio S/N $>3$ for all four lines.   The line ratios were computed from the line fluxes,  we did not try to correct for intrinsic reddening.   The fluxes and the equivalent width (EWs)  are available from the SDSS table {\tt galSpecLine} for the majority of the sources and were measured by us from the SDSS spectra for the rest. The mean ratios are listed in Table\,\ref{tab:gal_subsamples}.  The line ratios plotted in the  BPT diagram can be used to distinguish between high-excitation radio galaxies (HERGs) and low-excitation radio galaxies (LERGs). With the criterion of \citet{Buttiglione_2010} we find 23 LERGs and 4 HERGs among the PSS-CSO galaxies with $\mbox{S/N}>3$,  to be compared with 12 LERGs  and  6 HERGs for the ECS sample.  The ECS sample thus appears to contain a higher proportion of HERGs (Table\,\ref{tab:gal_subsamples}). However,  the ECS sample also contains a higher proportion of galaxies with low-S/N spectra that are not included in the diagram and  are most likely LERGs (see below).  All HERGs are classified as Seyferts in the BPT diagram.

\citet{Zhang_2018} proposed the so-called kinematics-excitation (KEx) diagram that uses the  [\ion{O}{iii}]  emission line dispersion, $\sigma_{\rm [\ion{O}{iii}]}$, in combination with the  [\ion{O}{iii}]/H$\beta$ line ratio and can be applied up to $z \sim 0.9$. The basic physics behind it is the difference  between AGNs and SF galaxies in the kinematics of the ionised gas in combination with the correlation between gas kinematics,  stellar mass, and  metallicity. 
(The [\ion{N}{ii}]/H$\alpha$ ratio  is known to correlate with the metallicity in SF galaxies.)  The KEx diagram for the galaxies from the PSS-CSO and the ECS samples is shown in the middle panel of Fig.\,\ref{fig:diagn_diagrams}.  Our data confirm the finding of previous studies  \citep[][and references therein]{Odea_2021} that compact sources have larger [\ion{O}{iii}]  line widths than extended ones.  According to the two-sided KS test with $\alpha = 0.05$, our PSS-CSO sample and the ECS sample have significantly different distributions of the line dispersion  $\sigma_{\rm [OIII]}$.  The mean value is 25\% higher for the former than for the latter  (Table\,\ref{tab:gal_subsamples}).   We applied the one-sided T test for two independent means to test 
${\rm H}^0: \overline{\sigma}_{\rm [OIII]}^{\rm \, PSS} = \overline{\sigma}_{\rm [OIII]}^{\rm \, ECS}$ against ${\rm H}^{\rm A}: \overline{\sigma}_{\rm [OIII]}^{\rm \, PSS} > \overline{\sigma}_{\rm [OIII]}^{\rm \, ECS}$ at a significance level of $\alpha = 0.05$. The result is that ${\rm H}^0$ must be clearly rejected in favour of ${\rm H}^{\rm A}$ ($p = 0.01$).

In both the BPT diagram and the KEx diagram, the two samples show similar distributions of the line ratios. The KS test shows that there is no reason to reject the null hypothesis that the PSS-CSO and the ECS samples  come from the same distribution  at a significance level $\alpha = 0.05$.  However, the selection constraint S/N $>3$ excludes a significant proportion of galaxies in both samples.  In the BPT diagram, 17\% of the PSSs and CSOs are missing because they have only weak emission lines, especially weak H$\beta$ lines. For the ECS sample, however, this proportion is about twice as high (39\%). This difference is significant according to the two-sided Z test ($p = 0.005$).

The bottom panel of Fig.\,\ref{fig:diagn_diagrams} shows the WHAN diagram \citep{CidFernandes_2011}, which uses only the EW of the H$\alpha$ line in combination with the line ratio [\ion{N}{ii}]/H$\alpha$.  It was developed as  a robust and economic classification scheme to cope with the large population of weak-line galaxies in SDSS.  Following \citet{CidFernandes_2011}, five different regions can be distinguished: pure star-forming galaxies (SF), strong AGNs (sAGN; e.g. Seyferts), weak AGNs (wAGN; e.g. LINERS), retired galaxies (RG; i.e. galaxies that have stopped SF and are ionised by evolved low-mass stars), and passive, basically lineless  galaxies.   The majority of the ECS galaxies, 67\%, are classified as retired or passive, compared to 14\% in the PSS-CSO sample. 
This difference is highly significant  and the KS test yields a significant difference between the EW(H$\alpha$) distributions of the two samples ($p < 10^{-5}$).

Next, we calculated the number of objects in the regions indicated by the demarcation lines of the three diagnostic diagrams.  The resulting histogram distributions are shown in Fig.\,\ref{fig:distribution_types}.  To find out whether these differences are significant or just by chance, we used Pearson's chi-square test for two independent samples.  There is no reason to reject the null hypothesis of no difference between the PSS-CSO and the ECS samples at $\alpha = 0.05$  for the  [\ion{N}{ii}]-based BPT diagram and the  KEx diagram. However, the WHAN diagram reveals a highly significant difference ($p = 4\cdot10^{-5}$) with a significantly larger proportion of weak-line galaxies in the ECS sample.

%
\section{Spectral energy distribution}\label{sect:SED}
%

\begin{figure*}[h]
\begin{center}
\includegraphics[width=0.43\textwidth,angle=270]{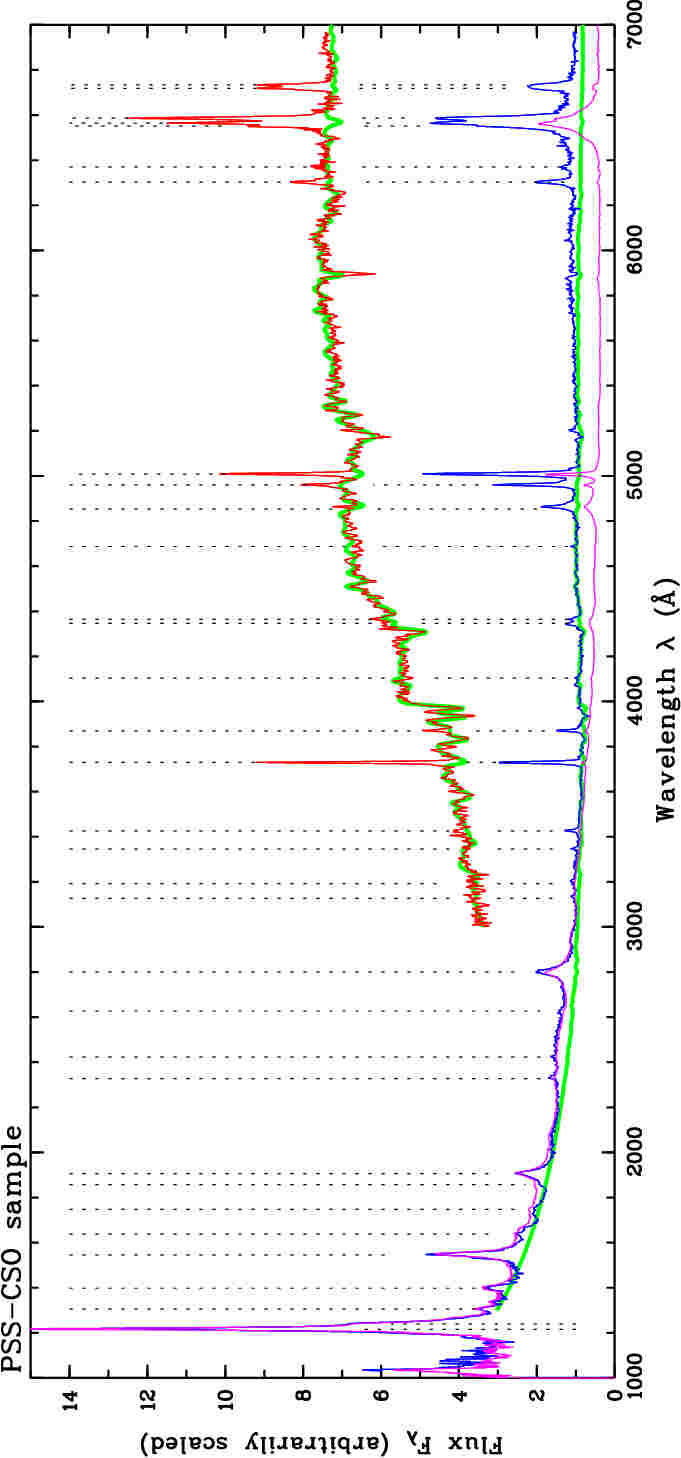} \\
\includegraphics[width=0.43\textwidth,angle=270]{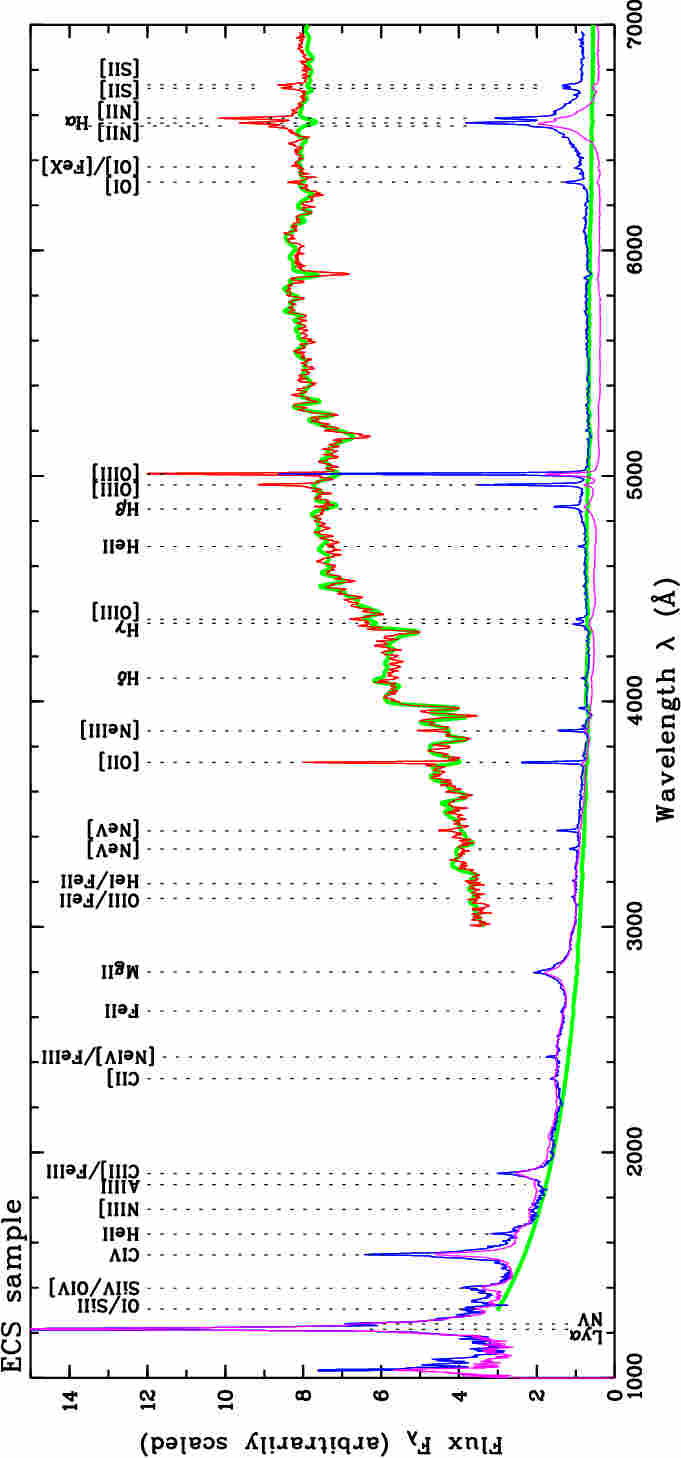} 
\end{center}
\vspace{3mm}
\caption{
Composite spectra for the PSS-CSO (top) and the ECS (bottom) sample.
The QSO subsamples are shown in blue, the galaxy samples in red.
The former were corrected for slight internal reddening using the SMC dust model to match 
the composite spectrum of the FBQS radio-loud quasars from \protect\citet{Brotherton_2001} (magenta). 
The spectra for the galaxy subsamples were shifted upwards by 2.5 flux units for clarity. 
The green spectra are best-fit models for the continuum (see text). 
The positions of the strongest emission lines are indicated by dotted vertical lines and labelled by ion in the lower panel.
}
\label{fig:composites}
\end{figure*}


\subsection{SDSS composite spectra}\label{sect:SDSS_comp}


\begin{table*}[htbp]
\caption{Diagnostic line ratios measured from the composite spectra.}
 \begin{tabular}{lccccc} 
\hline\hline 
\noalign{\smallskip}
sample 
                          & $\log\,[\ion{O}{ii}]/[\ion{Ne}{v}]$ 
                          & $\log\,[\ion{O}{ii}]/[\ion{O}{iii}]$     
                          & $\log\,[\ion{O}{iii}]/\mbox{H}\beta$                           
                          & $\log\,[\ion{N}{ii}]/\mbox{H}\alpha$
                          & FWHM [OIII]\,(\AA)
                          \\
\hline        
\noalign{\smallskip}
PSS-CSO-Q & $0.88 \pm 0.04$ &  $-0.33 \pm 0.03$ & \ \ $1.01 \pm 0.10$ & \ \ $0.26 \pm 0.07$    & $13.1 \pm 0.3$ \\         
ECS-Q         & $0.52 \pm 0.01$ &  $-0.65 \pm 0.02$  & \ \ $1.03 \pm 0.04$ & $-0.16 \pm 0.05$       & $8.1 \pm 0.2$  \\    
\hline
\noalign{\smallskip}
PSS-CSO-G & $1.09 \pm 0.09$ & \ \ $0.09 \pm 0.03$ & \ \ $1.29 \pm 0.24$ & \ \ $0.13 \pm 0.05$    & $7.6 \pm 0.2$ \\         
ECS-G         & $0.75 \pm 0.08$ &   $-0.11 \pm 0.06$ & \ \ $1.63 \pm 0.20$ & \ \ $0.08 \pm 0.05$    & $7.3 \pm 0.1$ \\  
\hline                                    
\end{tabular}             
\label{tab:line_ratios}                    
\end{table*}

Composite spectra provide insight into average spectral properties of AGN samples   \citep[e.g.][]{Francis_1991, Brotherton_2001, Vanden_Berk_2001, Reichard_2003, Pol_2017, Ren_2022}.  We followed the approach described in detail by \citet{Vanden_Berk_2001}.  The spectra were downloaded from SDSS DR16  and inspected individually to check the redshifts and spectral classification (see also Sect.\,\ref{sect:PSS_sample}). Next, the spectra were corrected for foreground extinction using $E(B-V)$ from \citet{Schlafly_2011} provided by IRSA's Galactic Dust Reddening and Extinction Service\footnote{https://irsa.ipac.caltech.edu/applications/DUST/docs/background.html}.  The corrected spectra were rebinned  to  a  common  wavelength  scale  in  their  restframe. Afterwards, the corrected spectra  were normalised,  i.e. the spectra were ordered by redshift, the flux density of the lowest-$z$ spectrum was arbitrarily scaled and the other spectra were scaled in order of redshift to the average of the flux density in the common wavelength region of the mean spectrum of all the lower-$z$ spectra.  Finally, the median flux density  and the standard deviation of all scaled spectra were determined in each wavelength bin.

It is useful to consider the composite spectra of quasars and galaxies separately.  Therefore, we subdivided  the PSS-CSO  sample into two sub-samples differentiated by spectral class (G = {\tt GALAXY} and Q = {\tt QSO}), i.e. PSS-CSO-Q and PSS-CSO-G.  In the same way,  the ECS sample is subdivided into the two subsamples ECS-Q and ECS-G. The final  arithmetic median composite rest-frame spectra are shown in Fig.\,\ref{fig:composites} with the two PSS-CSO composites at the top and the ECS composites at the bottom.  For comparison, the composite spectrum of  radio-loud quasars from the FIRST Bright QSO Survey \citep[FBQS;][]{Brotherton_2001} is overplotted (magenta). All composite spectra have been scaled such that the mean flux density has the value of 1 in the QSO continuum window $3030 - 3090$\,\AA, but the spectra for the galaxy subsamples were shifted upwards by 2.5 flux units for clarity.  The continua of the four composites were fitted by model spectra (see Sect.\,\ref{sect:continuum}), the best-fitting model spectra are also plotted (green). At first glance, the composites for the same spectral class are quite similar. The more subtle differences are discussed below.


\subsubsection{Emission lines and line ratios}\label{sect:lines}


The arithmetic median preserves the relative fluxes along the spectra, the composites can thus be used to measure  diagnostic line ratios. We measured the (arbitrarily scaled) line fluxes from our composite spectra after subtraction of a modelled continuum (Sect.\,\ref{sect:continuum}). The [\ion{N}{ii}]-H$\alpha$ complex was de-blended using a four-component Gaussian fit with a broad H$\alpha$ component and three narrow components for H$\alpha$,  [\ion{N}{ii}]\,$\lambda$\,6549\AA, and [\ion{N}{ii}]\,$\lambda$\,6585\AA. For the measurement of the H$\beta$ line, a two-component Gaussian fit was used for the narrow and the broad line components. Only the narrow Balmer components were used for the line ratios [\ion{N}{ii}]/H$\alpha$ and  [\ion{O}{iii}]/H$\beta$.  The results (Table\,\ref{tab:line_ratios}) can be compared with the diagnostic diagrams in Sect.\,\ref{sect:line_ratios}. For all four subsamples,  the line ratios from the composite spectra are clearly in the region of strong AGNs in the BPT and the KEx diagram. In accordance with Table\,\ref{tab:gal_subsamples}, the  ECS-G sample has a higher line ratio $[\ion{O}{iii}]/\mbox{H}\beta$  than the PSS-CSO-G sample.\footnote{We note that the line ratios $[\ion{O}{III}]/\mbox{H}\beta$ and $[\ion{N}{II}]/\mbox{H}\alpha$ from the composite spectra are larger than those in  Table\,\ref{tab:gal_subsamples}. This is only partly due to the different ways of averaging (logarithmic mean in  Table\,\ref{tab:gal_subsamples}, arithmetic median in Table\,\ref{tab:line_ratios}). The main difference lies in the inclusion of $S/N < 3$ spectra, and thus of objects with weak Balmer lines in the composite spectra. The difference is marginal for the QSO subsamples.} In the WHAN diagram, three subsamples are also in the region of strong AGNs, while the ECS-G subsample is close to the border between weak AGNs and retired galaxies, which reflects the weak H$\alpha$ line in the composite spectrum.

All four composites in Fig.\,\ref{fig:composites} show relatively strong forbidden lines [\ion{O}{iii}]\,$\lambda$\,5007\AA\ and [\ion{O}{ii}]\,$\lambda$\,3726\AA\ \footnote{This is actually a line doublet at 3725\,\AA\  and 3727\,\AA.}  (hereafter  [\ion{O}{iii}] and [\ion{O}{ii}]). 
\citet{Liao_2020} found a significant correlation between the  [\ion{O}{III}] line luminosity and the 5\,GHz luminosity in their PSS-CSO sample. A correlation of the relative strength of [\ion{O}{iii}], radio loudness and radio luminosity is known also from other studies \citep{Andika_2020, Wang_2022}.   With this in mind, strong lines are to be expected because the radio sources in our samples are radio-loud and belong to the upper end of the radio luminosity distribution at their redshift (Sect.\,\ref{sect:comp_sample}). Radio jets may play an important role in ionising the  narrow line region (NLR) gas and  the [\ion{O}{iii}] and [\ion{O}{ii}] lines could be significantly amplified by shocks caused by strong jet interaction with the line emitting gas clouds  \citep[e.g.][]{Best_2000, Labiano_2008, Kalfountzou_2012, Mullaney_2013, Andika_2020}. The higher $[\ion{O}{iii}]/\mbox{H}\beta$ ratio of the ECS-G sample compared with the PSS-CSO-G sample  is consistent with the idea  that the  [\ion{O}{iii}] line strength is  enhanced significantly by the jet expansion through the host galaxy \citep{Labiano_2008}. In this context, we also note that the blends of \ion{Fe}{ii} lines and Balmer continuum between $\sim 2000$ and 4000\,\AA\ (the `3000\,\AA\ bump') are relatively weak in both QSO composites of the present study, in agreement with the known anti-correlation between the [\ion{O}{iii}] line and the strength of the  \ion{Fe}{ii} emission \citep{Boroson_1992}.

In Sect.\,\ref{sect:line_ratios}, we found that the mean [\ion{O}{iii}] line dispersion  for the PSS-CSO galaxies is 25\% larger than for the ECS galaxies, but these galaxy samples are biased against spectra with weak H$\beta$ lines (and thus high [\ion{O}{iii}]/H$\beta$ ratios).  In the PSS-CSO-G composite spectrum, the  FWHM of the [\ion{O}{iii}] line  is also larger than that of the ECS-G composite, but the difference is now only $\sim 5$\% (Table\,\ref{tab:line_ratios}).

The host galaxies of radio-loud QSOs are known to have elevated [\ion{O}{ii}] with respect to hosts from radio-quiet QSOs.  The  [\ion{O}{ii}] line is stronger in the PSS-CSO composites, in agreement with the fact that its emission-line strength correlates inversely with  the  radio  source size \citep{Best_2000}.  The line ratio  [\ion{O}{ii}]/[\ion{O}{iii}] is sensitive to the ionisation parameter \citep{Penston_1990, Kewley_2002, Shirazi_2014}.   The ratio   [\ion{O}{ii}]/[\ion{O}{iii}]   for the ECS-Q subsample is close to that for the SDSS quasar composite spectrum \citep{Vanden_Berk_2001}. The value for the  PSS-CSO-Q subsample is  twice as high and thus indicates a lower ionisation. This is again in agreement with the  interpretation of the strength of the [\ion{O}{iii}] line.  The same trend is indicated by the comparison of the PSS-CSO-G and ECS-G composites. For the PSS-CSO-G subsample the line ratio [\ion{O}{ii}]/[\ion{O}{iii}] is slightly larger than 1, as typical of LINERs.

The [\ion{O}{ii}] line is often taken as a SF indicator, but it can also be emitted by low-density gas ionised by the AGN. On the other hand, 
the [\ion{Ne}{v}]\,$\lambda$\,3425\AA\ line is unambiguously related to AGN activity, either via direct photo-ionisation from the AGN or ionisation from AGN-driven shocks.  The [\ion{O}{II}]/[\ion{Ne}{v}] line ratio can thus be taken as an indicator of whether there is an excess of [\ion{O}{II}] emission that can be attributed to  radio jets acting within the host galaxies or to SF in the host galaxies \citep{Maddox_2018}.   This excess is higher in  the PSS-CSO than in the ECS  samples (Table\,\ref{tab:line_ratios}).

\begin{figure}[htbp] 
\includegraphics[viewport= 0 20 580 820,width=6.7cm,angle=270]{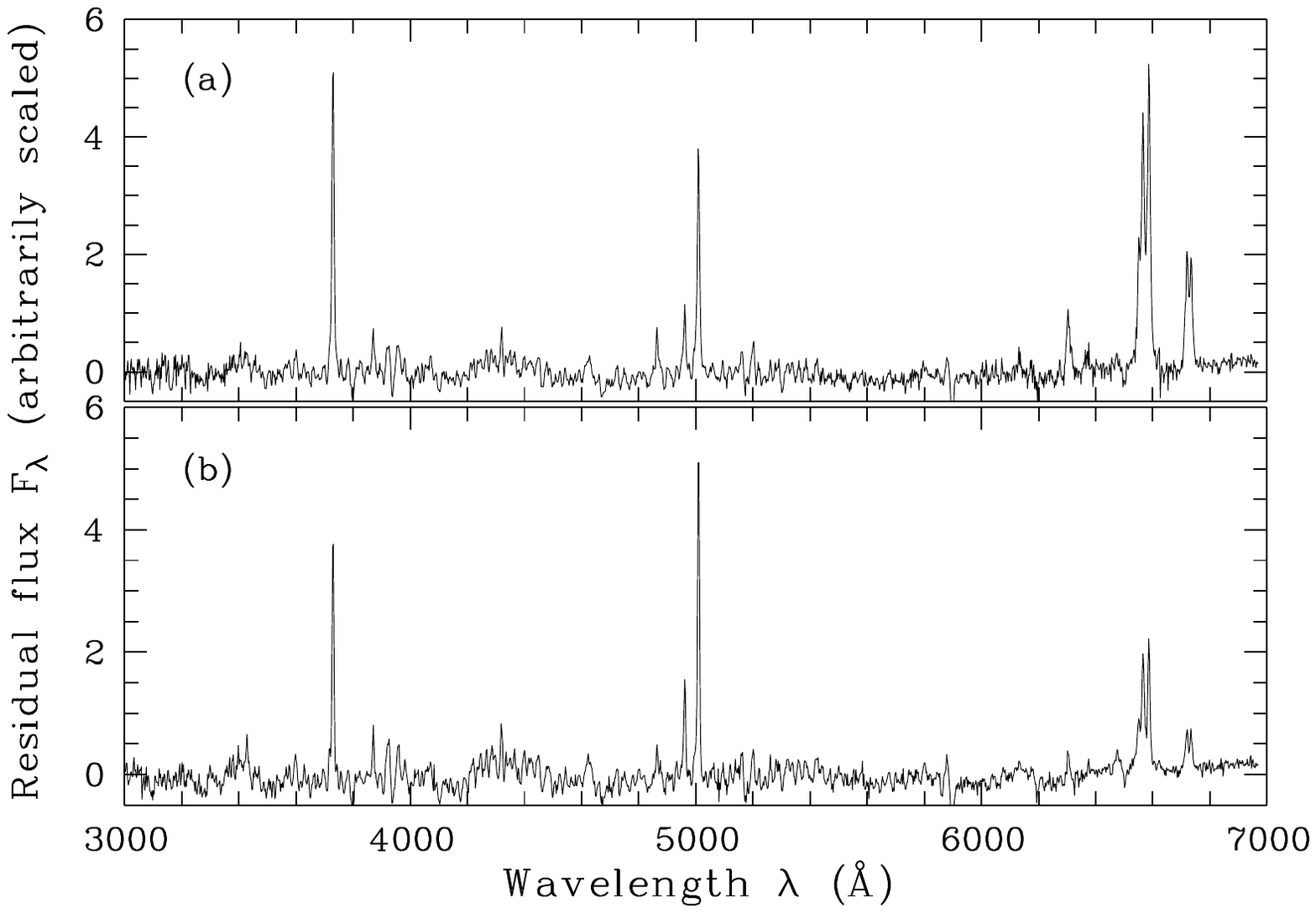}
\caption{
Residual spectra  (composite minus model) for the PSS-CSO-G (a) and the ECS-G (b) subsample.  The model spectra were arbitrarily scaled in such a way that the mean flux in the interval 4200-4800\,\AA\ is the same as in the corresponding observed composite spectrum. 
}
\label{fig:residuals}
\end{figure}


\subsubsection{Continuum and absorption lines}\label{sect:continuum}


The composite spectra of the galaxy subsamples are dominated by the light from an old stellar population.  There is, however,  a discernible difference  between the PSS-CSO-G and the ECS-G composite: the former has a flatter slope between $\sim 3000$ and 4500\,\AA,  indicating an additional blue component. We fitted the corresponding composites by model spectra consisting of two stellar components represented by the template spectra of a 13 Gyr old E galaxy (Ell 13) and an Sc galaxy from  \citet{Polletta_2007}, where their relative proportion is a free parameter. Another free parameter is the internal reddening $E(B-V)$ of the starlight assuming the reddening law for Milky Way dust  from \citet{Pei_1992}. In reality, the stellar populations of the host galaxies may show a much broader variety  and we cannot assume that such a simplified model  perfectly reflects the observed spectrum in all its details \citep[e.g.][]{Schmitt_1999}. Nevertheless, we find that  the model provides a reasonable fit to the continuum in the wavelengths range $\lambda \approx 3000$ to  7000\,\AA. In the best-fit model for the PSS-CSO-G sample, the Sc component  contributes $\sim 40\pm 3\,\%$ of the starlight at $3000-3200$\,\AA, compared to only $\sim 7\pm 3\,\%$ for the ECS-G sample. The internal reddening is slightly stronger in the PSS-CSO-G model, $E(B-V) = 0.07$ compared to 0.03 for  ECS-G. The differences between the observed composites and the model spectra are shown in Fig.\,\ref{fig:residuals}.  We note again the stronger [\ion{O}{ii}] line in the PSS-CSO-G sample, in accordance with Sect.\,\ref{sect:lines} and Table\,\ref{tab:gal_subsamples}.

In an alternative model, we described the blue component by a power-law $F_\lambda \propto \lambda^{\alpha_\lambda}$ for the  AGN continuum with $\alpha_\lambda = -0.5$ for radio-loud AGNs \citep{Brotherton_2001}.  Here, the fraction of the AGN continuum is a free parameter. This model provides a similarly good match when the AGNs contribute $\sim 20\%$ to the flux at $3000 - 3200$\,\AA\ in the PSS-CSO-G sample, but negligibly little in the ECS-G sample.  A moderate internal reddening of $E(B-V) \la 0.05$  cannot be excluded but does not significantly improve the fits.

Adding the blue component has the effect of reducing the discontinuity in  the model spectrum at $\sim 4000$\,\AA. For a single stellar population, the strength of this discontinuity increases with age, but it depends also on metallicity  \citep[e.g.][]{Bruzual_1983, Vazdekis_2015} and, due to the mass-metallicity relation \citep{Tremonti_2004}, on stellar mass \citep{Haines_2017}.  Because of the large stellar masses of the host galaxies in our samples (Table\,\ref{tab:gal_subsamples} and Sect.\,\ref{sect:SFR}), it may be that the 4000\,\AA\ discontinuity is stronger than in the Ell13 template spectrum, which would increase the necessary contribution from the blue component. However, this effect should be similar for both subsamples.

To summarise, the mean stellar populations of the hosts in the PSS-CSO-G and ECS-G subsamples appear to be slightly different. The flatter slope of the PSS-CSO-G composite at $\sim 3000 - 4500$\,\AA\  suggests a stronger contribution from a blue component,  which could be either a younger stellar population or AGN continuum. If PSSs and CSOs  are young, the former would suggest a link between triggering radio activity and star formation. The latter could also be understood as an evolutionary effect:  Let us assume that a percentage of $\sim 20$\% of the AGNs in the PSS-CSO-G sample are in an accretion mode with a standard accretion disk.  If the  lifetime  of such an AGN phase (i.e. the accretion disk) is smaller than the growth time of  the extended radio structure, we can assume that the percentage of such AGNs in the ECS sample is much smaller.

Because the QSOs in our samples are radio loud, the QSO composites can be compared with the  composite spectrum of the about 400 radio-loud quasars from the FBQS. 
Compared to the latter, our QSO composites have a slightly redder (i.e. flatter in Fig.\,\ref{fig:composites}) UV continuum.  The continuum slope can be influenced by several physical properties, such as intrinsic reddening, inclination of the accretion disk, accretion rate,  SMBH mass and spin \citep{Francis_1992, Richards_2002, Reichard_2003, Shankar_2016}, or by observational effects \citep[e.g.][]{Harris_2016, Milakovic_2021}.  Here, we assume that reddening by dust in the host galaxy is the most plausible cause,  although we cannot exclude that spectrum-to-spectrum variations in our small samples play a role.  We de-reddened our composite spectra using the SMC reddening curve \citep{Pei_1992} and found a good agreement of the UV continuum of both spectra with that of the FBQS composite if we assume an additional intrinsic reddening of $\Delta E(B-V) =$ 0.04. We note that Fig.\,\ref{fig:composites} shows the de-reddened spectra.

The change of the slope of the QSO composites at $\lambda \ga 4000$\,\AA\ can be attributed to the stellar light from the host galaxies,  as indicated by the easily recognisable stellar absorption lines  \ion{Ca}{ii} H and K,  \ion{Mg}{i} at 5190\,\AA,  and \ion{Na}{ii} at 5893\,\AA. Evidently, the composite spectra  can be described as a combination of an emission-line spectrum, a featureless continuum, and an absorption-line galaxy spectrum. We modelled the continuum plus stellar light in a similar way as described above for the galaxy composites and found an acceptable fit  up to $\sim 6000$\,\AA\ if we assume that the AGN continuum accounts  for $87\pm 3\,\%$ in the PSS-CSO-Q  and $93\pm 3\,\%$ in the ECS-Q composite  at 3000 to 3200\,\AA.  The uncertainty is a few percent, so the difference is probably not significant.  For both QSO subsamples,  the stellar continuum is well described by the Ell\,13 template, the contribution from the Sc template is nearly negligible (a few per cent).  At wavelengths $\ga 6000$\,\AA, the models underestimate the observed continuum flux.  Such a trend of a greater contribution  from starlight with increasing wavelength, which is also seen in other QSO composites, could be caused by emission from hot dust \citep[][and references therein]{Vanden_Berk_2001} and by the high mass and correspondingly high metallicity of the host galaxies \citep{Zhu_2010}.

Some particular `features'  in QSO spectra have sometimes been associated with early evolutionary phases of AGNs,  such as weak broad emission lines \citep[e.g.][]{Hryniewicz_2010, Meusinger_2014, Kumar_2022} or broad absorption lines \citep[BALs; e.g.][]{Priddey_2007, Bruni_2015, Zeilig-Hess_2020}.  For example, iron low-ionisation BALs (FeLoBALs) were considered to represent an early-stage of merger-driven accretion and a transition  between an obscured AGN and an ordinary optical QSO  \citep[e.g.][]{Voit_1993, Farrah_2010, Wethers_2021}.\footnote{For contrary results see e.g. \citet{Violino_2016, Villforth_2019}.} BAL properties (the `balnicity') and radio emission may be linked to the same underlying process \citep{Morabito_2019}. Strong iron emission bands provide another type of unusual spectral feature.  FeLoBAL QSOs often exhibit impressive \ion{Fe}{ii} and \ion{Fe}{iii} emission bands in the UV \citep{Hall_2002, Meusinger_2012}.   Strong optical \ion{Fe}{ii} emission is one of the classification characteristics of the rare type of Narrow Line Seyfert 1 (NLS1) galaxies \citep{Goodrich_1989, Veron_2001}. NLS1s harbour relatively low-mass SMBHs ($\mathcal{M}_\bullet \la 10^8  \mathcal{M}_\odot$) that accrete at high Eddington ratios and are interpreted as young or rejuvenated AGNs  \citep{Komossa_2006, Mathur_2012}. Radio-loud NLS1s show similarities with CSS sources \citep[e.g.][]{Yuan_2008, Caccianiga_2014, Berton_2017}.  \citet{Liao_2020} identified the CSS source \object{SDSS\,J133108.29+303032.9} as a NLS1,  \citet{Jaervelae_2022} found five new NLS1s that exhibit properties consistent with the CSS classification. We see in our PSS-CSO composite spectrum neither signs of conspicuously weak broad emission lines nor strong Fe multiplets in emission or absorption.  In addition, we note that there is no indication of a substantial unobscured post-starburst stellar population of an age of less $\sim 1$ Gyr,  which would be indicated by the Balmer absorption lines of A- and F-type stars, as has been observed in a few AGN galaxies \citep{Liu_2007, DePropris_2014}.

\begin{figure*}[htbp]
\begin{center}
\includegraphics[viewport= 0 30 590 790,width=4.7cm,angle=270]{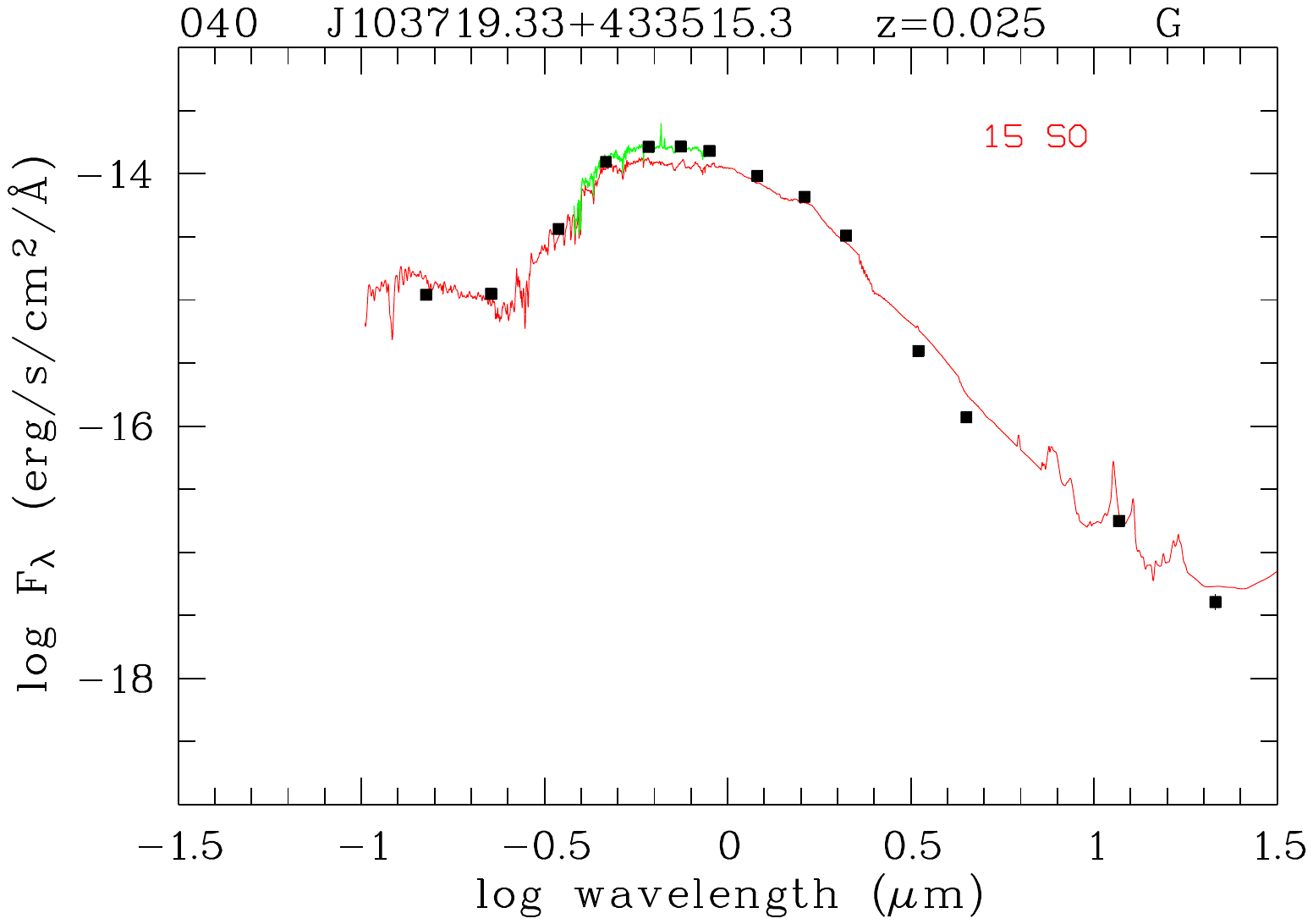}
\includegraphics[viewport= 0 30 590 790,width=4.7cm,angle=270]{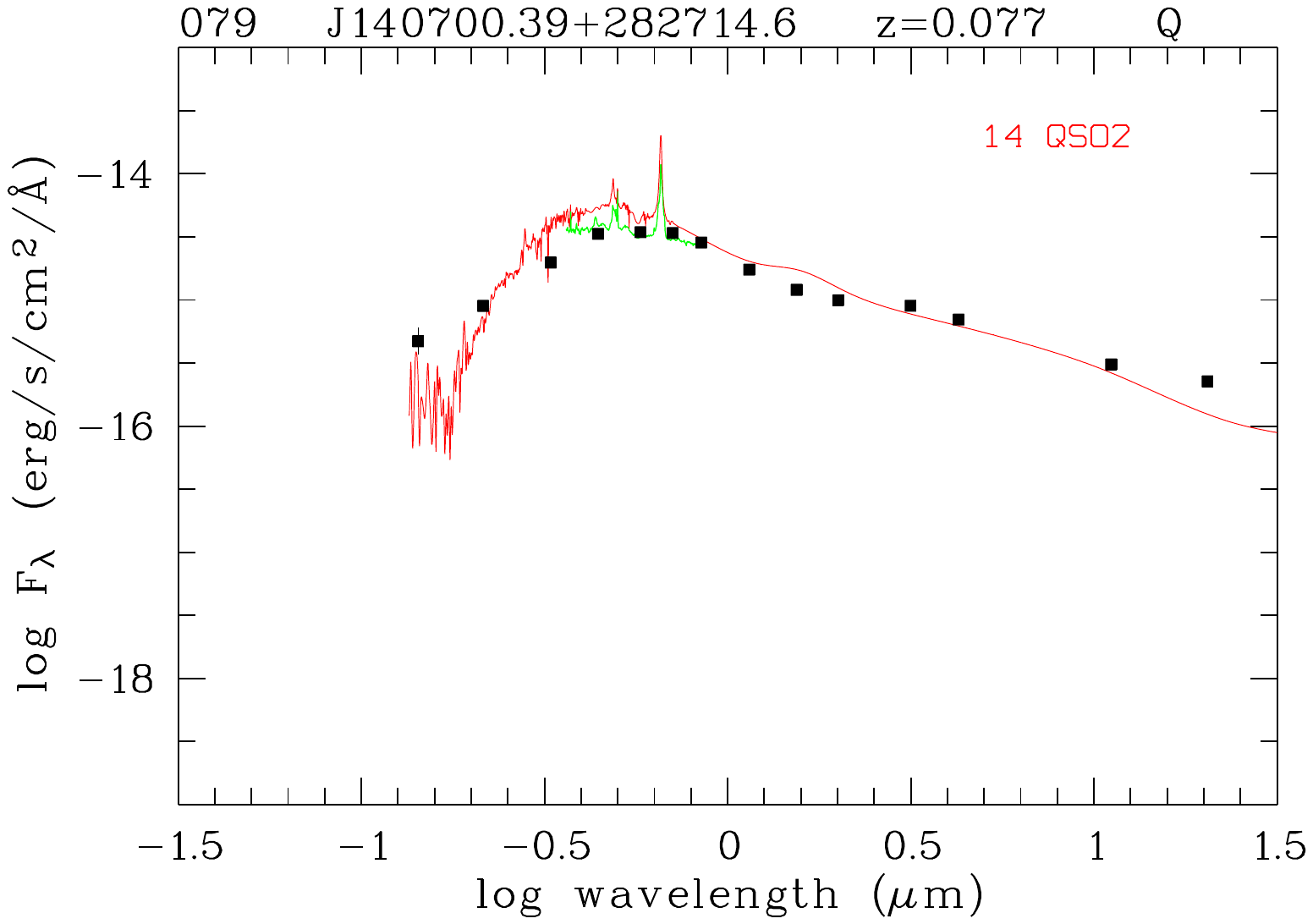}
\includegraphics[viewport= 0 30 590 790,width=4.7cm,angle=270]{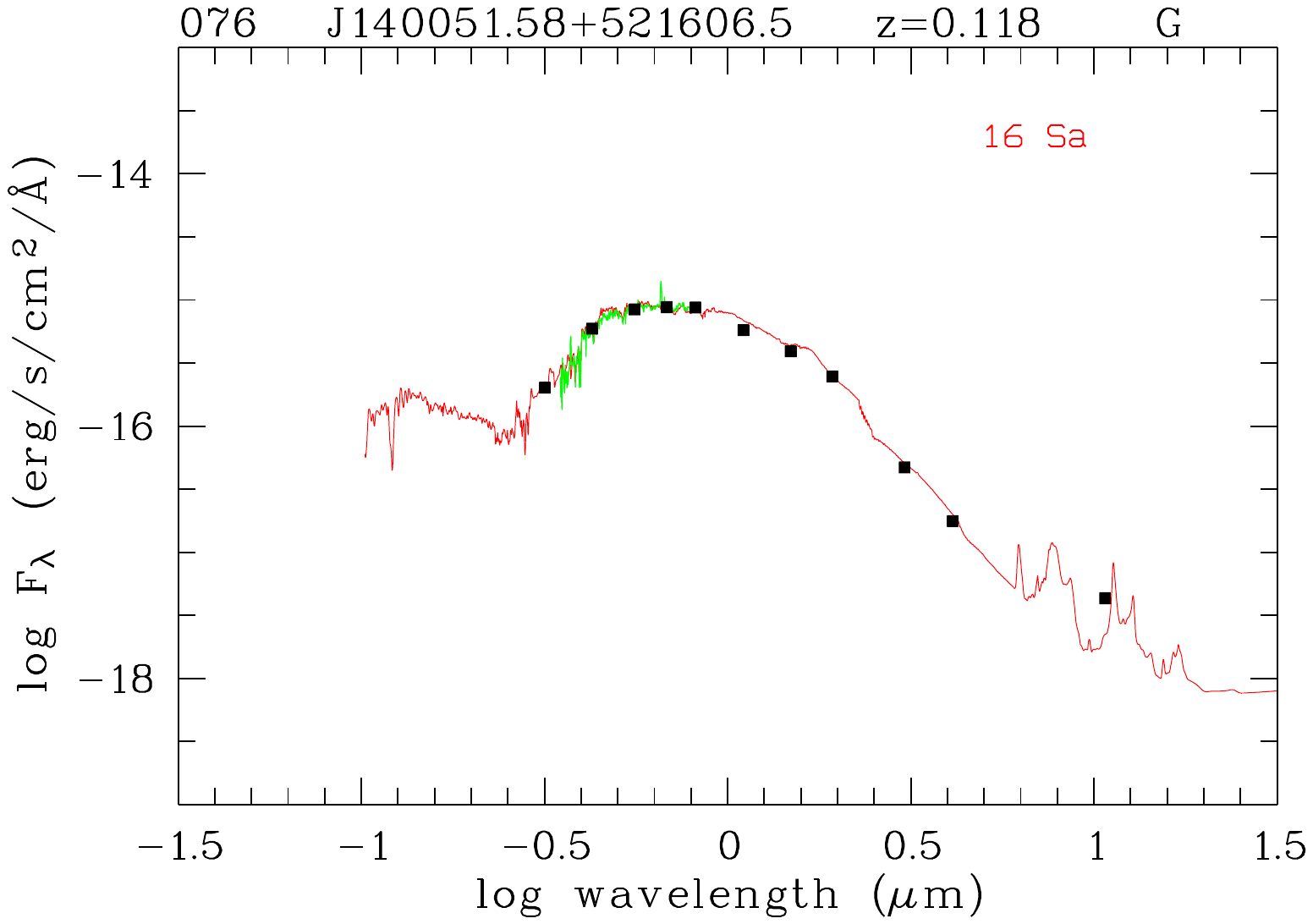}\\
\includegraphics[viewport= 0 30 590 790,width=4.7cm,angle=270]{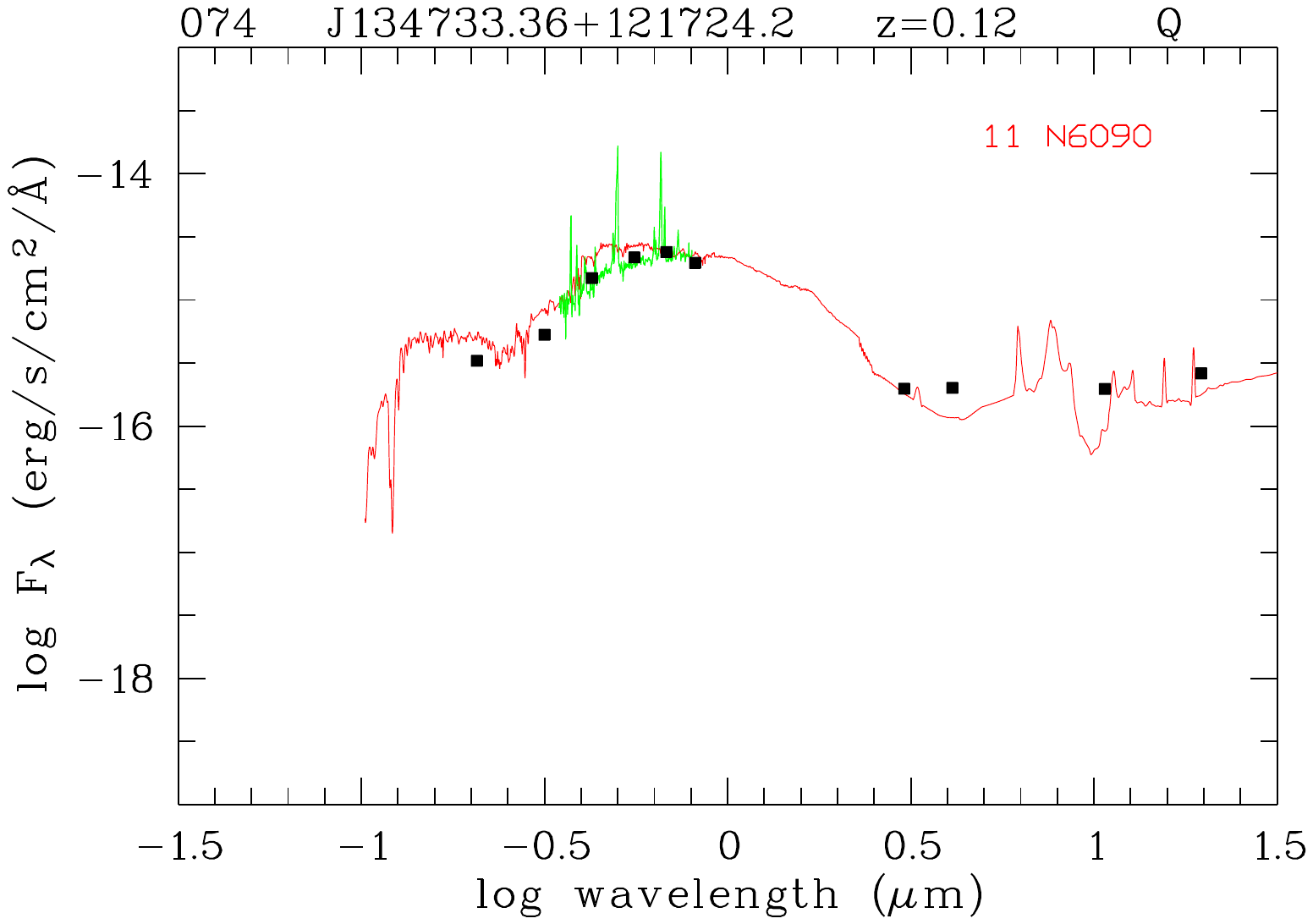}
\includegraphics[viewport= 0 30 590 790,width=4.7cm,angle=270]{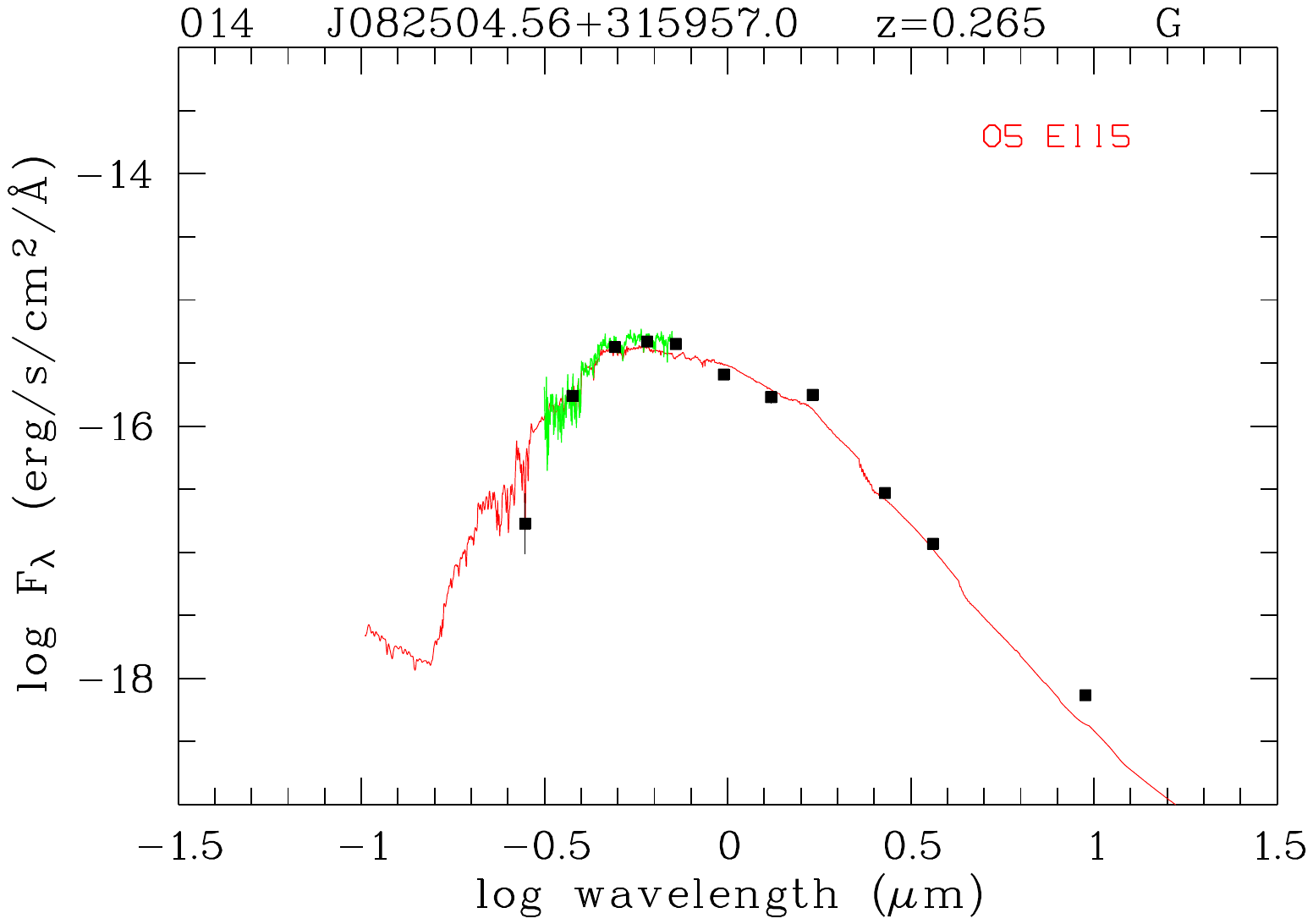}
\includegraphics[viewport= 0 30 590 790,width=4.7cm,angle=270]{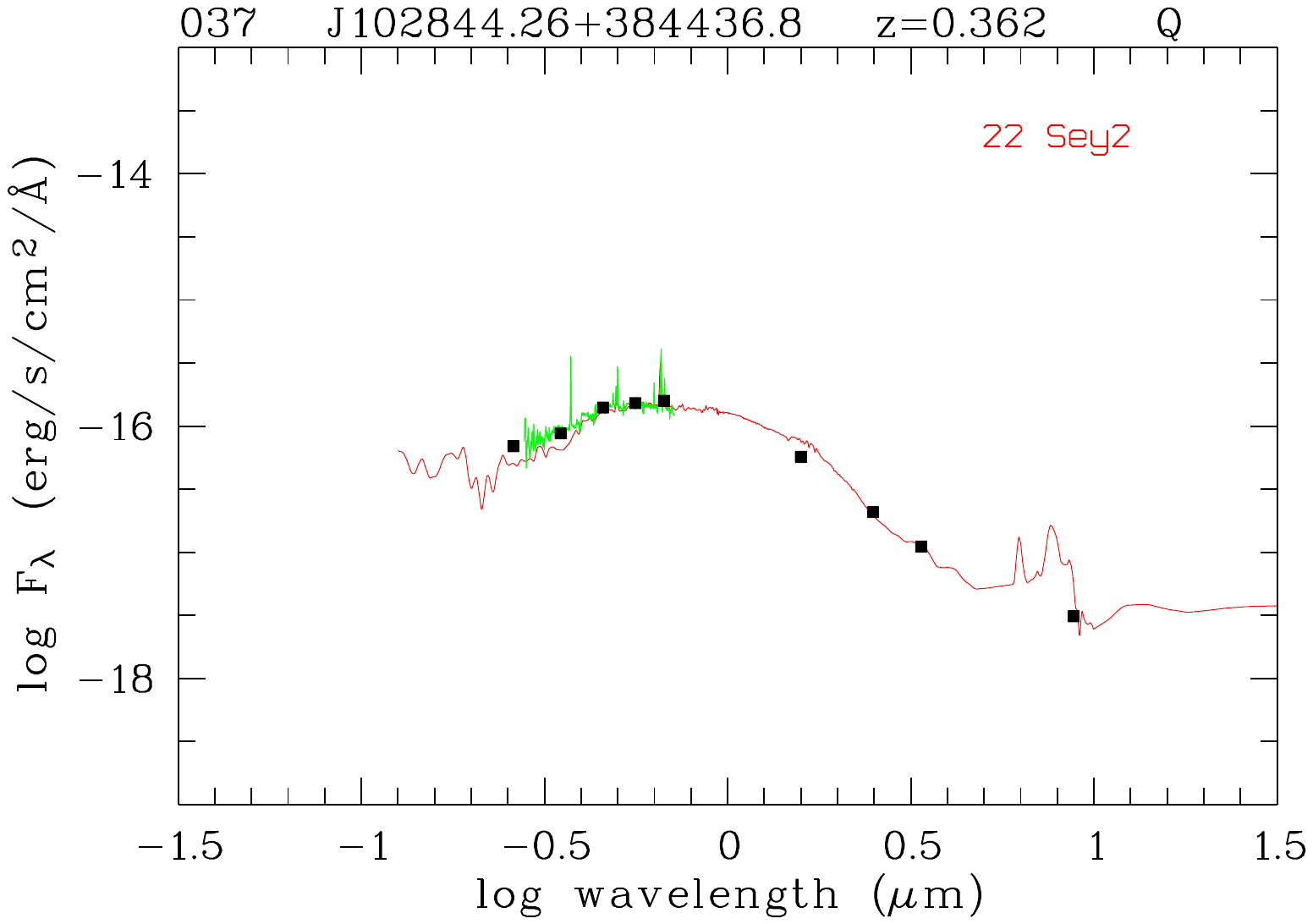}\\
\includegraphics[viewport= 0 30 590 790,width=4.7cm,angle=270]{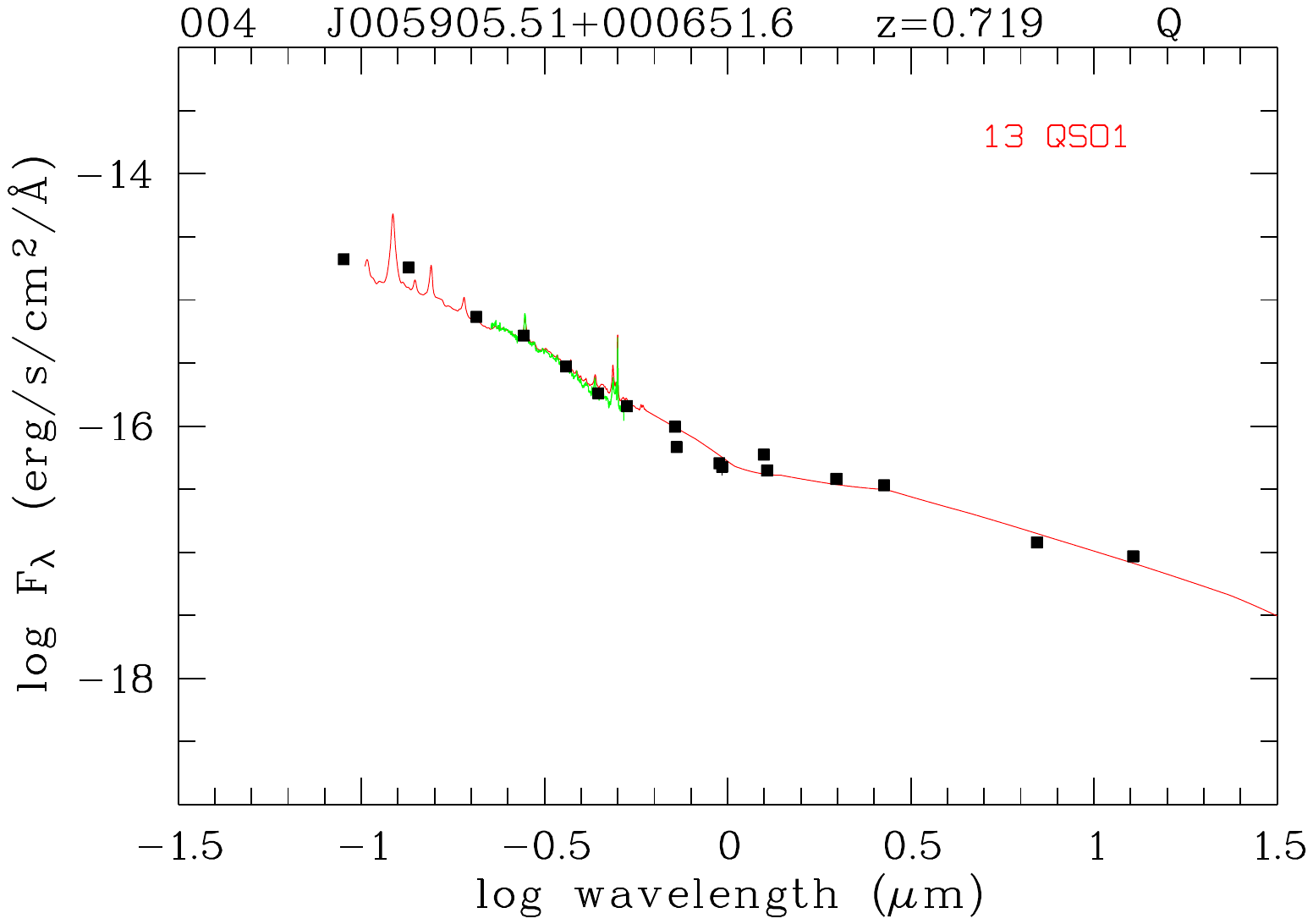}
\includegraphics[viewport= 0 30 590 790,width=4.7cm,angle=270]{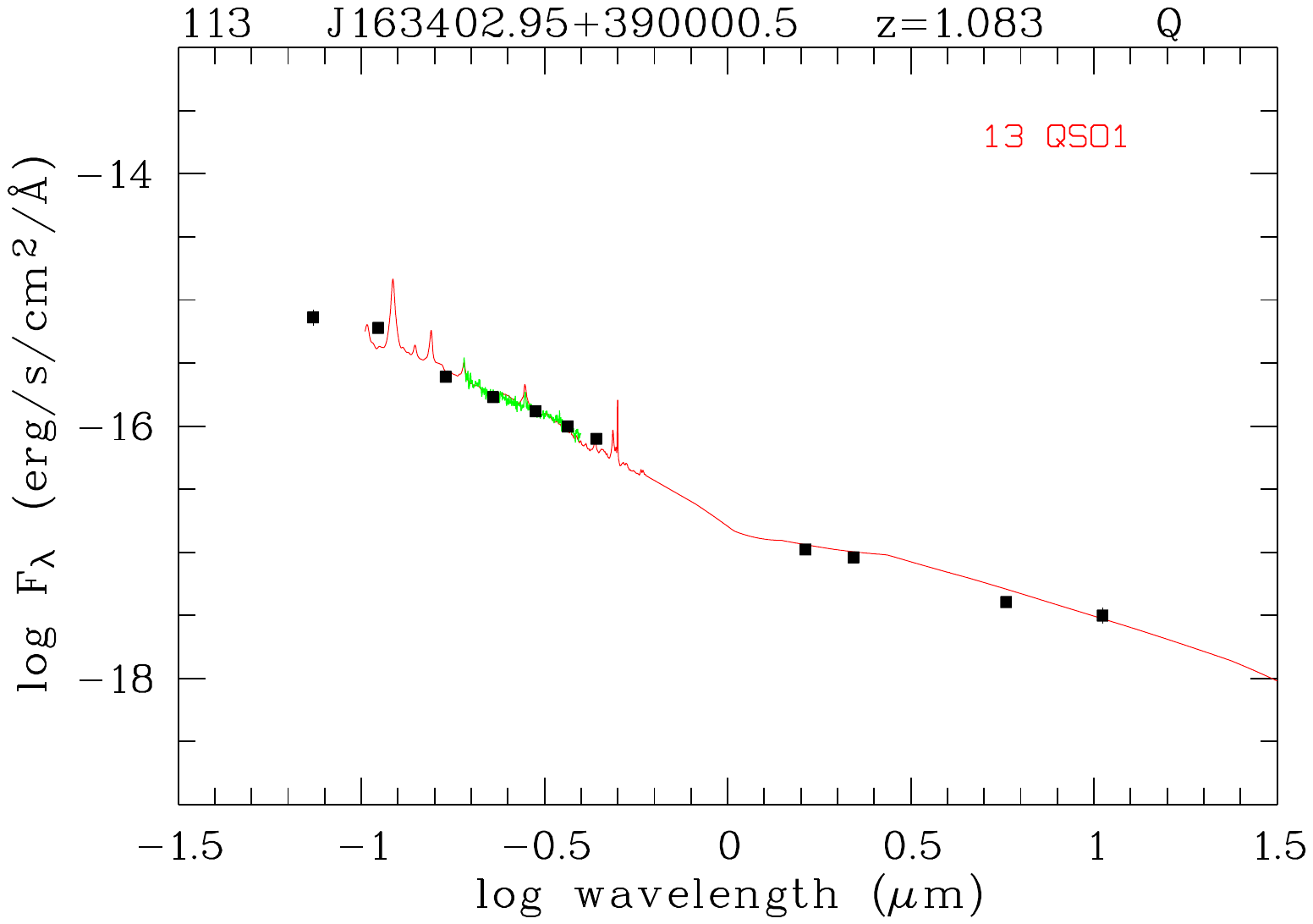}
\includegraphics[viewport= 0 30 590 790,width=4.7cm,angle=270]{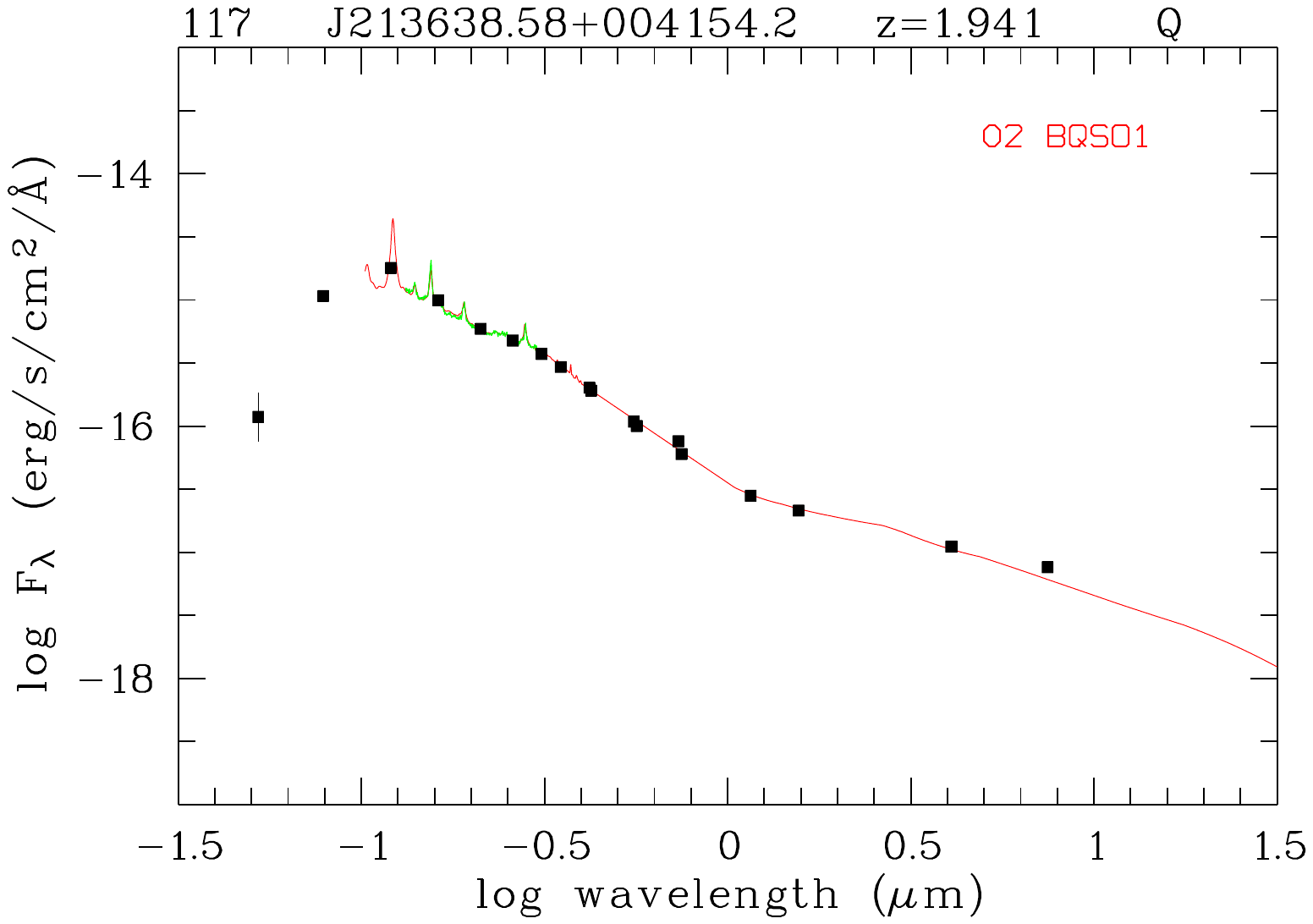}\\
\end{center}
\caption{
MBSEDs (in the rest frame)  for a selection of nine sources from the PSS-CSO sample (black symbols with error bars). 
Overplotted are the SDSS spectrum (green) and the best-fitting template (red).  The number and name of the best-fitting template are given in the top right corner of each panel.
} 
\label{fig:WBDEDs}
\end{figure*}  
\noindent


\subsection{Multi-band spectral energy distributions}\label{sect:MBSED}


We used photometric data from major sky surveys to derive the multi-band spectral energy distributions  (MBSEDs) from the far UV  to the MIR. 
Apart from the SDSS five-band photometry, we used data from the Galaxy Evolution Explorer  \citep[Galex;][]{Morrissey_2007} survey in the UV,  
the Two  Micron  All-Sky  Survey  \citep[2MASS;][]{Skrutskie_2006}, the  UKIRT  Infrared  Deep  Sky  Survey  \citep[UKIDSS;][]{Lawrence_2007, Hambly_2008} in the near-infrared (NIR), and the AllWISE  Data Release \citep{Cutri_2014} from the  Wide-Field Infrared Survey Explorer \citep[WISE;][]{Wright_2010} in the MIR. The magnitudes were extracted from the catalogues using  the  VizieR  Service  at  CDS\footnote{http://vizier.u-strasbg.fr/viz-bin/VizieR}  and the NASA/IPAC Infrared Science Archive (IRSA)\footnote{https://irsa.ipac.caltech.edu/applications/Gator/index.html}.  On average, an object from our PSS-CSO sample has measurements in 11  different bands,  95\% are measured in at least 8 bands and 68\% in at least 10 bands.  The magnitudes were corrected for Galactic foreground extinction using $E(B-V)$ from \citet{Schlafly_2011} and converted into fluxes in the rest frame.

The observed MBSEDs were compared with galaxy template spectra from the SWIRE library \citep{Lonsdale_2003, Polletta_2007}, which consists of 25 SEDs covering   the  wavelength range from 0.1 to 1000 $\mu$m.   It contains all major galaxy types, including ellipticals of different ages, early-type and late-type spirals, starburst (SB) galaxies,   type 1 AGNs,  type 2 AGNs,  and  composite types SB+AGN (Table\,\ref{tab:templates}).   The spectra for the ellipticals, the spirals and the SB galaxies were generated with a spectral evolution code.  The SB templates correspond to the observed SEDs of well-studied SB galaxies.  Templates of Seyfert 1.8 and 2 galaxies were obtained by combining models, broad-band photometric data and infrared spectra of a random sample of 28 Seyfert galaxies.  Another six AGN templates represent optically selected QSOs with different infrared-to-optical flux ratios (QSO1, TQSO1, and BQSO1) and two versions of type 2 QSOs (QSO2 and Torus).   The templates for composite types are empirical SEDs of objects that contain a powerful SB component, mainly responsible for their large infrared luminosities, and an AGN component that contributes to the MIR luminosities.

\begin{table}[htbp]
\caption{SED templates from \protect\citet{Polletta_2007} assigned to six types ($t_{\rm SED}$, see text).}
\begin{tabular}{ll cc} 
\hline\hline 
\noalign{\smallskip}
name
& explanation
& $t_{\rm SED}$\\
\hline              
\noalign{\smallskip}
Ell2   & 2 Gyr old elliptical galaxy & 1  \\                         
Ell5   & 5 Gyr old elliptical galaxy & 1  \\                         
Ell13  & 13 Gyr old elliptical galaxy & 1  \\                         
S0     & S0 galaxy & 1  \\                         
Sa     & Sa galaxy & 2  \\                         
Sb     & Sb galaxy & 2  \\                         
Sc     & Sc galaxy & 2  \\                         
Sdm    &Sdm galaxy & 2  \\                         
Sd     & Sd galaxy & 2  \\                         
Spi4   & Sc galaxy & 2  \\                                                               
I20551 & IRAS\,20551-4250, SB & 3  \\                         
I22491 & IRAS\,22491-1808, SB & 3  \\                         
M82    & M\,82, SB & 3 &  \\                         
N6090  & NGC\,6090, SB & 3  \\                         
N6240  & NGC\,6240, SB & 3  \\                         
Arp220 & SB, ULIRG & 4  \\                         
I19254 & IRAS\,19254-7245\,S, Sy\,2 + SB, ULIRG  & 4  \\                         
Mrk231 &BAL\,QSO + SB, Sy\,1, ULIRG & 4  \\
Sey18   &moderately luminous AGN, type Sy\,1.8& 5 \\                         
Sey2    &moderately luminous AGN, type Sy\,2& 5  \\                         
QSO2   &QSO of type 2& 5  \\                         
Torus    &heavily obscured QSO& 5 \\
BQSO1  &optically selected QSO  type 1 (faint)& 6  \\                         
QSO1   &optically selected QSO  type 1& 6  \\                         
TQSO1  &optically selected QSO  type 1 (bright)& 6  \\    
\hline
\end{tabular}                             
\label{tab:templates}                    
\end{table}

\begin{figure}[bhtp]
\begin{center}
\includegraphics[viewport= 0 0 822 580,width=8.5cm,angle=0]{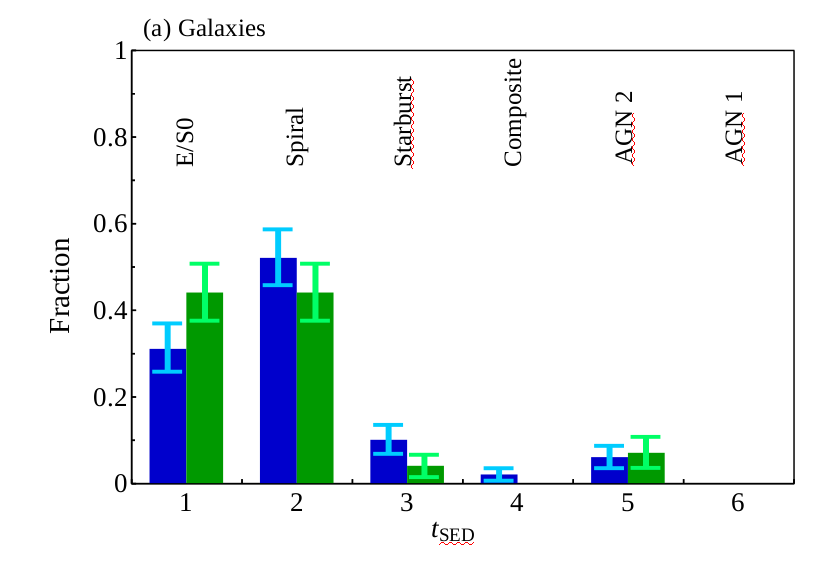} 
\includegraphics[viewport= 0 0 822 580,width=8.5cm,angle=0]{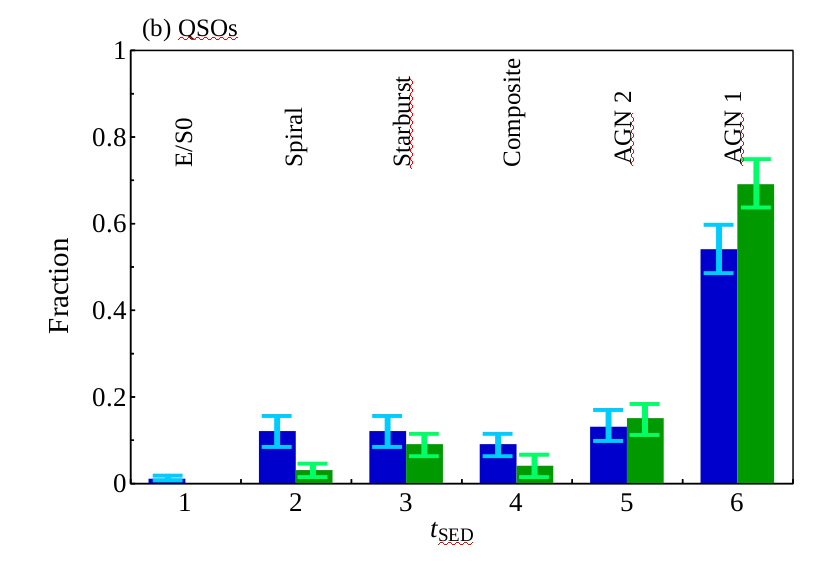}
\end{center}
\caption{
Histogram distributions of type $t_{\rm SED}$  for the G  (top) and the Q subsample (bottom) of the PSS-CSO (blue) 
and the ECS (green) sample. The vertical bars indicate the standard errors of the proportions.
}
\label{fig:hist_groups_g1}
\end{figure}

We fitted all template spectra from Table\,\ref{tab:templates} to the observed MBSEDs and determined the best-fitting template  using the least chi-square method.  Figure\,\ref{fig:WBDEDs} shows nine examples of resulting MBSEDs in the rest-frame.  The fluxes from the photometric data are shown as black squares with error bars.  In most cases, the error bars are smaller than the symbol size.  The best-fitting template is shown in red.  Also plotted is the SDSS spectrum corrected for Galactic foreground reddening, transformed into the rest image and scaled to the MBSED in the optical domain (SDSS bands g, r, and i).

To facilitate the use of the classification scheme, we have grouped the different SED templates into six groups (last column of Table\,\ref{tab:templates}).  The meaning of the group number code is  $t_{\rm SED} = 1$ for E and S0 galaxies (`early types'),  2 for spiral galaxies,  3 for SBs, 4 for SB-AGN composites (or transition objects),  5  for AGNs of type 1.8 to 2,  and 6  for type 1 QSOs.  The histograms in Fig.\,\ref{fig:hist_groups_g1} show the  distributions of these types in the PSS-CSO  (blue) and the ECS (green) sample, respectively,  with both samples subdivided again into the two subsamples G and Q. We also note that the histogram distributions from the second-best fit do not differ much from those in Fig.\,\ref{fig:hist_groups_g1}. As expected, there  is a clear and strong difference between the G and Q subsamples, both for the PSS-CSO sample and the ECS sample. The vast majority of objects assigned to spectral class QSOs based on SDSS spectra, i.e. subsample Q, also have MBSEDs of type 1 QSOs. On the other hand, most of the objects from the subsample G are found to show MBSEDs of normal galaxies (early type or spiral).   The proportion of early-type galaxies is higher in the ECS-G sample than in the PSS-CSO-G sample,  but overall Fig.\,\ref{fig:hist_groups_g1} shows a similar distribution of types in both samples. For a quantitative analysis, we performed Pearson's chi-square test for two independent samples to compare the distributions of types in the corresponding subsamples. The null hypothesis ${\rm H}^0$ states that the  subsamples PSS-CSO-G and ECS-G  have the same distributions of types, and likewise the subsamples PSS-CSO-Q and ECS-Q.  ${\rm H}^0$ was tested against the alternative hypothesis, ${\rm H}^{\rm A}$, that the two samples have different distributions.   The results indicate that there is no reason to reject  ${\rm H}^0$ in favour of  ${\rm H}^{\rm A}$ at $\alpha = 0.05$.

Next, we focus on the proportions  $f_{\rm sf}$ of SF galaxies. For this purpose, we have summarised the templates in a coarser scheme, distinguishing only between types with active SF  ($t_{\rm SED} = 2 - 4$) on the one hand and all other types (i.e. either dominated by an old stellar population or by an AGN) on the other. For this analysis, we included only sources with fluxes in at least eight wavelength bands, where at least two data points must be in the NIR or MIR.  The resulting samples consist of 115 PSSs and CSOs  and 116 ECSs, respectively. The fraction of SF galaxies is  $f_{\rm sf}^{\rm PSS} =  0.46\pm 0.05$  for the PSS-CSO sample and $f_{\rm sf}^{\rm ECS} = 0.28\pm 0.04$ for the ECS sample. We applied the one-tailed two-sample Z test of proportions where  the null hypothesis, ${\rm H}^0:  f_{\rm sf}^{\rm PSS} \le  f_{\rm sf}^{\rm ECS}$ is tested against the alternative hypothesis  ${\rm H}^{\rm A}:  f_{\rm sf}^{\rm PSS} >  f_{\rm sf}^{\rm ECS}$  at an error probability $\alpha$. The result is that ${\rm H}^0$ must be rejected in favour of ${\rm H}^{\rm A}$ at the 95\% confidence level ($\alpha = 0.05$),  i.e. we can assume that the proportion of SF galaxies in the PSS-CSO sample is larger than that in the comparison sample ($p = 0.006$). On the other hand, the two-sided Z-test for the proportions of AGNs (including AGN-SF composites) shows that there is no evidence of different AGN proportions in the PSS-CSO and ECS samples.

The formation of a spheroidal component seems to be important for the history of SF in galaxies. Late-type disk-dominated spiral galaxies (Sbc, Sc, Sd) tend to actively form stars, while E and S0 are usually quiescent. In terms of both morphology and SF indicators, Sa galaxies lie between the passive galaxies and SF galaxies \citep{Bendo_2002, Hameed_2005, Vika_2015, Gonzalez_Delgado_2016, Wang_2018}.  This is also evident in our two samples. A total of 21 sources were classified as Sa. For 10 of them, the MIR colours (Sect.\,\ref{sect:WISE}) correspond to spiral galaxies with moderate SFR, while 11 are passive galaxies, similar to E and S0.  Therefore, we have also performed the above analysis for an alternative classification scheme in which the Sa galaxies were classified as early. In this case, too, there is a clear difference between the two samples ($f_{\rm sf}^{\rm PSS} =  0.36\pm 0.05, f_{\rm sf}^{\rm ECS} = 0.23\pm 0.04, p = 0.014$).

The results from the analysis of the MBSEDs agree well with those derived from the composite spectra.  In particular, they support the interpretation of the differences between the composite spectra as a consequence of different contributions from younger stellar components (Sect.\,\ref{sect:continuum}). There is a stronger association with younger star formation in the PSS-CSO sample.


\subsection{WISE fluxes and colours}\label{sect:WISE}


Observational data in the MIR spectral range are particularly useful for studying the relationship between the AGN emission and the contribution from SF in the host galaxy. We used the WISE data in the four bands W1, W2, W3, and W4  at the central wavelengths 3.4, 4.6, 12, and 22\,$\mu$m, respectively \citep{Wright_2010}.

\begin{figure}[htbp]
\begin{center}
\includegraphics[viewport= 20 20 570 820,width=6.0cm,angle=270]{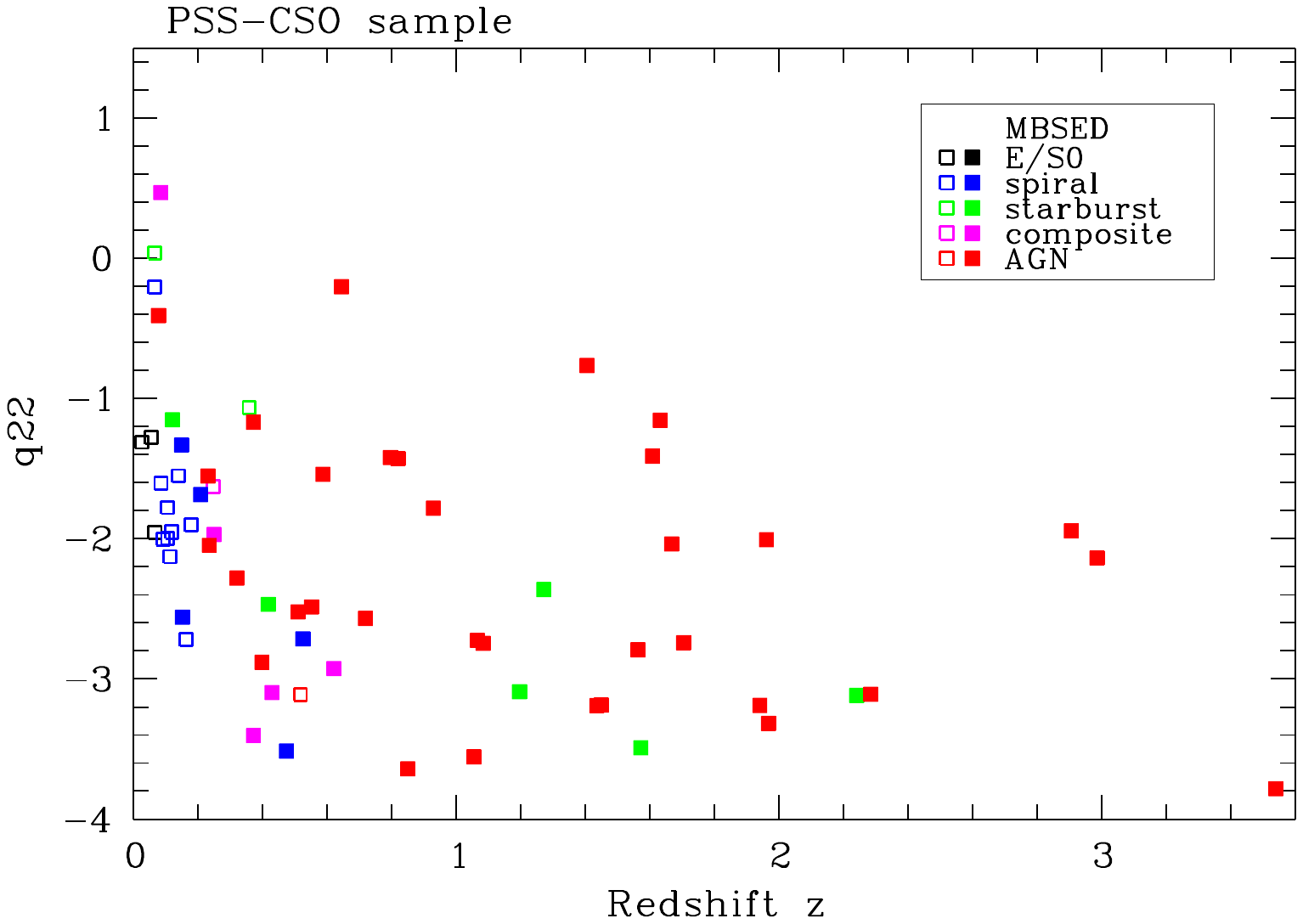} 
\includegraphics[viewport= 20 20 570 820,width=6.0cm,angle=270]{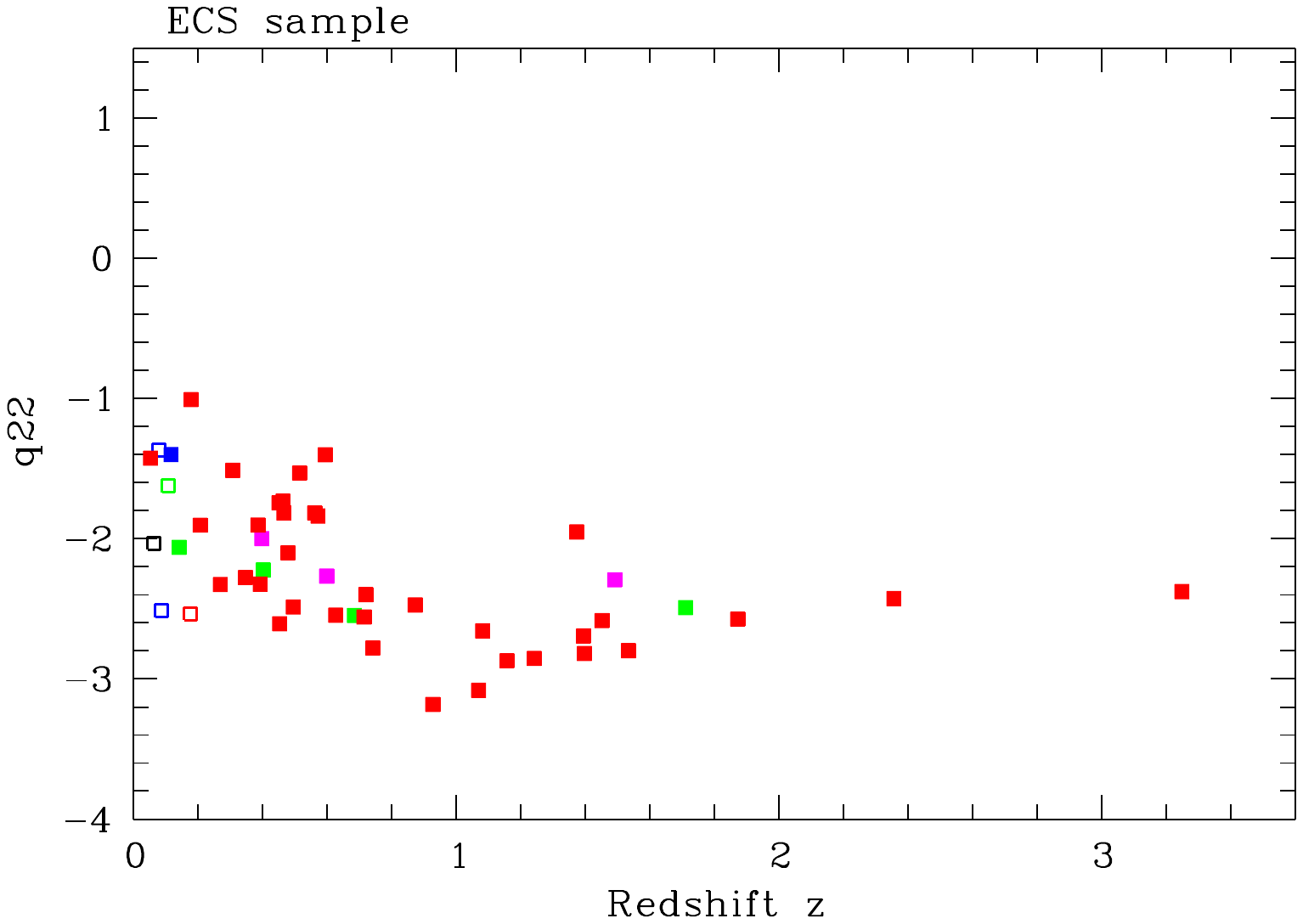}  
\end{center}
\caption{
MIR-to-radio flux density ratio $q22$ vs. redshift for the PSS-CSO  (top) and the ECS (bottom) sample; 
only sources with reliable type classification from the MBSED fitting and with  $\mbox{S/N} >3$ for the W4 flux density are plotted. Filled squares stand for the QSO subsample, open squares for the G subsample.  The colours indicate types from MBSED fitting (see inset). 
}
\label{fig:q22}
\end{figure}

\begin{figure}[htbp]
\begin{center}
\includegraphics[viewport= 20 20 570 820,width=6.0cm,angle=270]{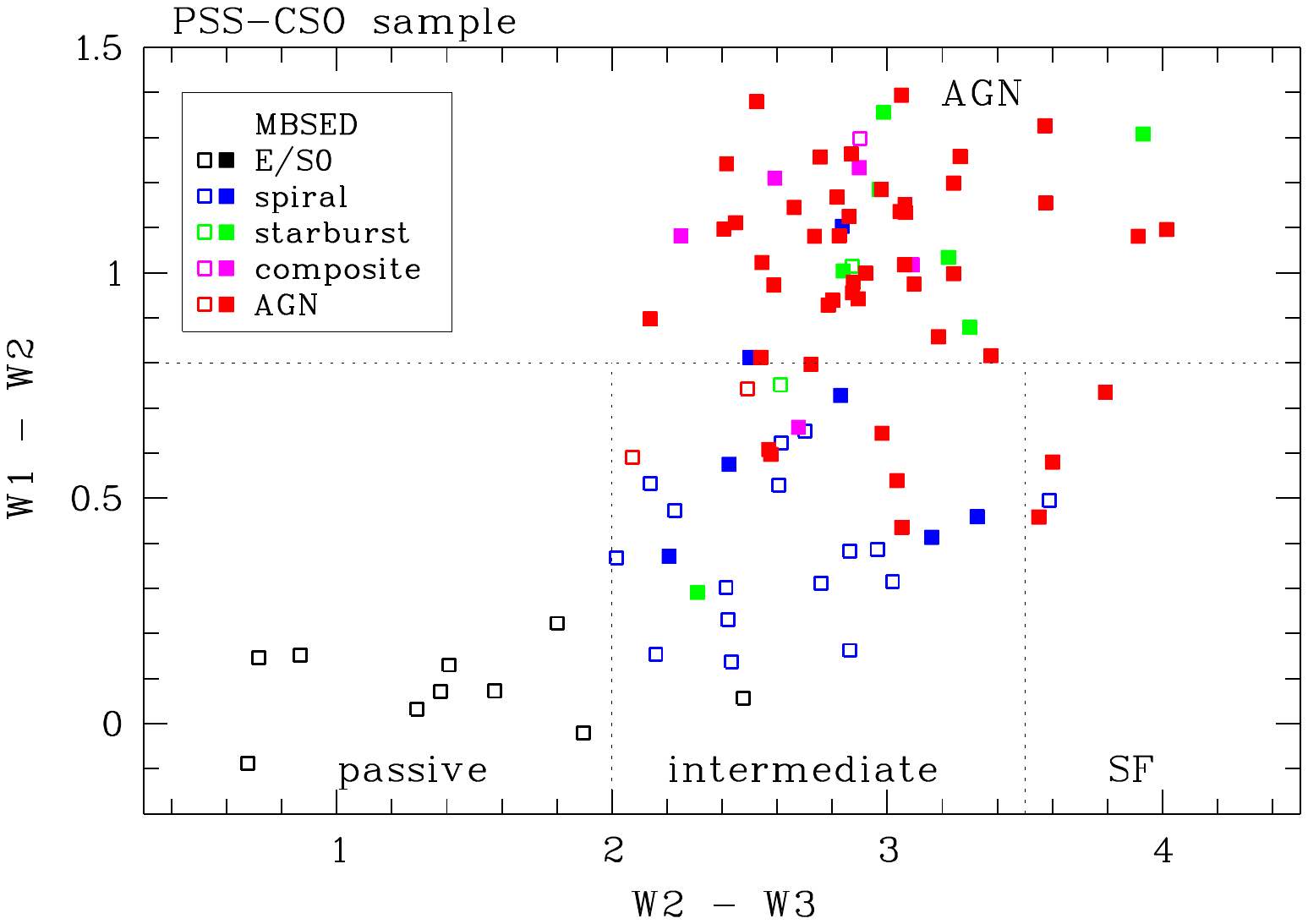}   
\includegraphics[viewport= 20 20 570 820,width=6.0cm,angle=270]{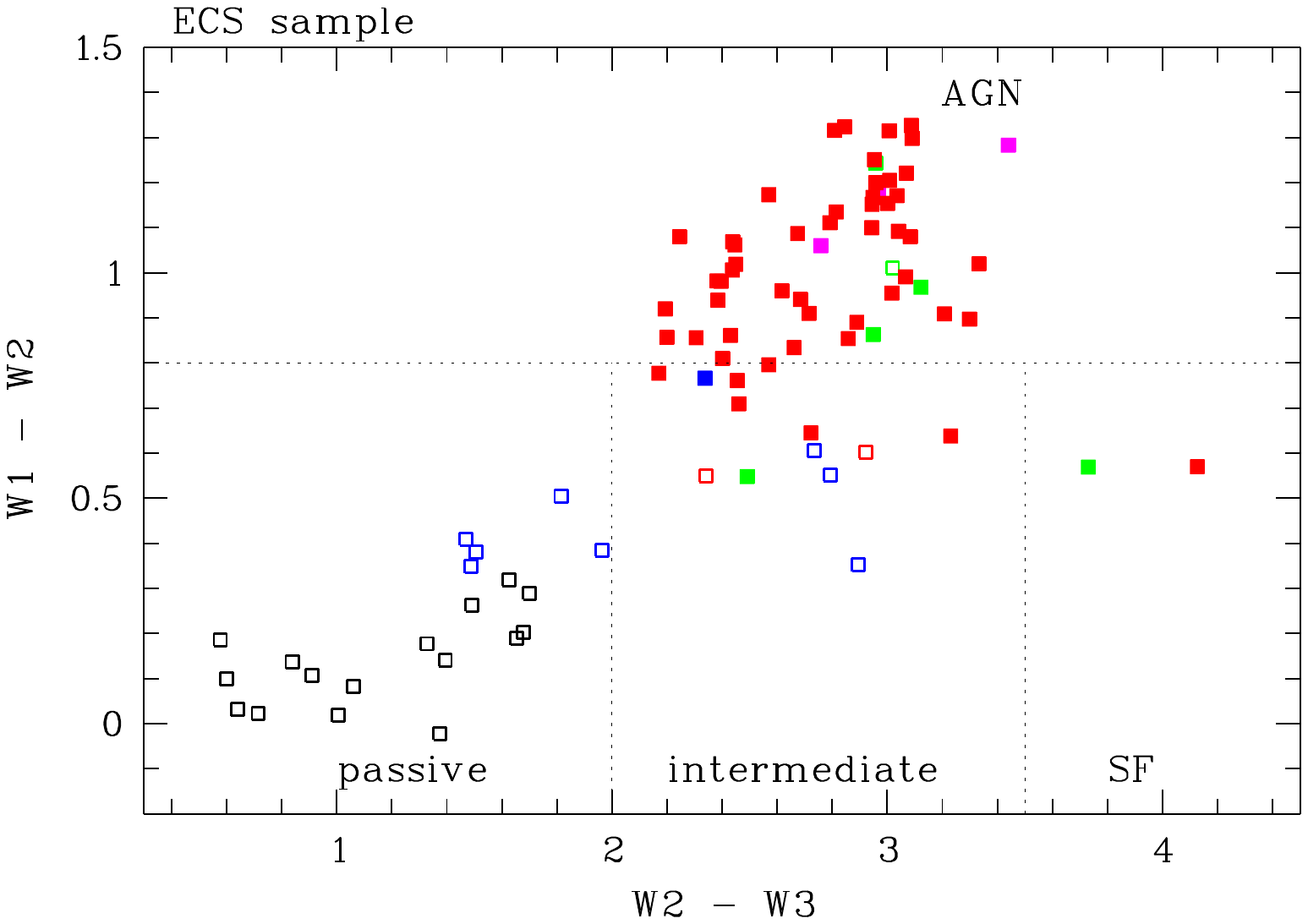}  
\includegraphics[viewport= 20 20 570 800,width=6.0cm,angle=270]{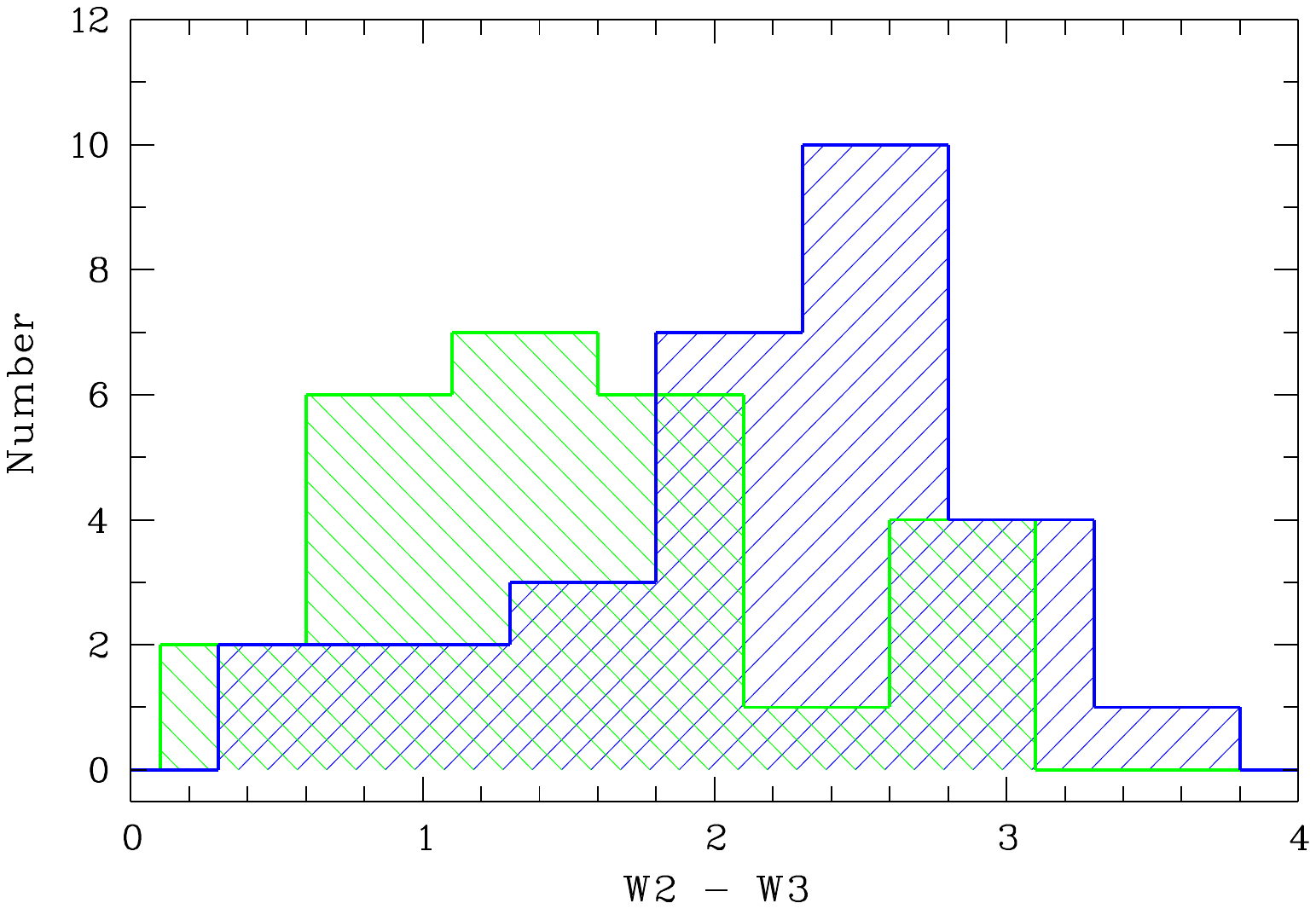}  
\end{center}
\caption{
WISE $W1-W2$ vs. $W2-W3$  diagram for the PSS-CSO  (top) and the ECS (middle) sample.   The symbols and colours have the same meaning as in Fig.\,\ref{fig:q22}.  The dotted vertical demarcation lines are from \protect\citet{Jarrett_2017}, the horizontal line marks the AGN threshold from \protect\citet{Stern_2012}. Bottom: Histogram distributions of $W2 - W3$ for the PSS-CSO-G (blue) and ECS-G (green)  galaxies with $W1-W2 < 0.8$. Only sources with $\mbox{S/N} > 5$ in the bands W1 and W2 and with $\mbox{S/N} > 3$  in the band W3 are plotted.
}
\label{fig:WISE_CCD}
\end{figure}

\begin{figure}[htbp]
\begin{center}
\includegraphics[viewport= 20 20 570 820,width=6.0cm,angle=270]{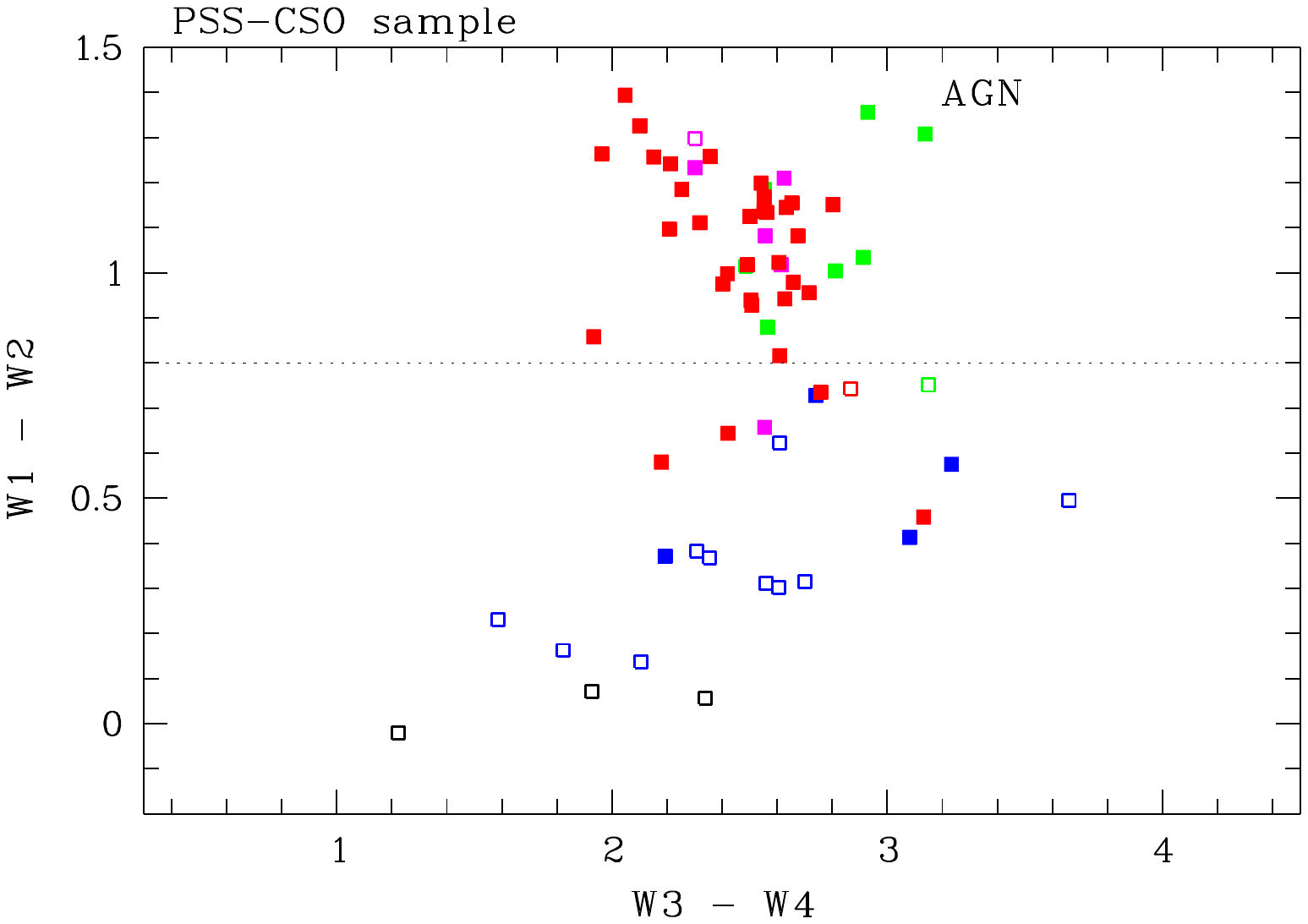}  
\includegraphics[viewport= 20 20 570 820,width=6.0cm,angle=270]{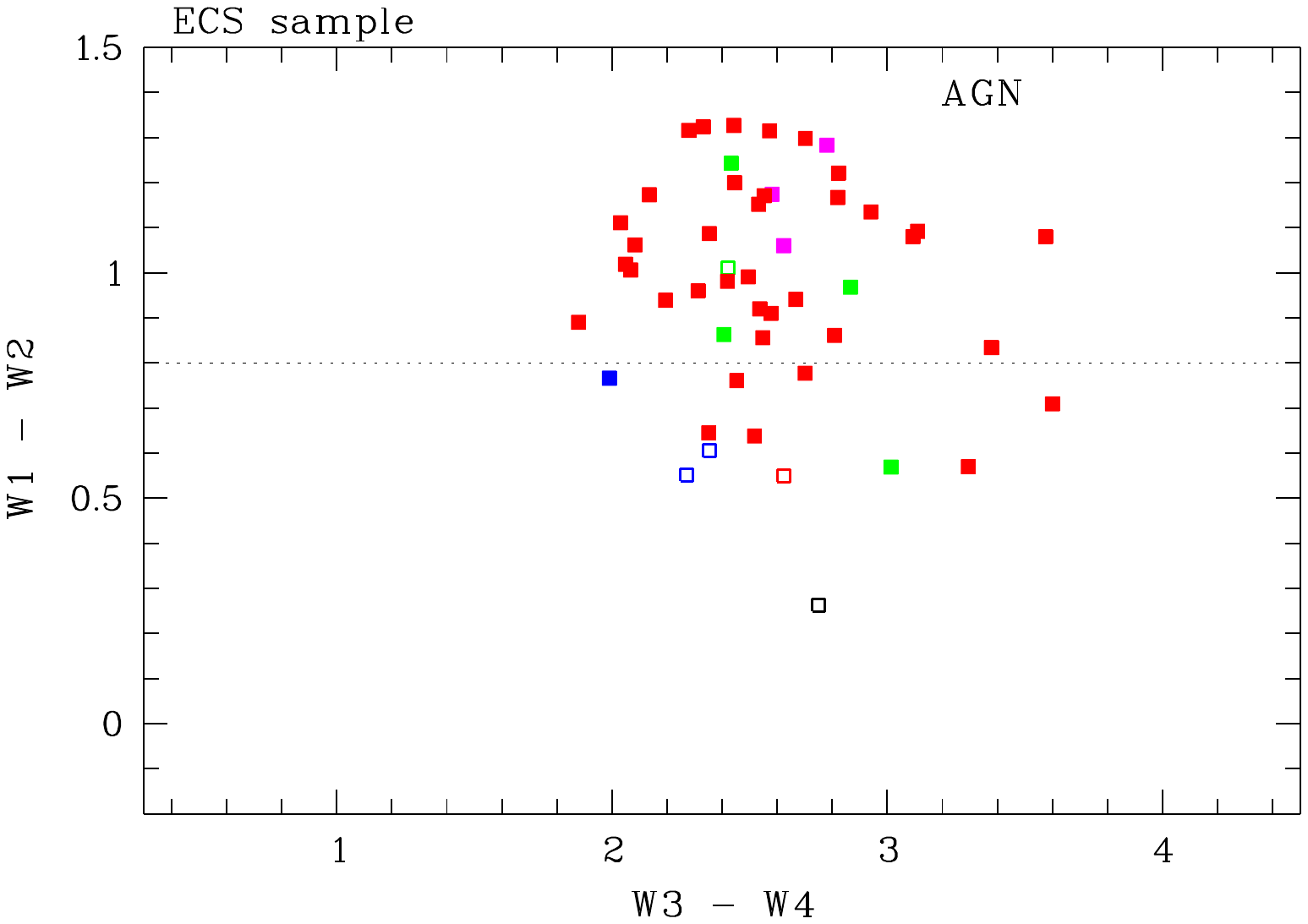} 
\includegraphics[viewport= 20 20 570 800,width=6.0cm,angle=270]{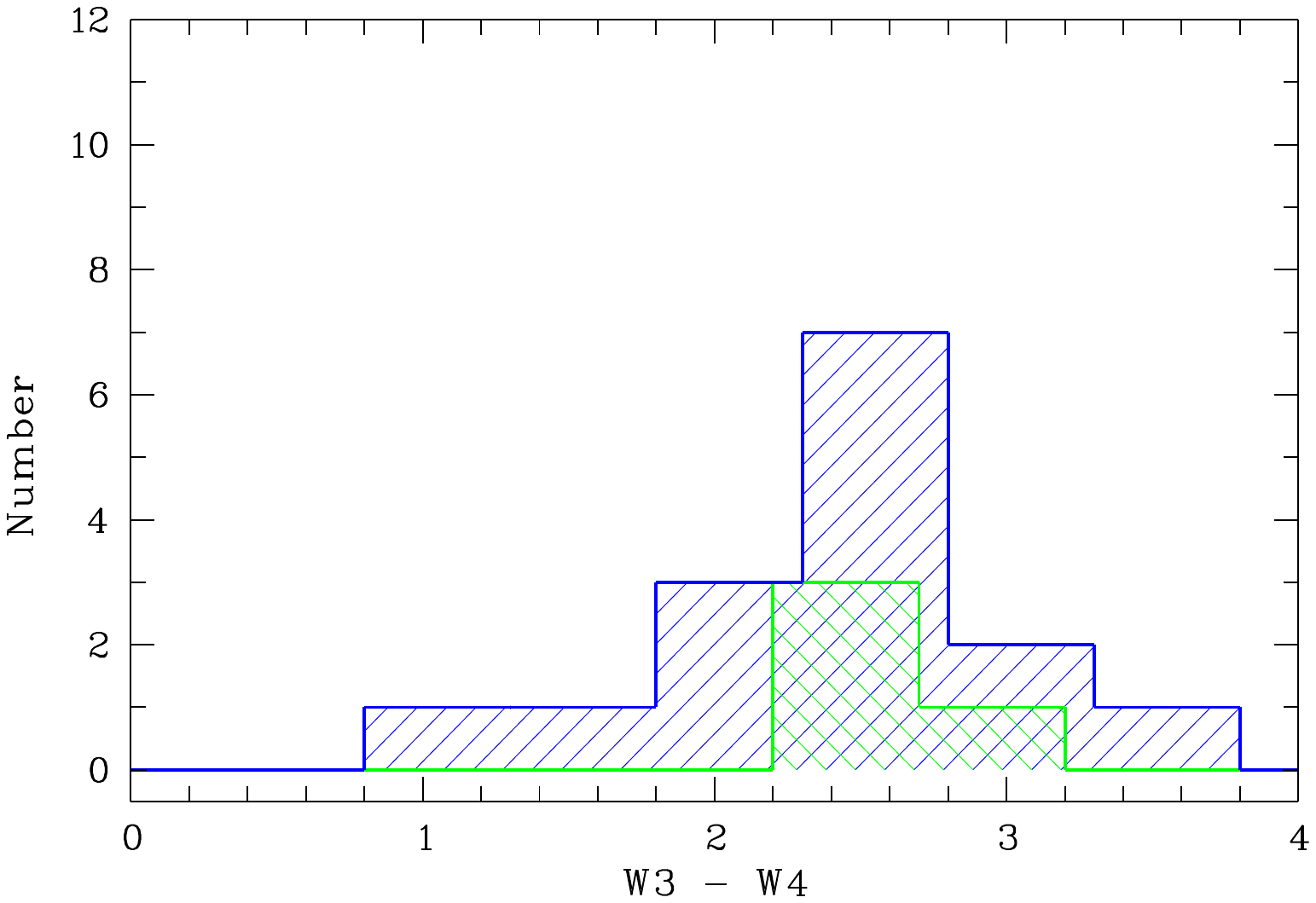}  
\end{center}
\caption{
The same as in Fig.\,\ref{fig:WISE_CCD}, but for $W1-W2$ vs. $W3-W4$.   Only sources with $\mbox{S/N} > 5$ in the bands W1 and W2 and with $\mbox{S/N} > 3$  in the bands W3 and W4 are plotted.
}
\label{fig:WISE_CCD2}
\end{figure}

Jet-dominated AGNs and SF galaxies differ in the ratios of their MIR flux to the radio flux. To quantify the MIR-to-radio ratio,  \citet{Caccianiga_2015} introduced the parameter  $q22 = \log\,(F_{\rm 22\,\mu m}/F_{\rm 1.4\,GHz})$, where $F_{\rm 22\,\mu m}$ is the flux density in the W4 band and $F_{\rm 1.4\,GHz}$ is the 1.4\,GHz flux density. Major star formation contribution to the radio emission can be expected in sources with $q22 > 1$ \citep{Caccianiga_2015}. We computed $q22$ with $k$-corrections using the individual spectral slopes between the 12 and 22 $\mu$m (observed frame) flux densities for $F_{\rm 22\,\mu m}$ and the radio slope $\alpha_{\rm r} = -0.7$ for $F_{\rm 1.4\,GHz}$. In both samples, the $q22$ values are distributed over a large range,  but all sources have $q22 \la 1$ and, except for two low-$z$ systems, $q22 < 0$ (Fig.\,\ref{fig:q22}).  SF activity is not a dominant source in the radio band,   the observed radio flux is most likely dominated by synchrotron emission from the relativistic jet. However, it must be taken into account that the plotted samples are highly incomplete.  Only 33\% of the PSS-CSO-G subsample and 11\% of the ECS-G subsample are detected in the W4 band and meet the requirement $\mbox{S/N} >3$ for $F_{\rm 22\,\mu m}$.  In particular, there is a bias against normal E, S0, and S galaxies ($t_{\rm SED} = 1$ or 2), as can be seen in Fig.\,\ref{fig:q22}.

With some cautions  \citep[e.g.][]{Gurkan_2014},  the WISE colours $W1-W2$ and  $W2-W3$ provide a useful tool for selecting AGNs and detecting SF activity in galaxy samples \citep[e.g.][]{Stern_2012, Jarrett_2017, Cluver_2017}. In the $W1-W2$ vs.  $W2-W3$  diagram, galaxies with little hot dust emission occupy a narrow sequence at $W1-W2 \approx -0.2 \ \mbox{to}  +0.5$ mag, where the colour index $W2-W3$ is a good indicator of the SF activity. Galaxies with high SFR, including luminous  infrared  galaxies  (LIRGs)  and  ultra-luminous  infrared galaxies  (ULIRGs) are found at the right-hand side  of this sequence at $W2-W3>3.5$ mag (`SF region'). Galaxies with little or absent SF populate the left-hand side at $W2-W3<2$ mag (`passive region').  The galaxy population in the intermediate area  is  attributed to  normal  spiral  disks with  moderate  SFR.   AGNs with hot dust populate the colour space above the threshold $W1-W2=0.8$ mag (`AGN region').

Figure\,\ref{fig:WISE_CCD} shows the $W1-W2$ vs.  $W2-W3$  diagram (Vega magnitudes)  for the PSS-CSO sample (top) and the ECS sample (middle), where only sources with reliable signal-to-noise ratios are plotted ($\mbox{S/N} > 5$ for the W1 and the W2 band,  $\mbox{S/N} > 3$ for W3). In the passive region, all PSSs and CSOs and the majority of the ECSs have MBSEDs of early-type galaxies.  In the AGN region, on the other hand, the vast majority of the sources belong to the QSO subsamples with MBSEDs classified as AGNs or composites. In both two samples, several objects with AGN-type MBSEDs have $W1-W2$ values below the AGN demarcation.  This is most likely due to a `dilution' of the AGN radiation by stellar light in the bands W1 and W2 \citep{Donley_2012}. The proportion of sources in the AGN region is similar in both samples ($53\pm5$\,\% for PSS-CSO and $59\pm5$\,\% for ECS).  In the non-AGN region, however,  the distribution along the $W2-W3$ axis is  different: For the PSS-CSO sample, we find $9\pm3$\,\% in the passive and $38\pm5$\,\% in the intermediate plus SF region,  compared to $24\pm5$\,\% and $17\pm4$\,\%  for the ECS sample. The bottom panel of Fig.\,\ref{fig:WISE_CCD} shows the distribution of $W2-W3$ for the galaxies (subsamples PSS-CSO-G and ECS-G) with $W1-W2 < 0.8$ and $z<0.6$.  The redshift limit was chosen because the strong  polycyclic aromatic hydrocarbon (PAH) features around $\sim 11\,\mu$m,  which are usually considered indicators for massive young stars, are no longer covered by the W3 band at higher redshifts. With only one exception, WISE measurements are available for all PSSs, CSOs, and ECSs with $z < 0.6$.  It can be seen that the PSS-CSO-G sample has a larger contribution from SF activity. We applied again the one-tailed two-sample Z test of proportions  to test the null hypothesis ${\rm H}^0: f_{\rm sf}^{\rm PSS} \le  f_{\rm sf}^{\rm ECS}$ against ${\rm H}^{\rm A}: f_{\rm sf}^{\rm PSS} >  f_{\rm sf}^{\rm ECS}$ and found that ${\rm H}^0$ can be rejected in favour of ${\rm H}^{\rm A}$ at a high significance level  ($p = 0.003$). The Z test for all sources at $z < 0.6$, i.e. without restricting S/N, gives the same result  ($p = 0.01$).

The WISE colour $W3-W4$ is particularly sensitive to the strength of SF because a high SFR leads to strong emission by a warm dust component, which contributes more to the W4 band than to the W3 band.   The colour-colour diagram $W1-W2$ vs.  $W3-W4$ is shown in Fig.\,\ref{fig:WISE_CCD2} for the sources with $\mbox{S/N} >5$ in the bands W1 and W2, and $\mbox{S/N}>3$ in W3 and W4. As in Fig.\,\ref{fig:WISE_CCD}, the `AGN cloud' at $W1-W2 > 0.8$ is clearly seen.  In the non-AGN region ($W1-W2 < 0.8$), the sources cover a wider interval of $W3-W4$ than the AGN cloud.  At first glance, the distribution of the PSSs and CSOs along the $W3-W4$ axis is similar to those in the colour-magnitude diagrams  presented by \citet{Nikutta_2014} for huge samples of QSOs, AGNs, and SF galaxies, respectively. They found the following median values $\mu$ of the $W3-W4$ distribution:  $2.45\pm 0.19$ for 14\,795 QSOs,  $2.54\pm 0.29$ for 4\,509 AGN galaxies, and $2.19\pm 0.33$ for their sample of 38\,092 SF galaxies. Here we derived the median values $\mu(W3-W4) = 2.55\pm 0.05$ for the PSS-CSO and $2.53\pm 0.05$  for the ECS  sample. These values are similar to those given by  \citet{Nikutta_2014} for their samples of AGN galaxies and QSOs.   The bottom panel of Fig.\,\ref{fig:WISE_CCD2} shows the histogram distribution of $W3-W4$ for the galaxies at $z<0.6$ in the PSS-CSO and the ECS samples. We note that reliable measurements of W4 are only available for a small number of galaxies.

The interpretation of the $W3-W4$ distribution in the non-AGN region is not entirely clear. When only the galaxies with $W1-W2 < 0.8$ and $z < 0.6$  are considered,  the resuting median values are $\mu(W3-W4) = 2.34\pm 0.16$ for the PSS-CSO subsample and $2.42\pm 0.08$ for the (very small) ECS subsample.  The blue PSSs and CSOs with $W3-W4 < 2$  all have MBSEDs classified as early-type galaxies ($t_{\rm SED} = 1$). On the other side, the very red sources with  $W3 - W4 > 3$ were classified as spiral or starburst galaxy.  The reddest galaxy in the top panel of Fig.\,\ref{fig:WISE_CCD2} is \object{SDSS\,J075756.71+395936.1} ($z = 0.066$) with $W3-W4 = 3.66$. Based on its MBSED, it was classified as an Sb galaxy, its SFR of $\sim 16,\mathcal M_\odot$\,yr$^{-1}$ \citep{Chang_2015} is the highest in our PSS-CSO sample  (see Sect.\,\ref{sect:SFR} and Table\,\ref{tab:low_z_PSS}). Investigating a sample of flat-spectrum NLS1s, \citet{Caccianiga_2015} argued that red colours $W3 - W4 > 2.5$ cannot be explained by AGN template spectra alone,  but require a significant SF component.  However, other effects may also contribute to a red $W3 - W4$ colour:   warm dust components heated by the AGN (polar dust and possible warm dust related to the dusty torus) or the suppression of PAHs by the AGN \citep[][and references therein]{Jaervelae_2022}.


\section{Star formation rates and stellar masses}\label{sect:SFR}


\begin{figure}[htp]
\begin{center}
\includegraphics[viewport= 20 20 570 820,width=6.0cm,angle=270]{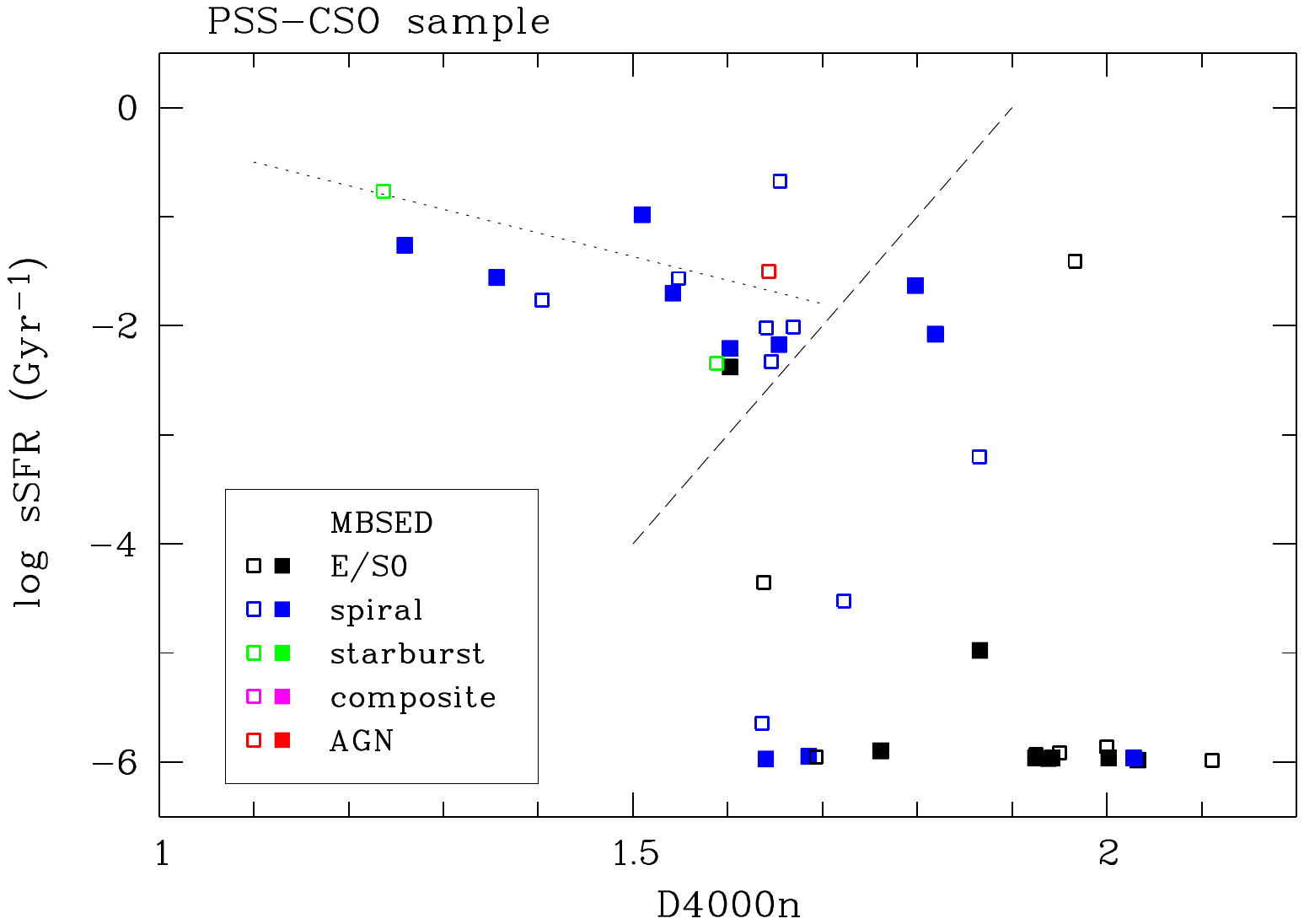}   \\  
\includegraphics[viewport= 20 20 570 820,width=6.0cm,angle=270]{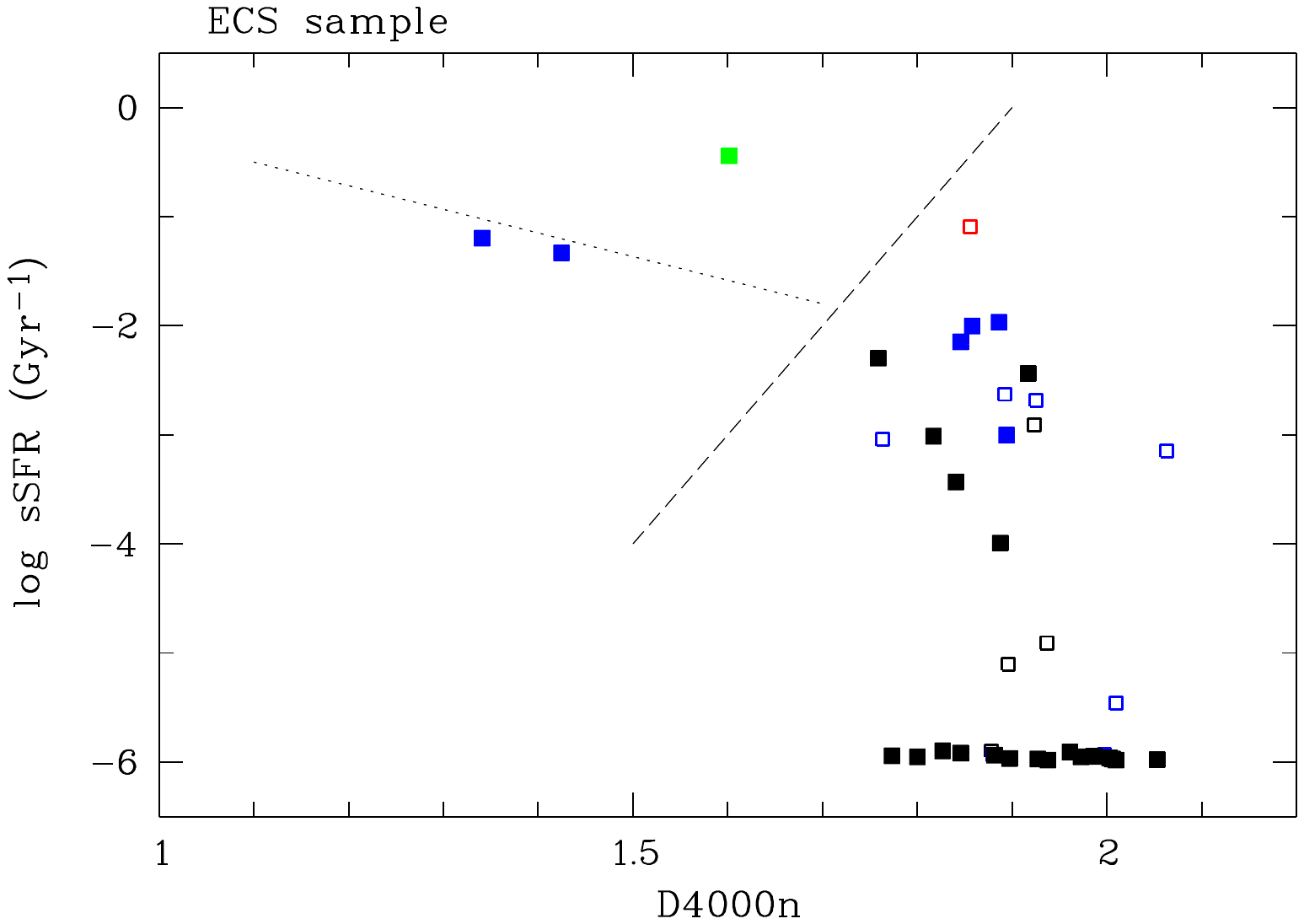}  \\ 
\includegraphics[viewport= 20 50 580 780,width=6.1cm,angle=270]{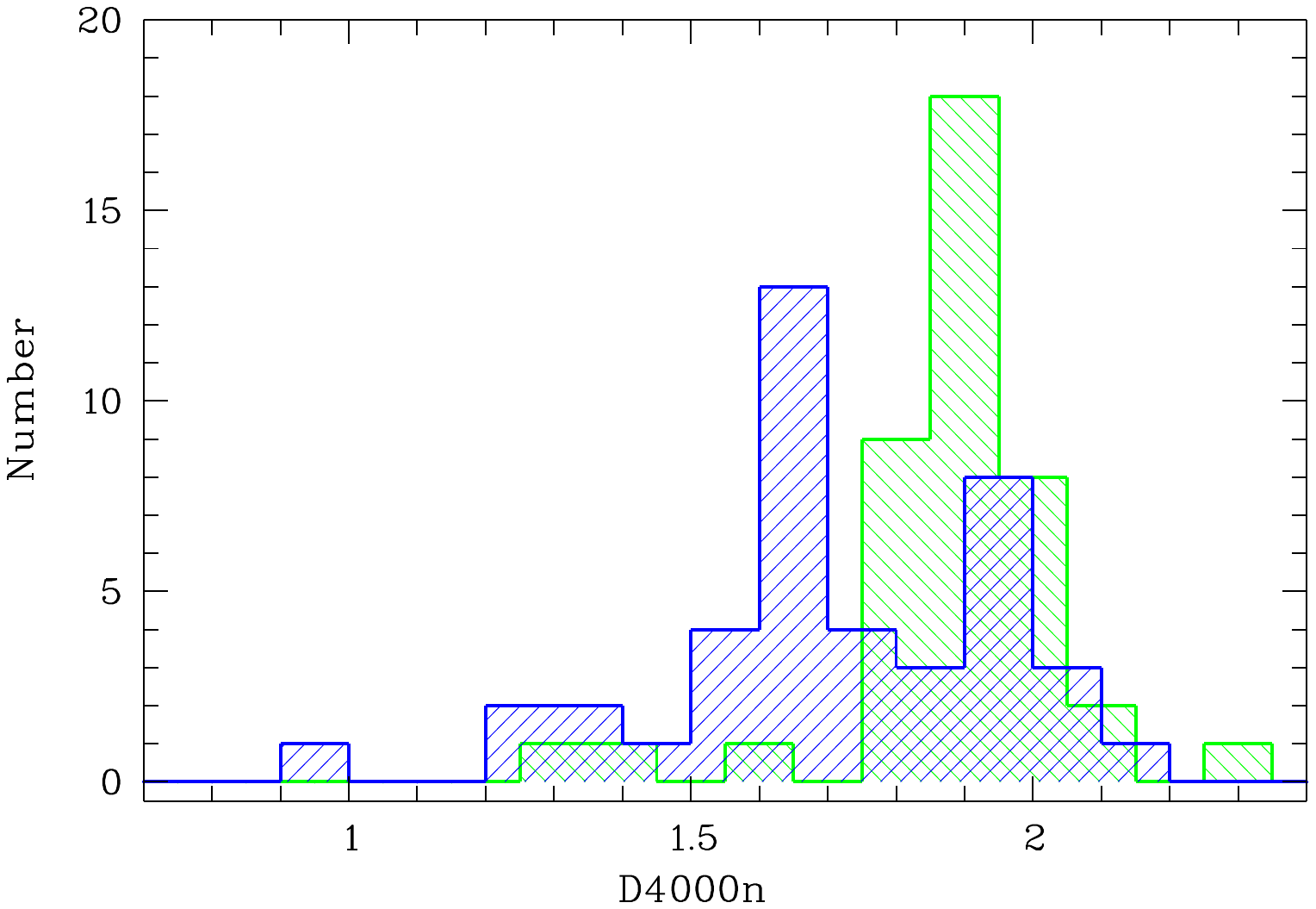} 
\end{center}
\caption{
D4000n index and sSFR. 
Top and middle: sSFR from \citet{Chang_2015} versus D4000n for the PSS-CSO and the ECS samples.  Filled squares signify sSFR data from good model fits ({\tt FLAG}=1).  The dashed line indicates the boundary between the SF sequence (above) and passive galaxies (below).  The dotted line is a power-law approximation of the mean values from \protect\citet[][their Fig. 11]{Brinchmann_2004}.  The spectral types $t_{\rm SED}$ from  the MBSED fit are colour coded as in Fig.\,\ref{fig:WISE_CCD}. Bottom: Histogram distributions of the D4000n index for the PSS-CSO (blue) and ECS (green) galaxies with $\mbox{S/N} > 3$.
}
\label{fig:D4000n_sSFR}
\end{figure}

\begin{figure}[htp]
\begin{center}
\includegraphics[viewport= 20 20 570 820,width=6.0cm,angle=270]{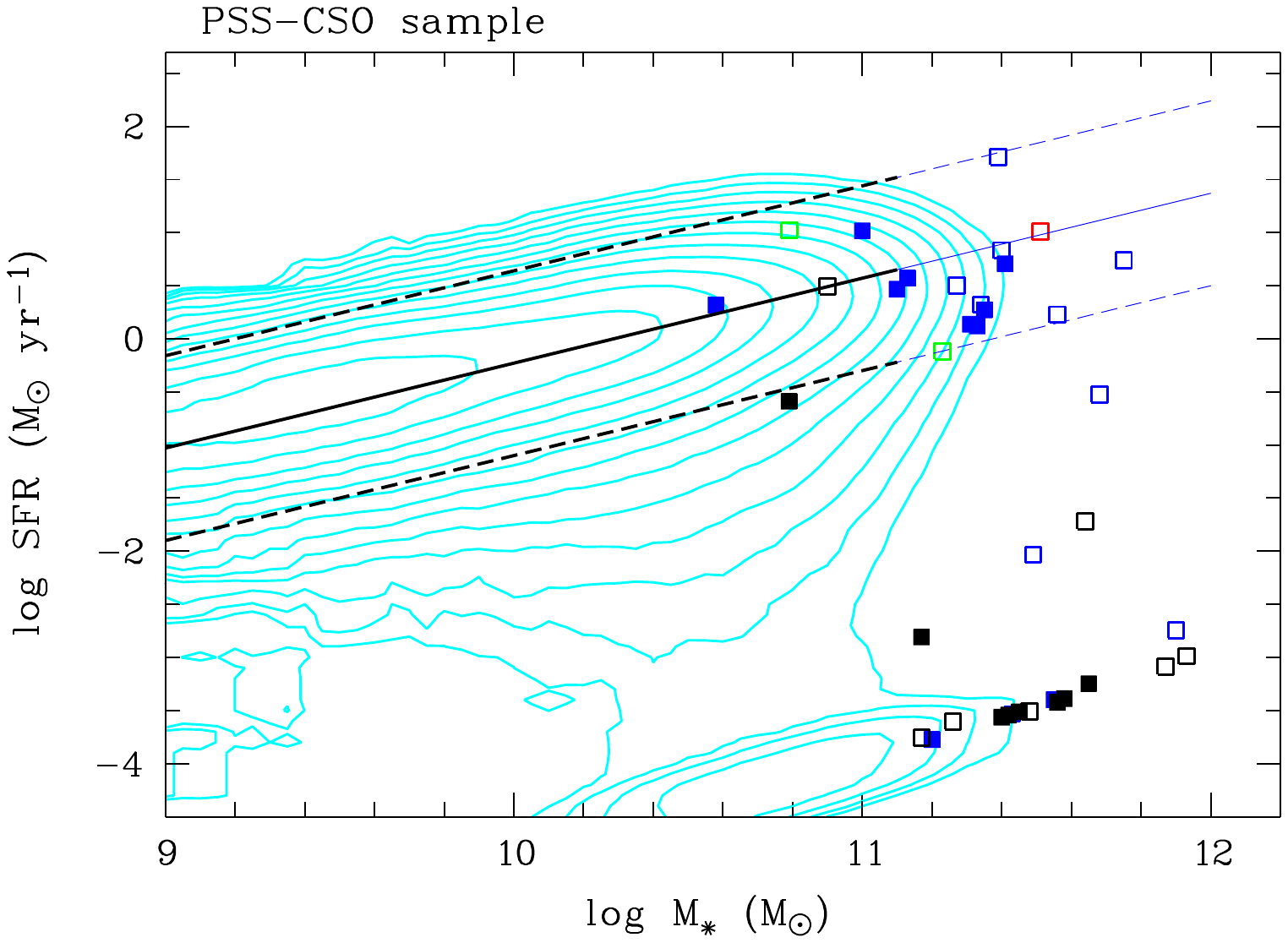} \\
\includegraphics[viewport= 20 20 570 820,width=6.0cm,angle=270]{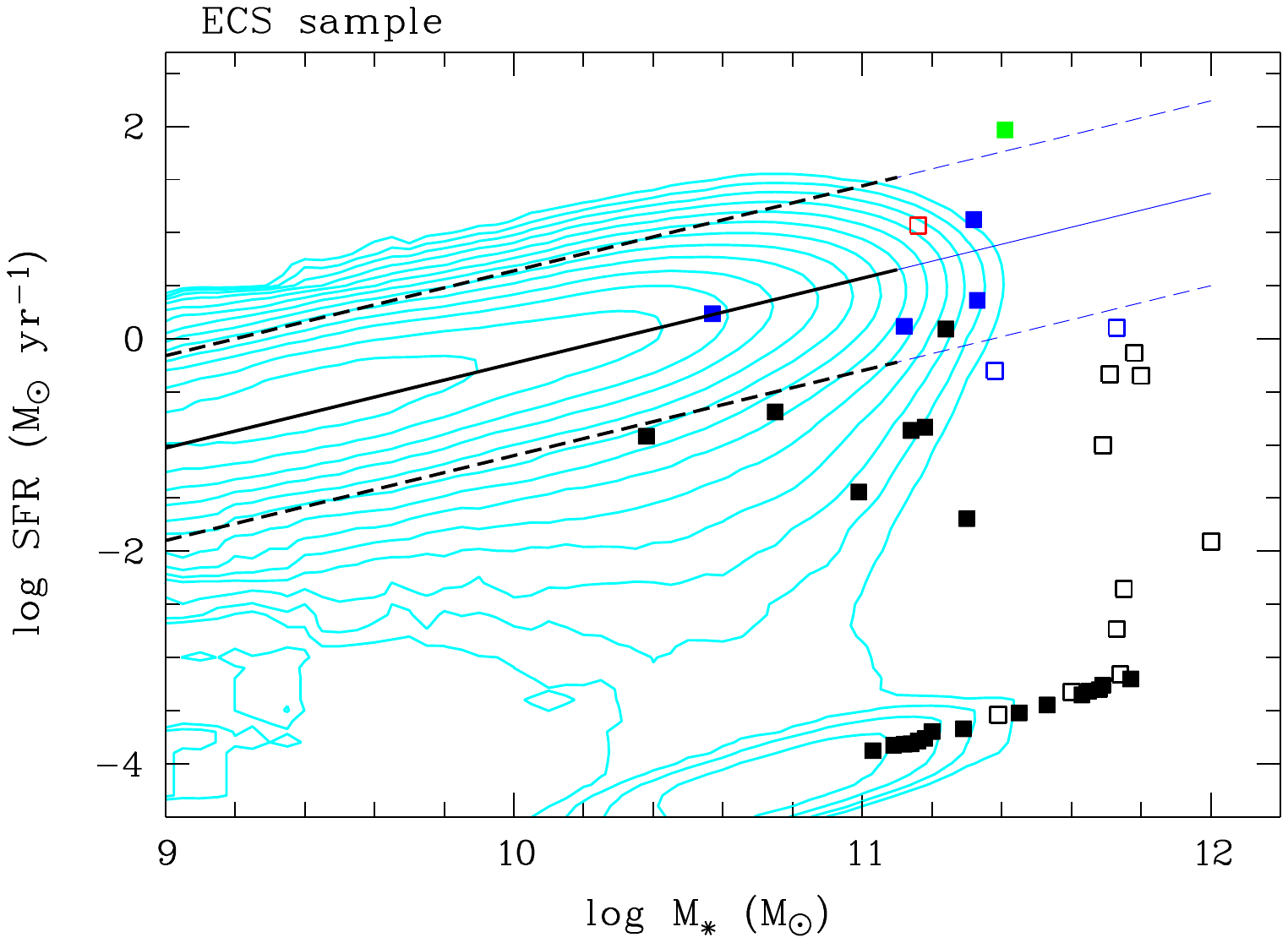} \\
\includegraphics[viewport= 25 20 585 820,width=6.15cm,angle=270]{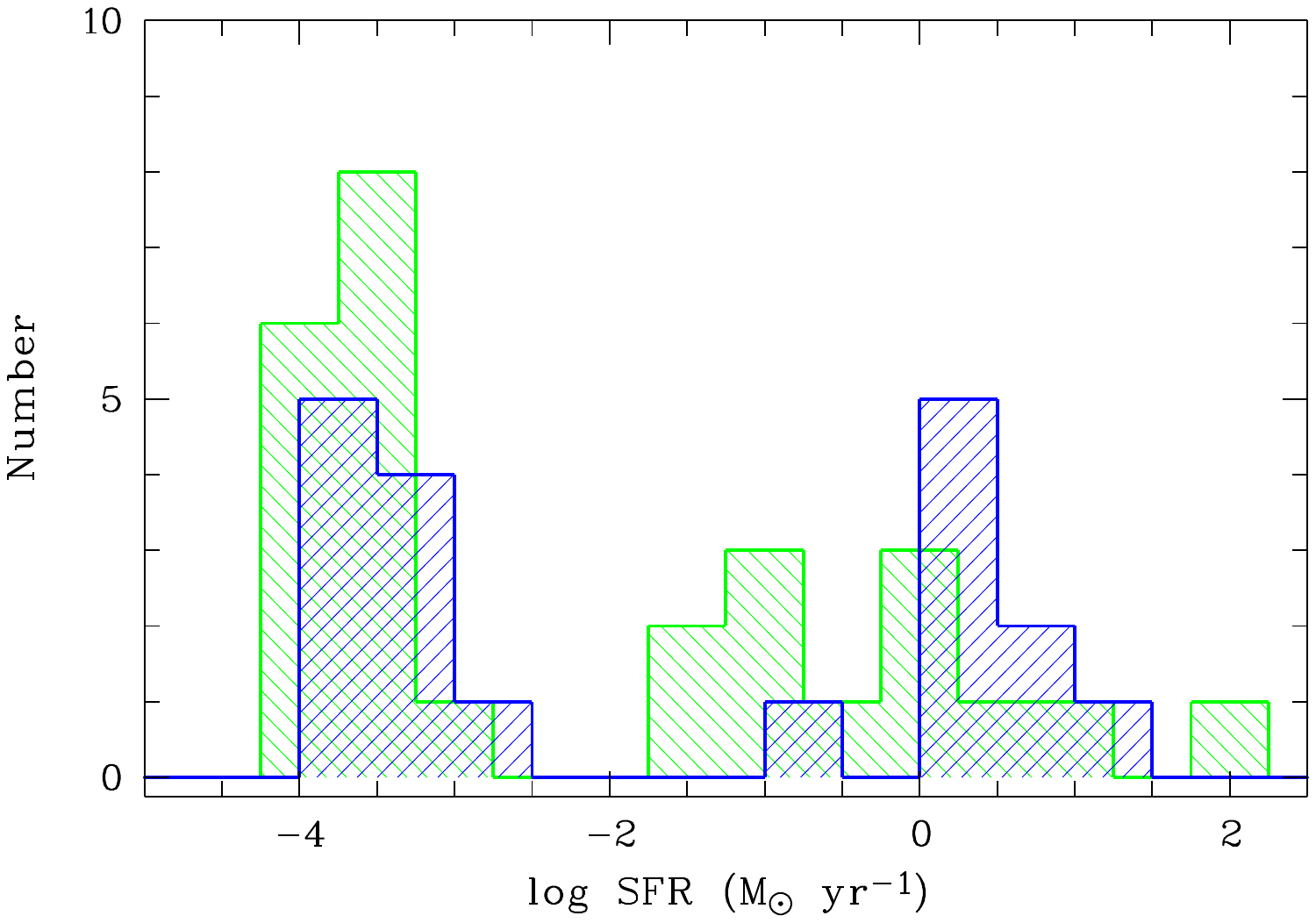} 
\end{center}
\caption{
Stellar mass and SFR.
Top and middle: SFR versus stellar mass from  \protect\citet{Chang_2015} for the PSS-CSO and the ECS samples. 
The colour coding of the symbols is the same as in Fig.\,\ref{fig:D4000n_sSFR}, filled squares signify galaxies with good model fits.   The cyan  curves are local density contours maps for $\sim 9\cdot10^4$ SDSS galaxies. The black lines indicate the SF sequence (solid) and the $3\sigma$ scatter (dashed), the blue lines are the corresponding extrapolations to $\log M_\ast/M_\odot = 12$. Bottom: Histogram distributions of the SFR for the PSS-CSO (blue) and ECS (green) galaxies with ${\tt FLAG} = 1$.
}
\label{fig:m_sfr}
\end{figure}

SFR values from simple model fits are unreliable for galaxies where the fluxes are likely to be significantly affected by  the AGN component. It has been known for a long time that the 4000\,\AA\ discontinuity in the spectra of galaxies correlates with the ratio of the present SFR to the past-averaged SFR and can thus be used to measure SF histories  \citep[e.g.][]{Bruzual_1983, Kauffmann_2003a, Gallazzi_2014, Haines_2017, Borghi_2021}, where it must be noted that there is also a correlation with metallicity \citep{Vazdekis_2015}.  \citet{Brinchmann_2004} suggested to use the 4000\,\AA\ break index D4000n  \citep{Balogh_1999} to estimate  the in-fibre SFR in composite and AGN galaxies. This index expresses the ratio of the average flux densities  $F_\nu$ in the narrow bands 4000 - 4100\,\AA\  and 3850 - 3950\,\AA. The values of D4000n are available from the MPA-JHU catalogues (SDSS table {\tt galSpecIndx}) for  $\sim 80\%$ of the galaxies (class = {\tt GALAXY}) in our samples.  For the other galaxies, we measured  D4000n from the SDSS spectra.  Using reliable data only  ($\mbox{S/N} > 3$),  the mean values of D4000n are $1.70 \pm 0.04$ for the PSS-CSO and $1.88 \pm 0.02$ for the ECS sample (Table\,\ref{tab:gal_subsamples}).  The larger value for the latter is consistent with  the stronger contribution of the old stellar population seen in the composite spectrum (Sect.\,\ref{sect:SDSS_comp}).

Robust data on SFRs and stellar masses for more than $8\,10^5$ galaxies of  the full SDSS spectroscopic galaxy sample have been provided by \citet{Chang_2015}.   Based on the combination of SDSS and WISE photometry,  the authors created SEDs that cover the wavelength range 0.4 - 22 $\mu$m. They employed MAGPHYS for SED modelling that consistently treats stellar emission along with absorption and re-emission by interstellar dust.  The differences between their data and the MPA-JHU data  is based on the addition of the MIR fluxes,  the use of  updated Galactic extinction corrections, and different dust attenuation laws.  Cross-correlation with the \citet{Chang_2015} catalogue yields 38 matches for our PSS-CSO-G sample  and 40 matches for the ECS-G sample.  Following the recommendation of \citet{Chang_2015}, we focus on objects with {\tt flag} = 1,  i.e. galaxies with reliable aperture corrections, good WISE photometry and good-quality SED fits. This reduces the samples to 19 (PSS-CSO) and 27 (ECS) galaxies, all at $z < 0.2$.  Among them, there is only one galaxy with $t_{\rm SED} > 2$, namely the ECS galaxy SDSS\,J094124.02+394441.8, which has been classified as a starburst galaxy ($t_{\rm SED} = 3$).  More than 90\% of the galaxies in both samples have SFRs $\la 5 \mathcal{M}_\odot\,{\rm yr}^{-1}$. Larger SFRs are found only for  SDSS\,J075756.71+395936.1  (10\,$\mathcal{M}_\odot\,{\rm yr}^{-1}$) from the PSS/CSS sample and the two sources SDSS\,J094124.02+394441.8 (93\,$\mathcal M_\odot$\,yr$^{-1}$) and SDSS\,J120732.92+335240.1 (13\,$\mathcal M_\odot$\,yr$^{-1}$) from the ECS sample.

The ratio of the present to the past-averaged SFR is expressed by the specific star formation rate sSFR = SFR/$\mathcal M _\ast$, where $\mathcal M_\ast$ is the stellar mass. The sSFR versus D4000n diagram for $\sim 10^5$ objects from the main SDSS galaxy sample with redshifts $z<0.2$  shows two clouds with the maximum population densities at  $\mbox{D4000n} \sim 1.3$ and  1.9, respectively  \citep{Kauffmann_2003a, Brinchmann_2004}.   The first cloud comprises SF galaxies, the second one consists of  old elliptical galaxies.  Figure\,\ref{fig:D4000n_sSFR} shows the sSFR versus D4000n diagrams for our samples.  The dotted line marks the mean relation for the SF sequence and the dashed line indicates the border between the SF and the passive region,  both taken from  \citet[][their Fig.\,11]{Brinchmann_2004}.  As expected, the population below the demarcation line is strongly dominated by galaxies classified as early type.  The diagrams look different for the two samples with more main-sequence galaxies in the PSS-CSO sample:  $37\pm 11$\,\% of the PSS-CSO galaxies with {\tt flag}=1 lie above the demarcation line, but only $11\pm 6$\,\% of the ECS galaxies ($42\pm 8$\,\% and  $8\pm 4$\,\% respectively for all galaxies).

The bottom panel of Fig.\,\ref{fig:D4000n_sSFR} shows the histogram distributions of the D4000n index. There is a strong peak at $\mbox{D4000n} \sim 1.9$ in the comparison sample, reflecting again the dominance of old stellar populations. This peak is also present in the PSS-CSO sample,  but there is a stronger peak at $\sim 1.6$, which reflects the clump at the end of the SF sequence seen in the top panel.  According to the two-sample KS test, the difference between both distributions is highly significant ($p = 3\cdot10^{-5}$).

Figure\,\ref{fig:m_sfr} displays the SFR vs. stellar mass diagrams. For comparative purposes, we  overplotted the distribution of the data from \citet{Chang_2015} for $\sim 9\cdot10^4$ galaxies from the SDSS spectroscopic sample with $z<0.2$ by equally spaced local density contours.\footnote{Following \citet[][their Fig.\,11]{Chang_2015},  we used only the  86\,923 galaxies above the (redshift-dependent) mass completeness limit for the contour map and applied the standard $1/V_{\rm max}$ weighting \citep{Schmidt_1968}. The contours were computed for a grid size of $(\Delta \log {\mathcal M}_\ast, \Delta \log\,\mbox{SFR}) = (0.05,0.1)$ and are spaced by a factor of 1.5, starting at 0.66.} Also plotted is a power law approximation of the SF sequence  and the $3\sigma$ limits with $\sigma = 0.39$ dex, both from \citet{Chang_2015}. Both the PSS-CSO and the ECS samples span a relatively narrow range of stellar masses at the high-mass end of the galaxy distribution. The derived masses are at the low-mass end of the mass distribution of the brightest cluster galaxies  \citep{Dalal_2021}, but are typical of luminous radio galaxies. On the other hand, the SFRs cover a wide range from close to the SF sequence down to almost zero.   The diagrams again indicate a lower proportion of SF galaxies in the ECS sample: $42\pm 10$\,\% of the PSS-CSO galaxies with {\tt flag}=1 lie above the lower limit of the extrapolated SF sequence, but only $22\pm 8$\,\% of the ECS galaxies ($45\pm 8$\,\% and $18\pm 6$\,\% for all galaxies). The histogram distributions of the SFR in the two samples are displayed in the bottom panel of Fig.\,\ref{fig:m_sfr} for the sources with good quality SFR data.  According to the two-sample KS test, the difference between the two distributions is significant ($p = 0.03$) when we include all galaxies, but not for the rather small samples of sources with {\tt flag}=1. A similar result is found for the sSFR distributions.  The mean values of D4000n, $\log\,M_\ast, \log\,\mbox{SFR}$, and  $ \log\,\mbox{sSFR}$ for the two samples are listed in Table\,\ref{tab:gal_subsamples}.

\begin{figure*}[htbp]
\includegraphics[viewport=0 0 540 540,width=6.0cm,angle=0]{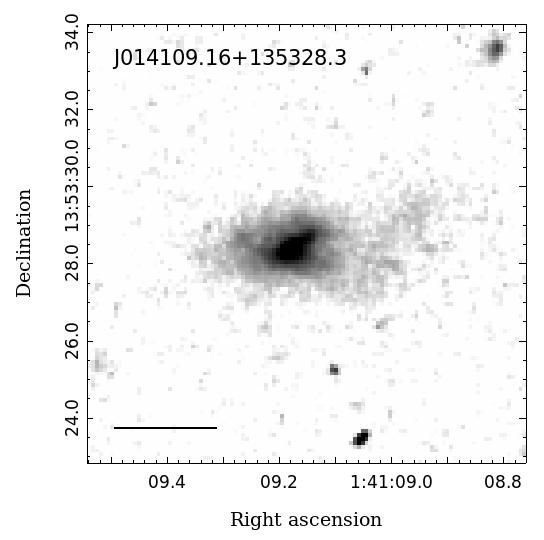}
\includegraphics[viewport=0 0 540 540,width=6.0cm,angle=0]{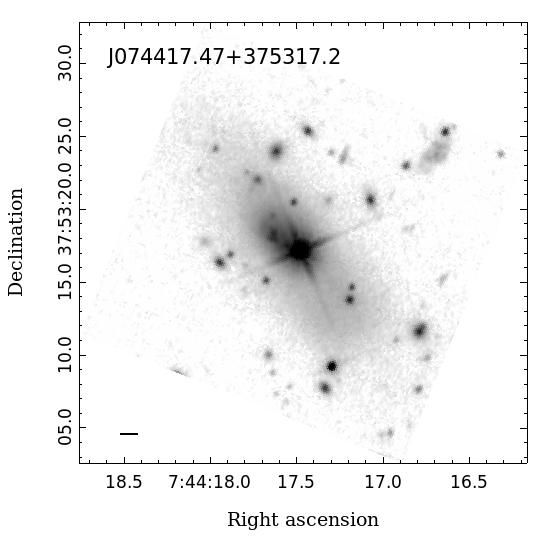}
\includegraphics[viewport=0 0 540 540,width=6.0cm,angle=0]{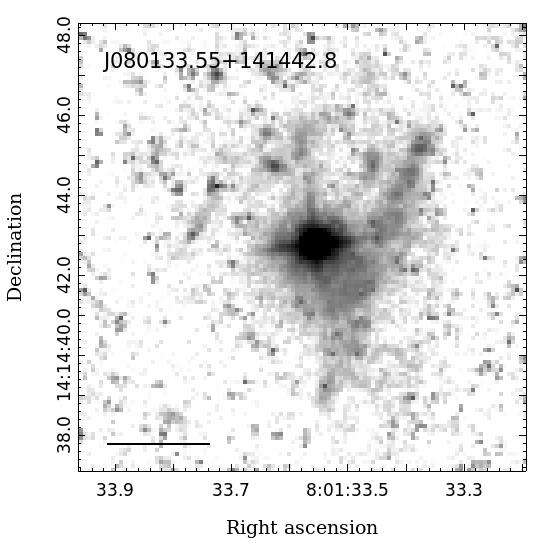}\\
\includegraphics[viewport=0 0 540 540,width=6.0cm,angle=0]{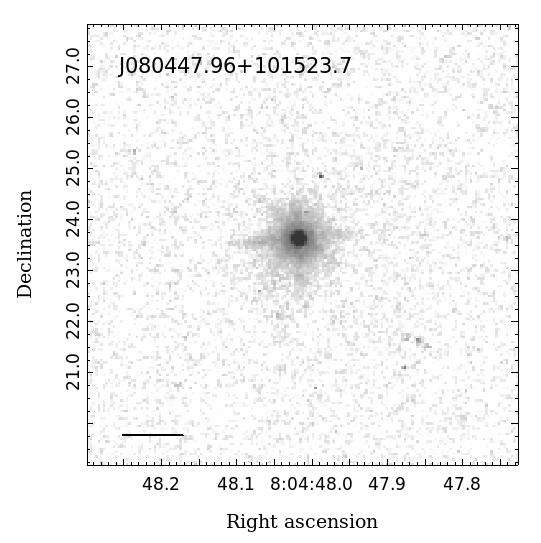}
\includegraphics[viewport=0 0 540 540,width=6.0cm,angle=0]{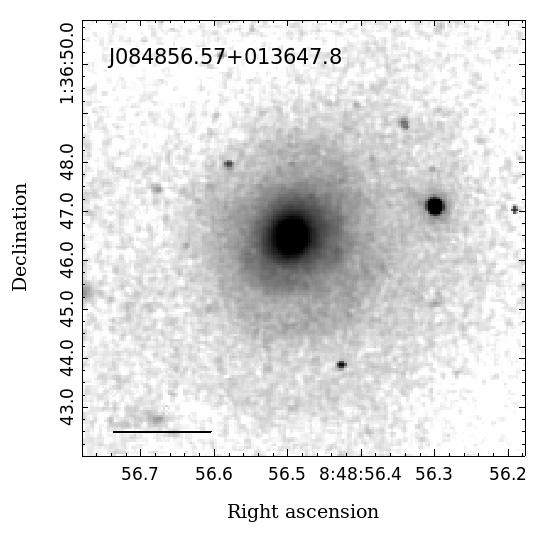}
\includegraphics[viewport=0 0 540 540,width=6.0cm,angle=0]{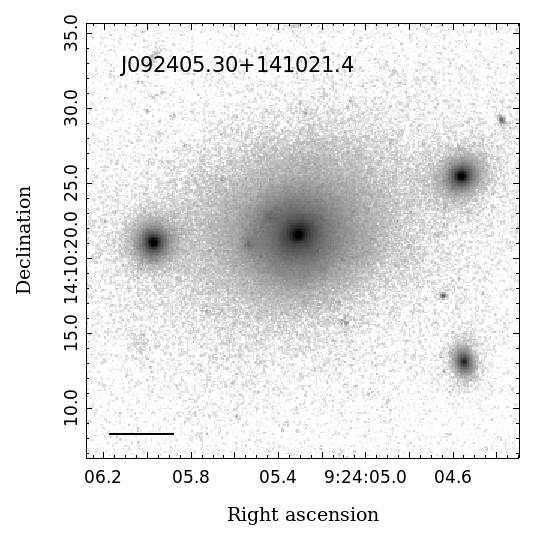}\\
\includegraphics[viewport=0 0 540 540,width=6.0cm,angle=0]{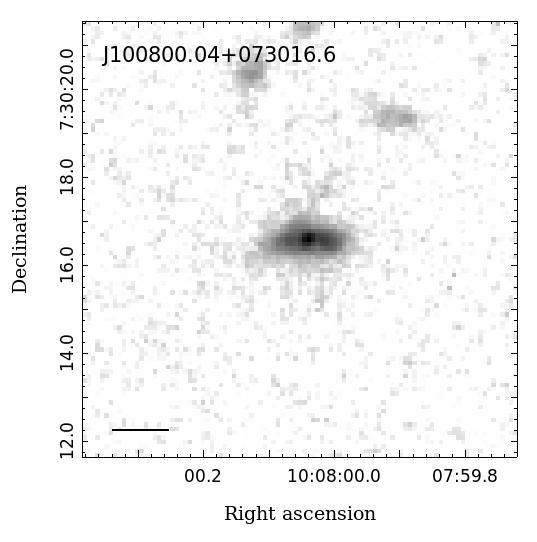}
\includegraphics[viewport=0 0 540 540,width=6.0cm,angle=0]{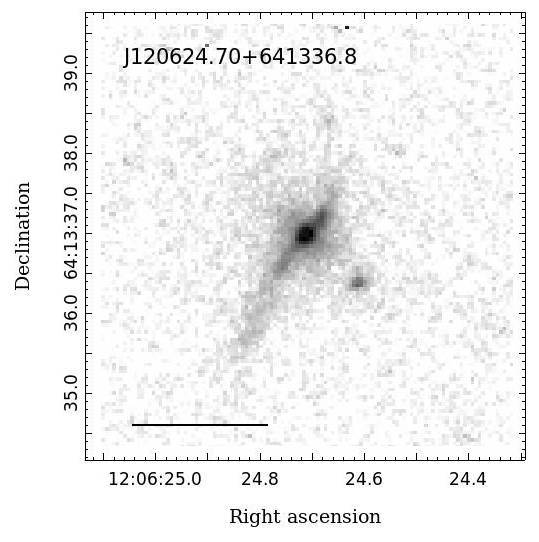}
\includegraphics[viewport=0 0 540 540,width=6.0cm,angle=0]{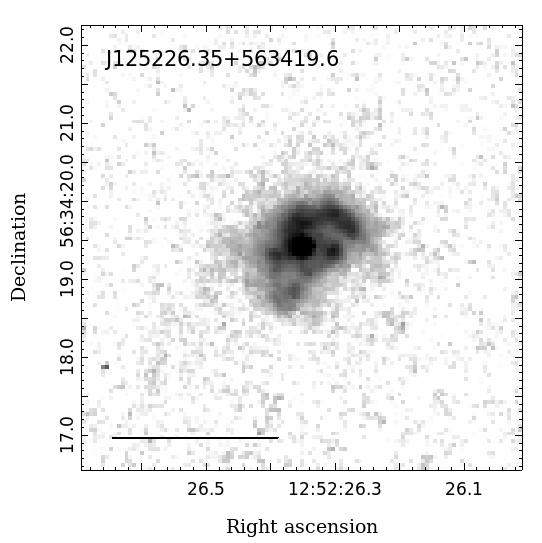}\\
\includegraphics[viewport=0 0 540 540,width=6.0cm,angle=0]{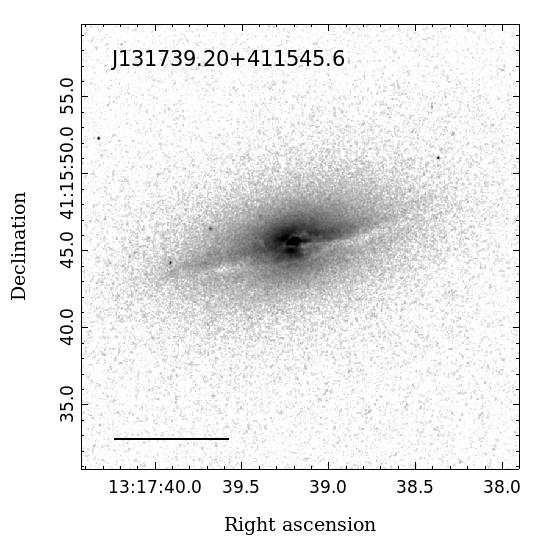} 
\includegraphics[viewport=0 0 540 540,width=6.0cm,angle=0]{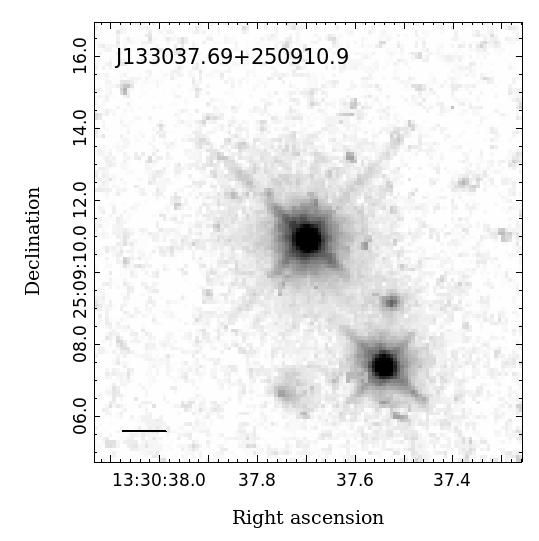}
\includegraphics[viewport=0 0 540 540,width=6.0cm,angle=0]{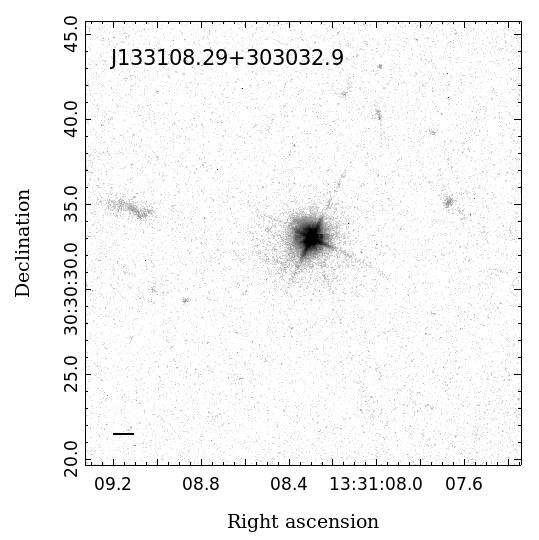}
\caption{
Image cutouts from the HST Legacy Archive (WFPC2, ACS, WFC3/UVIS with various filters) for 18 PSSs and CSOs  (always in the image centre).  The image size is adapted to the size of the ambient structure, the bar indicates 10 kpc at the distance of the source  (N up, E left).
}
\label{fig:HST}
\end{figure*}

\setcounter{figure}{15}
\begin{figure*}[htbp]
\includegraphics[viewport=0 0 540 540,width=6.0cm,angle=0]{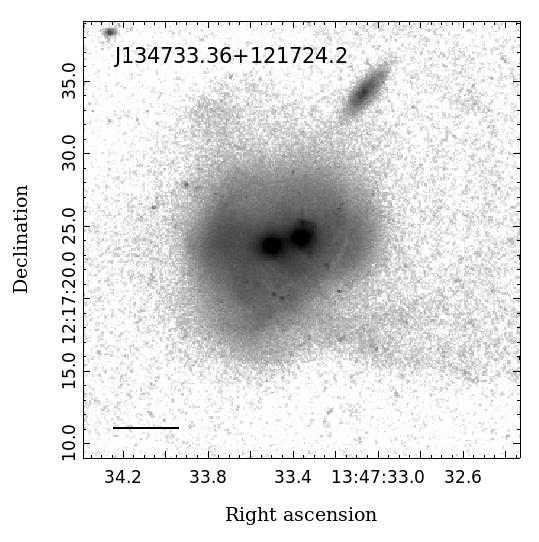}
\includegraphics[viewport=0 0 540 540,width=6.0cm,angle=0]{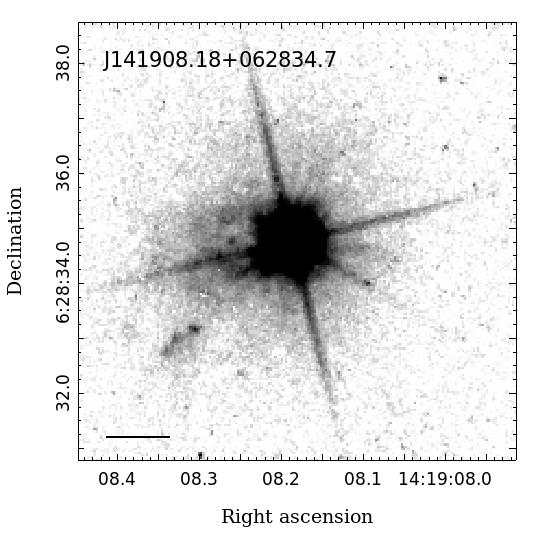}
\includegraphics[viewport=0 0 540 540,width=6.0cm,angle=0]{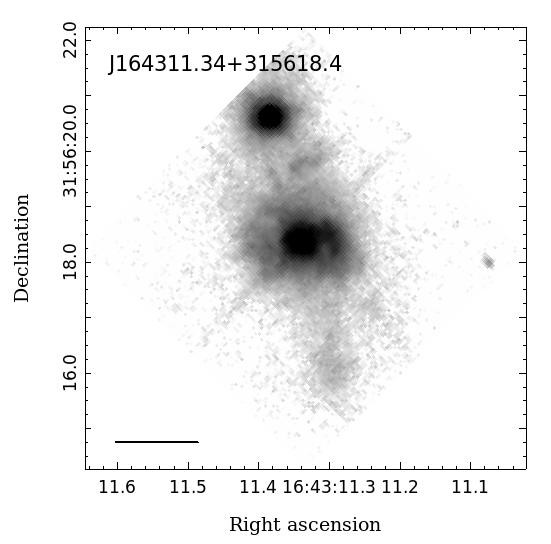}\\
\includegraphics[viewport=0 0 540 540,width=6.0cm,angle=0]{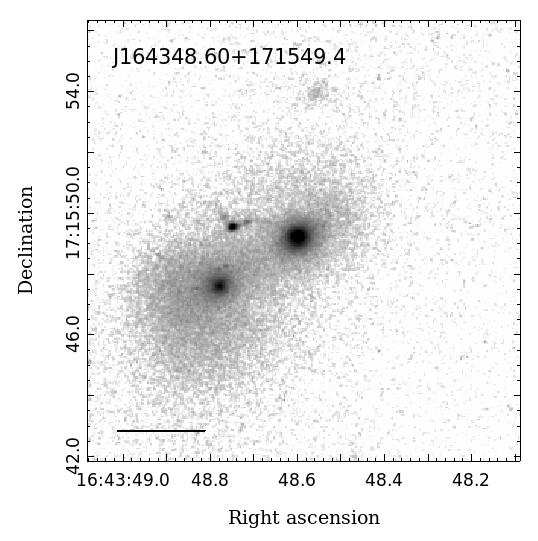}
\includegraphics[viewport=0 0 540 540,width=6.0cm,angle=0]{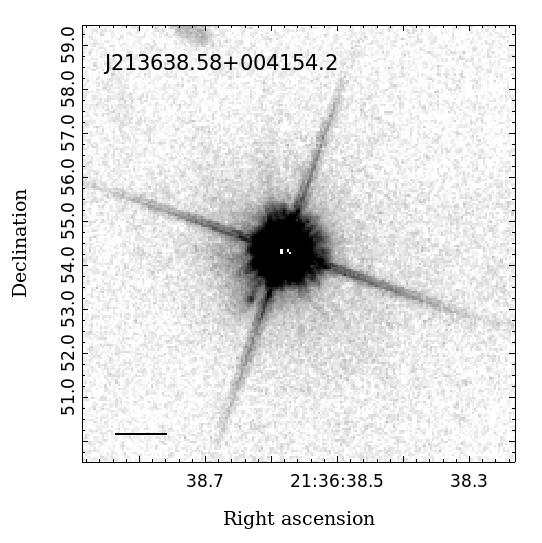}
\includegraphics[viewport=0 0 540 540,width=6.0cm,angle=0]{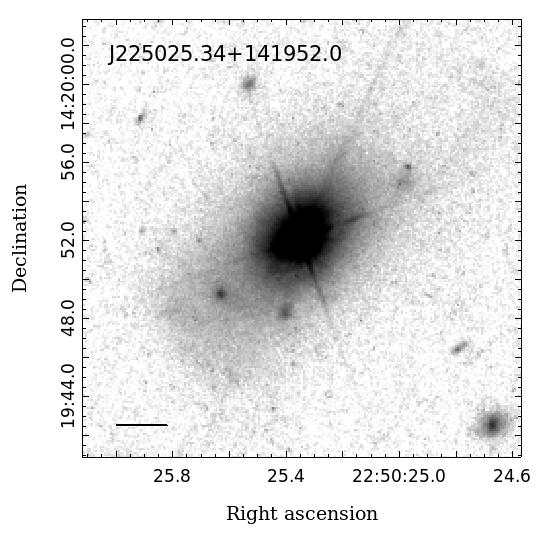}
\caption{
continued.
}
\end{figure*}

%
\section{Optical morphology}\label{sect:morph}
%

This section addresses the issue of interactions between galaxies and the merging of galaxies as one of the main drivers of core activity.  Extended tidal structures, shells, rings,  strong lopsidedness, or simply close galaxy pairs are usually taken as signposts of gravitational perturbation induced by a close encounter or merger.  Finding such structures in distant galaxies is a challenge.  Detecting extended low-surface brightness features requires deep observations and discovering peculiar structures in the inner regions of galaxies requires high  spatial resolution.  We have taken two approaches to achieve this goal.  First, we analysed archival images from the Hubble Space Telescope (HST)  for a subsample of 19 sources.  Second, we performed a statistical analysis of the subsamples of the low-$z$ galaxies on images from major galaxy surveys.  We started with the combination of the images from SDSS and  PanSTARRS-1 \citep[PS1,][]{Flewelling_2020} for the sources up to  $z=0.2$, and then we also analysed the images from the DESI Legacy Surveys \citep{Dey_2019} up to  $z=0.3$.

If the data set is not too large, the human visual system is still  the most efficient and reliable for recognising complex patterns. To determine the fraction of mergers or otherwise morphologically distorted galaxies, a classification system has to be defined.  Tidal structures are very diverse, dependent on the properties of the involved galaxies  and the merger parameters \citep[e.g.][]{Duc_2013, Ren_2020}. With regard to the subjectivity of an inspection by eye  we prefer here the simple classification scheme that was applied to post-SB galaxies by  \citet{Meusinger_2017}. It classifies galaxy morphology  into one of four types that are assigned to the numerical distortion flag $t_{\rm m}$ (index `m' for merger):
0 = no apparent peculiarities,
1 = weak indications of a structure that seems to be related to a neighbour galaxy (Messier 51 type),
2 = weak lopsidedness, tidal structures, such as streamers, tidal arms, fans, or shells,
3 = the same characteristics as for  $t_{\rm m} = 2$, but very pronounced.
Furthermore, $t_{\rm m}  = -1$ was provided if an evaluation is not possible, but this did not apply in any case.

A neighbouring galaxy (in the projection) is only considered to indicate an interaction if the two galaxies are either embedded in a common halo or connected by a light bridge. A lower value of $t_{\rm m}$ can, of course, be the result of factors that make it difficult to detect low-surface brightness structures,  such as a bright AGN (e.g. J080133.55+141442.8 in Fig.\,\ref{fig:HST}) or blending of the host by a nearby bright object. In general, it is difficult to distinguish minor mergers from old and faded major mergers. A value of $t_{\rm m}=2$ does therefore not necessarily mean that the distortions are intrinsically weaker than for $t_{\rm m}=3$.  In the following, we consider $t_{\rm m} \ge 2$ as indicative of strong tidal interactions and mergers.


\subsection{HST images}\label{Sect:HST_images}


The high spatial resolution of the HST is a great advantage for the detection of merger signatures, especially in the inner regions of the galaxies and at higher redshifts.  We searched the HST Legacy Archive\footnote{https://hla.stsci.edu/} for images of the PSS-CSO hosts from our sample.   We found images of 18 PSS-CSO galaxies from different observation programmes.  (In addition, the HST image of the discarded source \object{SDSS J101714.23+390121.1} is shown and discussed separately in Appendix\,\ref{sect:SDSSJ1017}.) Although this subsample is small and too inhomogeneous for the statistical analysis, it  gives an impression of the frequency of nearby neighbours and peculiar morphologies.

Figure\,\ref{fig:HST} shows the HST images, which were obtained with different cameras and through different filters. If several images from the same camera are available, they were stacked to increase the signal-to-noise ratio.  For the two most distant objects, \object{SDSS J080447.96+101523.7} ($z = 1.968$) and  \object{SDSS J213638.58+004154.2} ($z = 1.941$),  the host galaxies are not visible.  The same applies to \object{J133037.69+250910.9} ($z = 1.055$).\footnote{In SDSS DR16,  the other star-like object in the image is listed with AGN-typical WISE colours, which is based however on a misidentification with J133037.69+250910.9. The search in the IPAC Infrared Science Archive reveals the Spitzer colour $[3.6]-[4.5] = 0.1$, which is far away from the AGN wedge \citep{Stern_2005} and is consistent with stellar origin.  On the other hand, \object{J133037.69+250910.9} has AGN-typical MIR colours $[3.6]-[4.5] = 0.78$ and $[5.8]-[8.0]=1.13$.}  All three objects are spectroscopically classified as quasars. The HST image of \object{J131739.20+411545.6} ($z = 0.066$) strengthens the case for the existence of PSSs and CSOs in galaxies with significant disk components.   The edge-on view makes it difficult to detect possible morphological distortions in this galaxy.  All other host galaxies appear distorted or have nearby neighbours (\object{J092405.30+141021.4} and \object{J100800.04+073016.6}),  which indicates a high merger fraction of PSS-CSO galaxies.


\subsection{Images of low-$z$ galaxies from major surveys}\label{Sect:SDSS_images}


We applied the above classification  scheme to the subsamples of the low-$z$ host galaxies. According to our experience from previous studies \citep[e.g.][]{Meusinger_2017},  it is difficult to recognise tidal features from galaxy-galaxy interactions on SDSS images at redshifts larger than about 0.2.  Therefore, we restricted this analysis  to  $z < 0.2$,  which results in a list of altogether 67 galaxies: 33 from the PSS-CSO and 34 from the ECS sample.

One disadvantage of our method is, of course, the inevitable subjective character of the classification by eye.  Reliable conclusions about the relative frequency of distorted morphologies of PSS-CSO galaxies are only meaningful in relation to the comparison sample. In order to reduce the risk of a subjective bias towards over- or underestimation of peculiarities in the target sample,  it is important to ensure an unbiased assessment of the images of both samples in a `blind' approach.  To do so, we stored the images from both samples in the same archive and selected them randomly for the assessment, whereby the examiner could not know which sample they belonged to.  Simultaneous inspection of the SDSS and the PS1 i-band image\footnote{In general, PS1 is similar in depth to SDSS in g and r, but slightly deeper in i.} significantly reduces the risk of false detection of weak features due to noise or artefacts.  A peculiar structure is only considered recognised if it is visible on both images.  The images were evaluated in three independent runs.  In each run, a merger flag  $t_{\rm m}$ was assigned to each galaxy.  For 26 PSS-CSO galaxies (81\%), identical values were estimated in all three runs.

Based on these results, relative merger fractions  $f_{\rm m} = N_{\rm m} /N_{\rm tot}$ were computed,  where $N_{\rm m}$ is the number of mergers and $N_{\rm tot}$ is the total number of galaxies in the sample with $t_{\rm m} \ge 0$.  Using the merger criterion $t_{\rm m} \ge 2$, we find 16 mergers in the PSS-CSO but only 4 in the ECS sample.

Deeper optical images were recently provided by the DESI Legacy Imaging Surveys.  For our study, using deeper images not only has the advantage of improving the confidence in detecting peculiar morphological structures,  but also that the redshift range of the low-$z$ sample can be expanded and thus the statistics are improved. In the second step, we therefore analysed the image cutouts from the DESI Legacy Surveys DR10\footnote{https://www.legacysurvey.org/dr10/description/},  which completed three imaging projects on different telescopes in the g, r, and z bands and incorporates additional data in the i band.  The procedure is the same as before, the $z$ range has been expanded up to $z = 0.3$.  This new low-$z$ sample now consists of 43 galaxies from the PSS-CSO sample (Table\,\ref{tab:low_z_PSS}) and 41 ECS galaxies. 
Now we found 26 mergers in the PSS-CSO sample and only six in the ECS sample. The results  confirm the previous finding of a significantly higher merger fraction  for the PSS-CSO sample compared to  for the ECS sample.  The mean peculiarity index is $t_{\rm m} = 1.47\pm0.19$  for the former and  $t_{\rm m} = 0.45\pm0.14$  for the latter sample. In the following we only refer to the results from the Legacy Surveys images.
The mean $t_{\rm m}$ values from the three runs are listed in the last column of Table\,\ref{tab:low_z_PSS} along with other  properties of the sources from this low-$z$ subsample.\footnote{Because this study is focused on the host galaxies, we list the SDSS names in column 2, not the names of the radio sources.  The latter can be found in \citet{Liao_2020}.}

Examples of merger galaxies are shown and briefly described in Appendix\,\ref{sect:App_LS_images}. The resulting merger fraction $f_{\rm m}^{\rm \, PSS} =  0.61\pm 0.07$ for the PSS-CSO sample is in accordance with the range of 0.4 - 0.7 from previous systematic studies  \citep[][and references therein]{Odea_2021}.  We note however that the numbers are not directly comparable to each other because of the differences in the observational material, the evaluation methods and the merger criteria. The merger fraction of the ECS sample is much lower,  $f_{\rm m}^{\rm \, ECS} =   0.15\pm 0.06$. According to the one-tailed two-sample Z test, the difference is highly significant ($p < 10^{-5}$).

\begin{figure}[htbp]
\includegraphics[viewport= 0 0 822 580,width=8.4cm,angle=0]{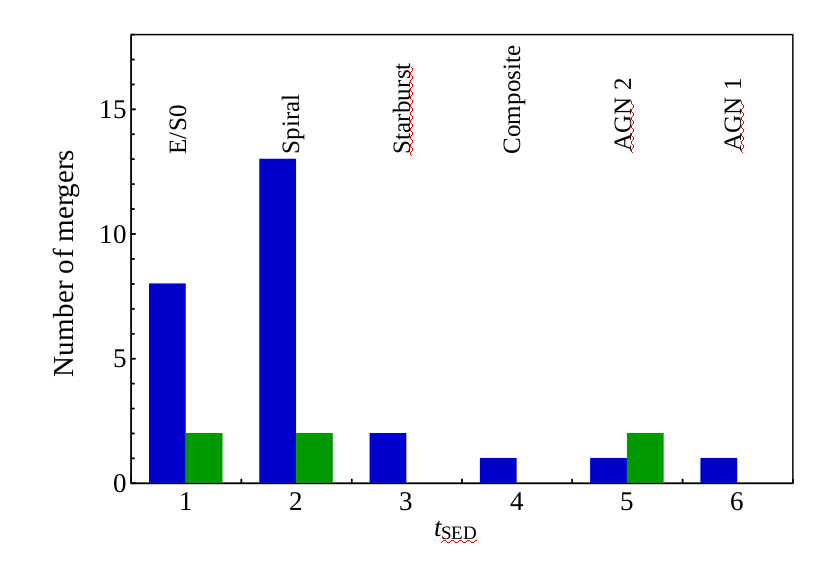}
\caption{
Number of mergers per MBSED type $t_{\rm SED}$ for PSS-CSO (blue) and ECS galaxies (green) with $z < 0.3$.
}
\label{fig:hist_number_merger}
\end{figure}

\begin{figure}[htbp]
\includegraphics[viewport= 0 0 822 580,width=8.4cm,angle=0]{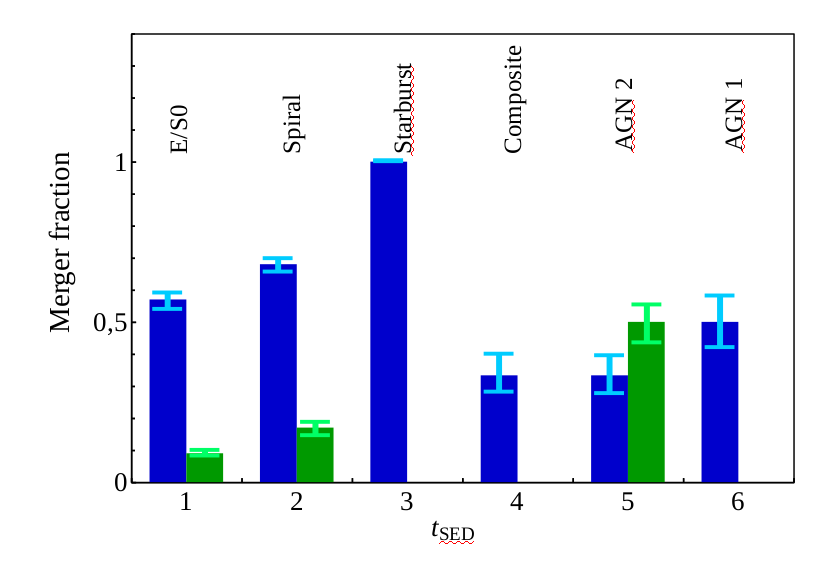}
\caption{
Merger fraction per MBSED type $t_{\rm SED}$ for PSS-CSO (blue) and ECS galaxies (green) with $z < 0.3$.
The vertical bars indicate the standard errors of the proportions.
}
\label{fig:hist_fraction_merger}
\end{figure}

It is well known that gravitational interactions and merging can elevate the SFR in galaxies.  This is also seen  in our data. The histograms in Figs.\,\ref{fig:hist_number_merger} and \ref{fig:hist_fraction_merger} show the distributions of the number of mergers across the different MBSED types and the merger fraction per type, respectively. $62\pm 9$\,\% of the merger galaxies in the PSS-CSO low-$z$ sample show SEDs of star forming galaxies ($t_{\rm SED} = 2, 3$ or 4).  For the non-merger galaxies, this proportion is smaller, $47\pm 12$\,\%. All four PSS-CSO host galaxies from Table\,\ref{tab:low_z_PSS} with SFR $> 10\,{\mathcal M}_\odot\,{\rm yr}^{-1}$  are mergers (Table\,\ref{tab:low_z_PSS}), although the reliability flag is set to 1 for only one them (SDSS J075756.71+395936.1). Conversely, not all disturbed galaxies with  $t_{\rm m} \ge 2$ have substantial SF, which is expected particularly for dry mergers.

\begin{table*}[htbp]
\caption{The subsample of low-redshift  ($z<0.3$) PSS-CSO galaxies. }
\begin{tabular}{rcccccclccclc} 
\hline\hline 
\noalign{\smallskip}
n 
& Name
& $z$
& $R_{\rm i}$
& Type
& Class
& BPT
& Bft
& $t_{\rm SED}$ 
& $t_{\rm WISE}$
& logSFR
& log$\mathcal{M}_\ast$
& $t_{\rm m}$\\
(1)
& (2)
& (3)
& (4)
& (5) 
& (6)
& (7)
& (8) 
& (9)
& (10)
& (11)
& \ \ (12)
& (13)\\
\hline
\noalign{\smallskip}
 1 & \object{J002833.42+005510.9} & 0.104 &  2.52 & CSS & LERG  & Sy  & Sc      & 2  &     inter  &  \  0.317 &    10.58 & 3.00 \\
 2 & \object{J075756.71+395936.1} & 0.066 &  1.42 & CSS & HERG  & Sy  & Sb      & 2  &     SF     &  \ 1.017 &    11.00 & 3.00 \\
 3 & \object{J081323.75+073405.6} & 0.112 &  2.17 & CSS & LERG  & L   & Sa      & 2  &     inter  &   -3.398 &    11.55 & 2.00 \\
 4 & \object{J082504.56+315957.0} & 0.265 &  1.94 & CSS &\ LERG: & \ L:  & Ell5    & 1  &     pass   &   -2.988: &    11.93: & 2.00 \\
 5 & \object{J083139.79+460800.8} & 0.131 &  1.76 & GPS & LERG  & L   & Ell13   & 1  &     pass   &   -3.388 &    11.58 & 0.33 \\
 6 & \object{J083411.09+580321.4} & 0.093 &  1.07 & GPS & LERG  & L   & Ell13   & 1  &     pass   &   -3.508: &    11.48: & 0.00 \\
 7 & \object{J083637.84+440109.6} & 0.055 &  1.46 & CSS & HERG  & S   & S0      & 1  &     inter  &   -2.808 &    11.17 & 2.00 \\
 8 & \object{J090615.53+463619.0} & 0.085 &  1.98 & GPS & LERG  & C   & Sa      & 2  &     inter  &  \  0.272 &    11.35 & 0.00 \\
 9 & \object{J092405.30+141021.4} & 0.136 &  2.38 & CSS & LERG  & L   & Sy2     & 5  &     pass   &          &         & 0.00 \\
10 & \object{J093430.68+030545.3} & 0.225 &  2.71 & CSS & LERG  & L   & Sb      & 2  &     pass   &  \  0.322 &    11.34 & 3.00 \\
11 & \object{J093609.36+331308.3} & 0.076 &  0.99 & GPS & LERG  & L   & Ell13   & 1  &     pass   &   -3.513 &    11.45 & 2.00 \\
12 & \object{J094525.90+352103.6} & 0.208 &  2.54 & CSS & QSO   & S   & Sdm     & 2  &     inter  &          &         & 2.00 \\
13 & \object{J100955.51+140154.2} & 0.213 &  3.22 & CSS & LERG  & L   & Sy18    & 5  &     inter  &  \  1.007: &    11.51: & 0.00 \\
14 & \object{J102618.25+454229.3} & 0.152 &  2.00 & CSO & LERG  & L   & Sa      & 2  &     pass   &   -3.533 &    11.43 & 0.00 \\
15 & \object{J103719.33+433515.3} & 0.025 &  0.61 & CSS & LERG  & L   & S0      & 1  &     pass   &   -3.758: &    11.17: & 3.00 \\
16 & \object{J104029.94+295757.7} & 0.091 &  2.11 & CSS & LERG  & L   & Sa      & 2  &     inter  &  \  0.572 &    11.13 & 3.00 \\
17 & \object{J105731.17+405646.1} & 0.025 &  0.01 & GPS & LERG  & C   & Ell5    & 1  &     pass   &   -3.603: &    11.26: & 3.00 \\
18 & \object{J114339.59+462120.4} & 0.116 &  2.01 & CSS & LERG  & L   & Ell13   & 1  &     pass   &   -3.248 &    11.65 & 0.00 \\
19 & \object{J115000.08+552821.3} & 0.139 &  2.17 & CSS &\ LERG: & L   & Sb      & 2  &     inter  &  \  0.122 &    11.33 & 2.00 \\
20 & \object{J115727.60+431806.3} & 0.230 &  2.73 & CSS & QSO   & C   & QSO1    & 6  &     QSO    &          &         & 0.00 \\
21 & \object{J120902.79+411559.2} & 0.096 &  1.56 & CSS &\ LERG: & L   & Ell13   & 1  &     pass   &          &         & 2.33 \\
22 & \object{J124419.96+405136.8} & 0.249 &  2.85 & CSS & QSO   & Sy  & Arp220  & 4  &     inter  &          &         & 0.00 \\
23 & \object{J124733.31+672316.4} & 0.107 &  2.19 & CSO & LERG  & L   & Ell13   & 1  &     pass   &   -3.563 &    11.40 & 0.00 \\
24 & \object{J131739.20+411545.6} & 0.066 &  1.47 & GPS & LERG  & L   & Ell13   & 1  &     pass   &   -3.423 &    11.56 & 0.00 \\
25 & \object{J132419.67+041907.0} & 0.263 &  2.08 & CSS & LERG  & C   & Ell2    & 1  &     pass   &   -3.088: &    11.87: & 0.00 \\
26 & \object{J132513.37+395553.2} & 0.076 &  0.97 & GPS & LERG  & L   & Ell5    & 1  &     pass   &   -3.543 &    11.42 & 2.00 \\
27 & \object{J134035.20+444817.3} & 0.065 &  1.63 & GPS & HERG  & Sy  & I20551  & 3  &     QSO    &  \  1.022: &    10.79: & 3.00 \\
28 & \object{J134733.36+121724.2} & 0.120 &  3.02 & GPS & QSO   & Sy  & N6090   & 3  &     QSO    &          &         & 3.00 \\
29 & \object{J140051.58+521606.5} & 0.118 &  2.01 & CSS & LERG  & L   & Sa      & 2  &     inter  &  \  0.137 &    11.31 & 0.00 \\
30 & \object{J140700.39+282714.6} & 0.077 &  2.10 & GPS & QSO   & C   & QSO2    & 5  &     QSO    &          &         & 3.00 \\
31 & \object{J140942.44+360415.9} & 0.149 &  2.15 & CSS & QSO   & Sy  & Sb      & 2  &     inter  &          &         & 2.00 \\
32 & \object{J141327.22+550529.2} & 0.282 &  2.63 & CSS & \ LERG: & L   & Sc      & 2  &     inter  &  \  0.502: &    11.27: & 2.00 \\
33 & \object{J141558.81+132023.7} & 0.247 &  3.89 & CSO & \ HERG: & \ Sy: & Mrk231  & 4  &     QSO    &          &         & 0.00 \\
34 & \object{J143521.67+505122.9} & 0.100 &  1.83 & CSS & LERG  & L   & Sb      & 2  &     inter  &   -3.773 &    11.20 & 2.33 \\
35 & \object{J144712.76+404744.9} & 0.195 &  2.98 & CSS & LERG  & Sy  & Sb      & 2  &     inter  &  \  0.467 &    11.10 & 0.00 \\
36 & \object{J151141.26+051809.2} & 0.084 &  1.62 & HFP & QSO   & Sy  & I19254  & 4  &     QSO    &          &         & 2.00 \\
37 & \object{J152349.34+321350.2} & 0.110 &  2.29 & CSS & HERG  & Sy  & Ell5    & 1  &     pass   &   -0.588 &    10.79 & 2.00 \\
38 & \object{J155927.67+533054.4} & 0.179 &  2.56 & CSS & LERG  & Sy  & Sc      & 2  &     inter  &  \  1.712: &    11.39: & 2.00 \\
39 & \object{J160246.39+524358.3} & 0.106 &  2.24 & CSS & LERG  & L   & Sa      & 2  &     inter  &  \  0.707 &    11.41 & 2.00 \\
40 & \object{J160335.16+380642.8} & 0.241 &  3.12 & CSS & HERG  & Sy  & Sb      & 2  &     inter  &          &         & 0.00 \\
41 & \object{J161148.52+404020.9} & 0.151 &  2.77 & CSS & QSO   & L   & Sa      & 2  &     inter  &          &         & 0.67 \\
42 & \object{J164348.60+171549.4} & 0.163 &  3.21 & CSS & G     & SF: & Sb      & 2  &     inter  &          &         & 2.33 \\
43 & \object{J225025.34+141952.0} & 0.235 &  3.31 & CSS & QSO   & C:  & QSO1    & 6  &     QSO    &          &         & 2.00 \\  
\hline                 
\noalign{\smallskip}
\end{tabular}      
\tablefoot{
(1) running number;
(2) SDSS name;
(3) redshift; 
(4) radio loudness parameter  (see Sect.\,\ref{sect:Samples});
(5) radio source type from \protect\citet{Liao_2020};
(6) optical spectral class;
(7) spectral type based on BPT diagram (Sy = Seyfert, L = LINER, C = composite, SF = star forming); 
(8) best-fitting SWIRE template;
(9) spectral type from MBSED (Table\,\ref{tab:templates}); 
(10) type based on WISE $W1-W2$ vs. $W2-W3$ diagram; 
(11) SFR from \protect\citet{Chang_2015} in units of $\mathcal{M}_\odot$\,yr$^{-1}$; 
(12) stellar mass from \protect\citet{Chang_2015} in units of $\mathcal{M}_\odot$;
(13) merger index based on images from the DESI Legacy Surveys. 
A colon marks uncertain data.}
\label{tab:low_z_PSS}                    
\end{table*}


\subsection{Discussion of the different merger fractions}\label{Sect:discussion}


The merger-AGN connection is the subject of a long-running debate.  Both theoretical arguments and numerical simulations suggest  that  `wet' galaxy mergers  (i.e. with at least one gas-rich component) are capable of efficiently triggering streams of cold gas towards the innermost regions of galaxies, thus in principle creating the conditions for triggering nuclear activity by feeding the central SMBH \citep[e.g.][]{Toomre_1972, Sanders_1996, Kauffmann_2000, DiMatteo_2005, Hopkins_2008, Holt_2009, Blecha_2018, Weigel_2018, McAlpine_2020, Zhang_2021}. However, despite intensive efforts, the observational evidence is inconsistent. Several studies do not point to a close connection between AGNs and (observable evidence for) recent galaxy mergers \citep{Malkan_1998, McLure_1999, Grogin_2005, Georgakakis_2009, Cisternas_2011, Schawinski_2011, Hewlett_2017, Villforth_2017, Lambrides_2021, Sharma_2021}.  In contrast, the view that the phenomenon of luminous AGNs is essentially related to interactions between galaxies is supported by  other studies \citep[][]{Hutchings_1983,  Gehren_1984, Keel_1985,  Heckman_1986, Stanghellini_1993, Bennert_2008, Karouzos_2010, Ramos_2011, Bessiere_2012, Villar-Martin_2012, Sabater_2013, Chiaberge_2015, Weston_2017, Goulding_2018, Gao_2020}. In particular, galaxy mergers may play a role in triggering powerful radio AGNs \citep{Ramos_2012, Chiaberge_2015}.  These seemingly contradictory results could be partly due to differences in the depth of the images used to classify mergers \citep{Bennert_2008, Bessiere_2012} and due to different techniques of selecting the AGN sample  \citep[e.g.][and references therein]{Ellison_2019}.  Furthermore, strong AGN variability  could have an impact if the variability time scale is much shorter than the characteristic lifetime of merger relics   \citep{Hickox_2014, Goulding_2018, French_2023}.

Another controversial connection is that between the triggering of an AGN and the quenching of SF in the host galaxy.  Because of  the strong energetic and mechanical feedback,  AGNs are  assumed to play a decisive role in this process \citep{Fabian_2012, Heckman_2014}.  However, only a few post-SB galaxies show signs of AGNs in the form of strong AGN emission lines or bright emission  in the MIR, X-ray or radio domain \citep{Nielsen_2012, Meusinger_2017, Greene_2020}.  Recently, \citet{Wu_2023} reported a high occurrence rate of AGN-driven [\ion{O}{iii}]\,5007\AA\ emission derived from ultra-deep spectra of a small sample of post-starburst galaxies, suggesting a higher abundance of AGNs than previously thought.

It is not our aim to evaluate the merger fractions from Sect.\,\ref{Sect:SDSS_images} in the context of the debate on the merger-AGN-SF connection.  However, we think it is important to understand the difference between the rates for the PSS-CSO and the comparison sample. 
If the PSS-CSO sample represents young radio galaxies in which a fraction  $f_{\rm m}^{\rm \, PSS}$ was triggered by mergers,  it seems reasonable to explain the lower fraction  in the comparison sample by evolutionary or environmental differences in the observability of the tidal structures. In the following, we examine in more detail the extent to which these two explanations could apply to our samples.

%
\subsubsection{Evolutionary effects}\label{sect:evolution}
%

A simple evolutionary scenario is based on the fact that the tidal structures disappear and become invisible after a certain time. If a significant proportion of young radio galaxies is associated with galaxy mergers and the radio sources are still present for a considerable period after the tidal structures are no longer detectable, a sample of young radio galaxies can be expected to have a higher percentage of mergers than a comparison sample of extended sources.  The main requirements for this scenario are that (i) the comparison sample represents the more evolved stages of the PSSs and CSOs and  (ii) the radio source lifetime is comparable to or longer than the lifetime of the observable morphological merger signatures.

Observations suggest that the merger-AGN association is particularly clear for luminous AGNs obscured by dust \citep{Urrutia_2008, Satyapal_2014, Kocevski_2015, Fan_2016, Weston_2017, Ricci_2017, Goulding_2018, Steinborn_2018, Ellison_2019}, which are predominantly found in the late stages of mergers. \citet{Patil_2020} studied radio galaxies with extremely red MIR-optical colour ratios and found that these sources are 
`consistent with a population of newly triggered, young jets caught in a unique evolutionary stage in which they still reside within the dense gas' and might become GPS or CSS sources. Based on the analysis of cosmological hydrodynamical simulations,  \citet{McAlpine_2020} investigated the connection between galaxy-galaxy mergers and enhanced SMBH growth.  They concluded that, at low redshift, the enhancement of AGNs peaks at $\sim 0.3$\,Gyr after coalescence.  In the late stage of mergers, tidal structures disappear quickly.  \citet{Barnes_2016} computed the evolution of the tidal fraction $f_{\rm tid}$, i.e. the fraction of disk particles in the tidal tails,  in numerical simulations of major mergers of disk/bulge/halo systems for a set of different merger parameters.  From their Fig.\,14, we derive a mean decline rate of $\Delta f_{\rm tid}/f_{\rm tid} = -1.5 \pm 0.13$ per Gyr in the late stage of the mergers, where we adopted the value $\Delta t_{\rm p12} = 0.5$ Gyr for the mean time between the first and the second pericentric passage \citep{Mihos_1996}.  We assume that the tidal fraction can be taken as a proxy of the observability of the tidal features. The typical lifetimes of FR\,II galaxies are usually assumed  to be $\sim 10^6 - 10^8$\,yr \citep{Allen_2006, Odea_2009, Bird_2008, Antognini_2012, Garofalo_2018}.  Thus, if we assume a mean age of $\la 0.05$\,Gyr for the evolved radio galaxies in the comparison sample, the merger fraction of the comparison sample is expected to be  only less than 8\% smaller than that of the PSS-CSO sample,  which is far from sufficient to account for the observed difference between $f_{\rm m}^{\rm \, PSS} = 0.61$ and $f_{\rm m}^{\rm \, ECS} = 0.15$.

The above evolution scenario is based on the assumption that all PSSs and CSOs evolve into extended radio galaxies represented by the ECS sample. However,  if not all PSSs and CSOs evolve \citep[][and references therein]{Odea_2021}, the merger fraction of evolved radio sources may be lower than that of the young sources due to another effect: If a significant fraction of the merger-triggered PSSs and CSOs get stuck in the dense interstellar medium of the merger core and suffers the `dying type 3 source' fate described by  \citet{An_2012}, the counterparts of these merger galaxies are missing from the comparison sample.  Let us assume that the proportion $f_{\rm m}^{\rm \, y}$ of the young sources (upper index `y' for young) are triggered by mergers and only a fraction  $f_{\rm m,1}^{\rm \, y} <  f_{\rm m}^{\rm \, y}$ can evolve, while the fraction 
$f_{\rm m,2}^{\rm \, y} = f_{\rm m}^{\rm \, y} - f_{\rm m,1}^{\rm \, y}$ are episodic sources or `impostors' which do not evolve. The ratio of the merger fraction to the non-merger fraction for the evolved sources is  $R_1 =  f_{\rm m}^{\rm \, e}/f_{\rm nm}^{\rm \, e} = f_{\rm m,1}^{\rm \, y}/f_{\rm nm}^{\rm \, y}$,  and thus the merger fraction of the ECS sample results in  
$f_{\rm m}^{\rm \, ECS} = f_{\rm m,1}^{\rm \, y} + f_{\rm m,2}^{\rm \, y}\cdot R_1/(1+R_1)$. Using the abbreviations $R_2 := f_{\rm m,2}^{\rm \, y}/f_{\rm m,1}^{\rm \, y}$ and $S_1 := R_1/(1+R_1)$, we get  $f_{\rm m}^{\rm \, ECS} = f_{\rm m}^{\rm \, y}\cdot (1+ R_2\cdot S_1)/(1+R_2)$. If we set $f_{\rm m}^{\rm \, y}=0.61\pm0.07$, i.e. equal to the observed merger fraction of the PSS-CSO sample, we find that the low observed merger fraction of the comparison  sample,  $f_{\rm m}^{\rm \, ECS} = 0.15\pm0.06$, can be reproduced if $R_2 = 7.5\pm1.5$.
This means that $88\pm3\,\%$ of the merger-triggered sources, or at least $\sim54\pm2\,\%$ of all young radio sources, do not evolve to extended sources represented by the ECS sample.

%
\subsubsection{Cluster environment}\label{sect:cluster}
%

Current evidence suggests that both the persistence of tidal structures and the properties of AGNs depend on the environment of the galaxies \citep[e.g.][]{Mihos_2004, Sabater_2013, Marshall_2018}. In particular, tidal structures tend to be weaker and disappear more rapidly in the dense environment of  galaxy cluster cores.   As a result, the merger fraction of a sample of cluster galaxies would be smaller than that of field galaxies.  \citet{Odea_2021} summarised previous studies on small samples of powerful GPSs and CSSs. They concluded that their environment appears to be similar to that of powerful large radio sources and suggested that follow-up studies be carried out on larger samples.

We have used the following large catalogues of galaxy clusters to check the cluster membership. The GMBCG Galaxy Cluster Catalog \citep{Hao_2010} is a catalogue of galaxy clusters selected from SDSS DR7 by identifying  the red-sequence plus the brightest cluster galaxies.   It contains 55\,424 rich galaxy clusters  in the redshift range $0.1 < z < 0.55$. \citet[][WHL]{Wen_2012} identified 132\,684 clusters with $0.05 \la z \la 0.75$ on the basis of photometric redshifts of galaxies from SDSS DR8. \citet{Wen_2015} published an updated version of the WHL catalogue, using the SDSS DR12 spectroscopic data and additionally a list of 25\,419 complementary clusters  at high redshifts around bright galaxies.  For about 89\% of the identified clusters in the SDSS spectroscopic survey region, spectroscopic redshifts are available in the SDSS DR12. The cluster redshifts cover the range $0.05 < z < 0.75$. \citet{Banerjee_2018} presented a galaxy cluster catalogue constructed from the SDSS DR9, which comprises 46\,479 galaxy clusters at $z \la 0.64$.  Complementary to the above data sources,  the galaxy cluster catalogues of \citet{Burenin_2017},  \citet{Wen_2018}, and \citet{Kirkpatrick_2021} were included in our search. The extension of the Planck galaxy cluster catalogue by \citet{Burenin_2017} comprises $\sim 3\,000$  galaxy clusters detected in the Planck all-sky Compton parameter maps  and identified using data from the WISE and SDSS surveys.  \citet{Wen_2018} presented a compilation of $\sim 47\,600$ clusters identified from photometric data of 2MASS, WISE, SuperCOSMOS, and detection by  ROSAT and XMM-Newton.  Finally, the  catalogue of \citet{Kirkpatrick_2021} was included, which is the largest compilation of visually confirmed X-ray selected spectroscopic galaxy clusters.  The latter lists  2\,740 galaxy clusters spectroscopically confirmed using SDSS DR16 spectroscopic data for 33\,340 individual galaxy members.  In all the catalogues used, the redshift distributions of the clusters drop significantly at $z \sim 0.6$.

We consider a galaxy as belonging to a cluster if  (a) the projected distance from the catalogued cluster position is $d_{\rm c} \le 1$ Mpc at the redshift of the cluster and  (b)  the radial component of the relative velocity with respect to the cluster is $v_{\rm rel} \le 10^3$\,km\,s$^{-1}$.   The latter was determined from the relation $v_{\rm rel}/c = (z_{\rm g} - z_{\rm cl})/(1 + z_{\rm cl})$  \citep{Harrison_1974},  where $z_{\rm g}$ and $z_{\rm cl}$ are the redshifts of the galaxy and the cluster, respectively.   We identified host clusters  in all used catalogues, with the sole exception of the X-ray cluster catalogue by  \citet{Kirkpatrick_2021}. If a nearby cluster was identified in two or more catalogues, we used the following priority list: (1) \citet{Wen_2015}, (2) \citet{Banerjee_2018},  (3) \citet{Burenin_2017},  (4) \citet{Hao_2010}, (5) \citet{Wen_2018}.   The catalogue by  \citet{Wen_2015} contains by far the most sources from our samples. The order of the other catalogues was chosen arbitrarily.

The criterion of cluster membership is fulfilled by 15 sources from the PSS-CSO  and 13 from the ECS sample. This corresponds to $19\pm 4$\,\% (PSS-CSO) and $17\pm 4$\,\% (ECS) cluster members for the redshift range $z < 0.6$. The difference is not significant according to the two-sided Z-test ($p = 0.67$). More revealing is the proportion of sources located in the cluster core regions, where the strongest effects are to be expected.  Of the PSSs and CSOs, $15\pm 4$\% (12/78) are found within $d_{\rm c} = 0.2$\,Mpc,  and a similar percentage of $12\pm 4$\,\% (9/78) is found  for the ECS sample. With only two exceptions, all identified clusters are listed in the \citet{Wen_2015} catalogue, where the estimated cluster radius $r_{500}$ is also given. The mean values are $r_{500} = 0.76\pm 0.04$\,Mpc for the PSS-CSO  and $0.72\pm 0.09$\,Mpc for the ECS sample. These results suggest that the cluster environment does not play a decisive role in the difference  between the merger fractions of the two samples.

Finally, we note that the merger fraction among cluster galaxies is significantly lower than in general.  Among the PSS-CSOs in clusters, we find four mergers ($f_{\rm m} = 0.27\pm0.11$), among the ECS none at all.  The mean merger index is $t_{\rm m} = 0.64\pm1.08$ (PSS-CSO) and $0.15\pm 0.38$ (ECS), respectively.

%
\section{Summary and conclusions}\label{sect:conclusion}
%

We compared a sample of compact and presumably young extragalactic radio sources with a sample of evolved, extended radio galaxies and QSOs.
To construct the former, we used the sample from  \citep{Liao_2020} with available SDSS spectra, here referred to as the `PSS-CSO sample'. The extended comparison sources (ECSs) were selected from a larger sample of extended radio galaxies and QSOs  with steep radio spectra.  The selection was made  in such a way that each object from the PSS-CSO sample has one counterpart in the ECS sample with a similar redshift, 1.4\,GHz luminosity, optical luminosity, and SDSS spectral class, where  the linear size,  measured in VLASS, must be greater than 50 kpc.

This study focuses in particular on spectra, SEDs and morphological properties and is based on  archival data in the spectral range from the far UV to MIR.  We divided both samples into the subsamples of galaxies and QSOs and generally compared the corresponding subsamples with each other using statistical tests.  The main results are the following.

(i) 
We compared the distributions of the radio galaxies in three different diagnostic diagrams.  There are no significant differences in the [\ion{N}{ii}]-based BPT diagram.  In the  KEx diagram, the PSS-CSO sample shows a higher mean line dispersion of the [\ion{O}{iii}] line. Both diagrams suffer from a significant bias because a substantial portion of galaxies have weak or missing H$\beta$ lines. The more complete WHAN diagram, on the other hand, shows a significantly higher percentage of passive and retired galaxies   in the ECS compared to the PSS-CSO sample.

(ii)  
In the SDSS composite spectra of the galaxy subsamples, the continuum is sufficiently well described by a mixture of  old early-type galaxies and SF galaxies, with a significantly larger contribution from SF galaxies in the PSS-CSO sample.  Such a difference is not visible between the continua of the QSO composites.  All four composite spectra  show relatively strong forbidden emission lines [\ion{O}{iii}] and [\ion{O}{ii}] in combination with weak Fe blends, as expected for luminous radio galaxies.  The PSS-CSO composites show a larger line ratio [\ion{O}{ii}]/[\ion{Ne}{v}] which can be attributed to SF or to the interaction of the radio jets with the dense interstellar medium of the host galaxy.

(iii) 
We classified the individual MBSEDs using a set of template spectra. $44\pm 7$\,\% of the galaxies from the ECS-G sample are best matched by templates of early-type galaxies,  compared to only $31\pm 6$\,\%  for the PSS-CSO-G sample. The latter is largely in agreement with \citet{Nascimento_2022}. The proportion of SF galaxies is significantly larger for PSSs and CSOs than for ECSs, both in the entire sample and in the two subsamples of galaxies and QSOs.  The colour-colour diagram based on the first three WISE bands  also shows a significantly lower proportion of SF galaxies in the  ECS compared to the  PSS-CSO sample.

(iv) 
The distributions of the D4000n index are different for PSS-CSO and ECS galaxies.  The mean values are D4000n = $1.70\pm 0.04$ for the PSS-CSO-G and $1.88 \pm 0.02$ for the ECS-G sample.  The larger value for the latter is consistent with the stronger contribution of old stellar populations and is consistent with the previous points. The distributions of the SFR and the sSFR from  \citet{Chang_2015} also differ for the two samples in such a way that they indicate a higher proportion of SF galaxies in the PSS-CSO sample. In the SFR - stellar mass diagram,  $42\pm 10$\,\% of the PSS-CSO galaxies with available SFR data lie on the main sequence of SF galaxies,  compared to only $22\pm 8$\,\% for the ECS-G sample.  For the vast majority, the  SFRs and stellar masses of PSS-CSO-G host galaxies cover the ranges from $\sim 0$ to 5\,$\mathcal{M}_\odot$\,yr$^{-1}$  and $3\cdot10^{11}$ to $10^{12} \mathcal{M}_\odot$, respectively, in good agreement with \citet{Nascimento_2022}.

(v) 
We analysed archival HST images of  18 PSSs and CSOs from our sample. In all cases where the host galaxy can be studied, we found indications of gravitational interactions and mergers.  Motivated by this high merger fraction,  we used the DESI Legacy Surveys images to analyse the morphologies of all objects with  $z \le 0.3$ from both samples. We found a higher proportion of mergers in the PSS-CSO  ($0.61\pm 0.07$) compared with the ECS sample ($0.15\pm 0.06$).  This highly significant difference ($p < 10^{-5}$) can be explained by assuming that 61\% of the PSSs and CSOs were triggered by mergers,  but $88\pm3$\,\% of the merger-triggered or $54\pm2$\,\% of all young radio sources  do not evolve and are thus not represented in the comparison sample of extended radio sources. This interpretation is in line with the recent finding by \citet{Kiehlmann_2023} that most CSOs  belong to a distinct class of jetted AGNs.

(vi) 
To look for possible differences in the environments of the PSSs-CSOs and the ECSs, we matched the positions and redshifts of the sources with $z < 0.6$ from both samples with several large galaxy cluster catalogues.   The results do not suggest a significant difference.

(vii) 
We provide further evidence that the host galaxy of the CSS source  \object{4C\,+39.29} is most likely not \object{SDSS J101714.23+390121.1} at $z=0.211$  but \object{SDSS J101714.10+390123.8} at $z=0.526$ as suggested already by \citet{Gandhi_2006}.   The observations are consistent with an ongoing jet–cloud interaction in that source.

We conclude that our PSS-CSO sample consists of both young radio sources that evolve into extended radio structures and frustrated or hindered sources that do not. 
Galaxy-galaxy interactions and mergers appear to play an important role in triggering the radio AGN. These results need to be verified with a larger PSS-CSO sample.

\begin{acknowledgements}

We thank the referee for useful feedback, which has improved this manuscript. We would like to thank G\"ulay G\"urkan for comments and suggestions on the manuscript  and the other members of the TLS Working Group on Extragalactic Radio Astronomy for  numerous discussions.

This research has made use of data products from the Sloan Digital Sky Survey (SDSS). Funding for the SDSS and SDSS-II has been provided by
the Alfred P. Sloan Foundation, the Participating Institutions (see below), the National Science Foundation, the National Aeronautics and Space Administration, the U.S. Department of Energy, the Japanese Monbukagakusho, the Max Planck Society, and the Higher Education Funding Council for England. The SDSS Web site is http://www.sdss.org/. The SDSS is managed by the Astrophysical Research Consortium (ARC) for the Participating Institutions. The Participating Institutions are: the American Museum of Natural History, Astrophysical Institute Potsdam, University of Basel, University of Cambridge (Cambridge University), Case Western Reserve University, the University of Chicago, the Fermi National
Accelerator Laboratory (Fermilab), the Institute for Advanced Study, the Japan Participation Group, the Johns Hopkins University, the Joint Institute for Nuclear Astrophysics, the Kavli Institute for Particle Astrophysics and Cosmology, the Korean Scientist Group, the Los Alamos National Laboratory, the Max-Planck-Institute for Astronomy (MPIA), the Max-Planck-Institute for Astrophysics (MPA), the New Mexico State University, the Ohio State University, the University of Pittsburgh, University of Portsmouth, Princeton University, the United States Naval Observatory, and the University of Washington. \\

The Legacy Surveys consist of three individual and complementary projects: the Dark Energy Camera Legacy Survey (DECaLS; Proposal ID \#2014B-0404; PIs: David Schlegel and Arjun Dey), the Beijing-Arizona Sky Survey (BASS; NOAO Prop. ID \#2015A-0801; PIs: Zhou Xu and Xiaohui Fan), and the Mayall z-band Legacy Survey (MzLS; Prop. ID \#2016A-0453; PI: Arjun Dey). DECaLS, BASS and MzLS together include data obtained, respectively, at the Blanco telescope, Cerro Tololo Inter-American Observatory, NSF’s NOIRLab; the Bok telescope, Steward Observatory, University of Arizona; and the Mayall telescope, Kitt Peak National Observatory, NOIRLab. Pipeline processing and analyses of the data were supported by NOIRLab and the Lawrence Berkeley National Laboratory (LBNL). The Legacy Surveys project is honored to be permitted to conduct astronomical research on Iolkam Du’ag (Kitt Peak), a mountain with particular significance to the Tohono O’odham Nation. NOIRLab is operated by the Association of Universities for Research in Astronomy (AURA) under a cooperative agreement with the National Science Foundation. LBNL is managed by the Regents of the University of California under contract to the U.S. Department of Energy. This project used data obtained with the Dark Energy Camera (DECam), which was constructed by the Dark Energy Survey (DES) collaboration. Funding for the DES Projects has been provided by the U.S. Department of Energy, the U.S. National Science Foundation, the Ministry of Science and Education of Spain, the Science and Technology Facilities Council of the United Kingdom, the Higher Education Funding Council for England, the National Center for Supercomputing Applications at the University of Illinois at Urbana-Champaign, the Kavli Institute of Cosmological Physics at the University of Chicago, Center for Cosmology and Astro-Particle Physics at the Ohio State University, the Mitchell Institute for Fundamental Physics and Astronomy at Texas A\&M University, Financiadora de Estudos e Projetos, Fundacao Carlos Chagas Filho de Amparo, Financiadora de Estudos e Projetos, Fundacao Carlos Chagas Filho de Amparo a Pesquisa do Estado do Rio de Janeiro, Conselho Nacional de Desenvolvimento Cientifico e Tecnologico and the Ministerio da Ciencia, Tecnologia e Inovacao, the Deutsche Forschungsgemeinschaft and the Collaborating Institutions in the Dark Energy Survey. The Collaborating Institutions are Argonne National Laboratory, the University of California at Santa Cruz, the University of Cambridge, Centro de Investigaciones Energeticas, Medioambientales y Tecnologicas-Madrid, the University of Chicago, University College London, the DES-Brazil Consortium, the University of Edinburgh, the Eidgenossische Technische Hochschule (ETH) Zurich, Fermi National Accelerator Laboratory, the University of Illinois at Urbana-Champaign, the Institut de Ciencies de l’Espai (IEEC/CSIC), the Institut de Fisica d’Altes Energies, Lawrence Berkeley National Laboratory, the Ludwig Maximilians Universitat Munchen and the associated Excellence Cluster Universe, the University of Michigan, NSF’s NOIRLab, the University of Nottingham, the Ohio State University, the University of Pennsylvania, the University of Portsmouth, SLAC National Accelerator Laboratory, Stanford University, the University of Sussex, and Texas A\&M University. BASS is a key project of the Telescope Access Program (TAP), which has been funded by the National Astronomical Observatories of China, the Chinese Academy of Sciences (the Strategic Priority Research Program “The Emergence of Cosmological Structures” Grant \# XDB09000000), and the Special Fund for Astronomy from the Ministry of Finance. The BASS is also supported by the External Cooperation Program of Chinese Academy of Sciences (Grant \# 114A11KYSB20160057), and Chinese National Natural Science Foundation (Grant \# 12120101003, \# 11433005). The Legacy Survey team makes use of data products from the Near-Earth Object Wide-field Infrared Survey Explorer (NEOWISE), which is a project of the Jet Propulsion Laboratory/California 
Institute of Technology. NEOWISE is funded by the National Aeronautics and Space Administration. The Legacy Surveys imaging of the DESI footprint is supported by the Director, Office of Science, Office of High Energy Physics of the U.S. Department of Energy under Contract No. DE-AC02-05CH1123, by the National Energy Research Scientific Computing Center, a DOE Office of Science User Facility under the same contract; and by the U.S. National Science Foundation, Division of Astronomical Sciences under Contract No. AST-0950945 to NOAO.

This publication has made use of the VizieR catalogue access tool, CDS, Strasbourg, France, and of the NASA/IPAC Infrared Science Archive (IRSA), operated by the  Jet Propulsion Laboratories/California Institute of Technology, founded by the National Aeronautic and Space Administration. In particular,  this publication makes use of data products from the Wide-field Infrared Survey Explorer,  which is a joint project of the University of California, Los Angeles,  and the Jet Propulsion Laboratory/California Institute of Technology, 
funded by the National Aeronautics and Space Administration.  In addition, we used data products from the Two Micron All Sky Survey, which is a 
joint project of the University of Massachusetts and the Infrared Processing and Analysis Center/California Institute of Technology, funded by the National Aeronautics  and Space Administration and the National Science Foundation. We also used observations made with the NASA Galaxy Evolution Explorer, GALEX, which is operated for NASA by the California Institute of Technology under NASA contract NAS5-98034. We also made use of data products from the United Kingdom Infrared Telescope (UKIRT) Infrared Deep Sky Survey (UKIDSS). UKIRT is operated by the Joint Astronomy Centre on behalf of the Science and Technology Facilities Council of the U.K..\\

This work is also based in part on observations made with the NASA/ESA Hubble Space Telescope,  obtained from the Hubble Legacy Archive, which is a joint project of the Space Telescope Science Institute (STScI),  the Space Telescope European Coordinating Facility (ST-ECF), and the Canadian Astronomy Data Centre (CADC). We also used observations made with Subaru Telescope which is operated by the National Astronomical Observatory of Japan  and obtained from the SMOKA, which is operated by the Astronomy Data Center, National Astronomical Observatory of Japan. We also  used observations made with Hectospec at the MMT Observatory, a joint facility of the Smithsonian Institution and the University of Arizona, and obtained from the CfA Optical/Infrared Science Archive.\\

We also used images from Pan-STARRS1 (PS1).  The PS1 Surveys and the PS1 public science archive have been made possible through contributions by the Institute for Astronomy, the University of Hawaii, the Pan-STARRS Project Office, the Max-Planck Society and its participating institutes, the Max Planck Institute for Astronomy, Heidelberg and the Max Planck Institute for Extraterrestrial Physics, Garching, The Johns Hopkins University, Durham University, the University of Edinburgh, the Queen's University Belfast, the Harvard-Smithsonian Center for Astrophysics, the Las Cumbres Observatory Global Telescope Network Incorporated, the National Central University of Taiwan, the Space Telescope Science Institute, the National Aeronautics and Space Administration under Grant No. NNX08AR22G issued through the Planetary Science Division of the NASA Science Mission Directorate, the National Science Foundation Grant No. AST-1238877, the University of Maryland, Eotvos Lorand University (ELTE), the Los Alamos National Laboratory, and the Gordon and Betty Moore Foundation.\\

This research has made use of the CIRADA cutout service at URL cutouts.cirada.ca, operated by the Canadian Initiative for Radio Astronomy Data Analysis (CIRADA). CIRADA is funded by a grant from the Canada Foundation for Innovation 2017 Innovation Fund (Project 35999), as well as by the Provinces of Ontario, British Columbia, Alberta, Manitoba and Quebec, in collaboration with the National Research Council of Canada, the US National Radio Astronomy Observatory and Australia’s Commonwealth Scientific and Industrial Research Organisation.

\end{acknowledgements}


\bibliographystyle{aa} 
\bibliography{literature} 


\begin{appendix}

%
\section{The counterpart of 4C\,+39.29}\label{sect:SDSSJ1017}
%

\begin{figure}[htbp]
\includegraphics[viewport= 50 20 570 820,width=6.1cm,angle=270]{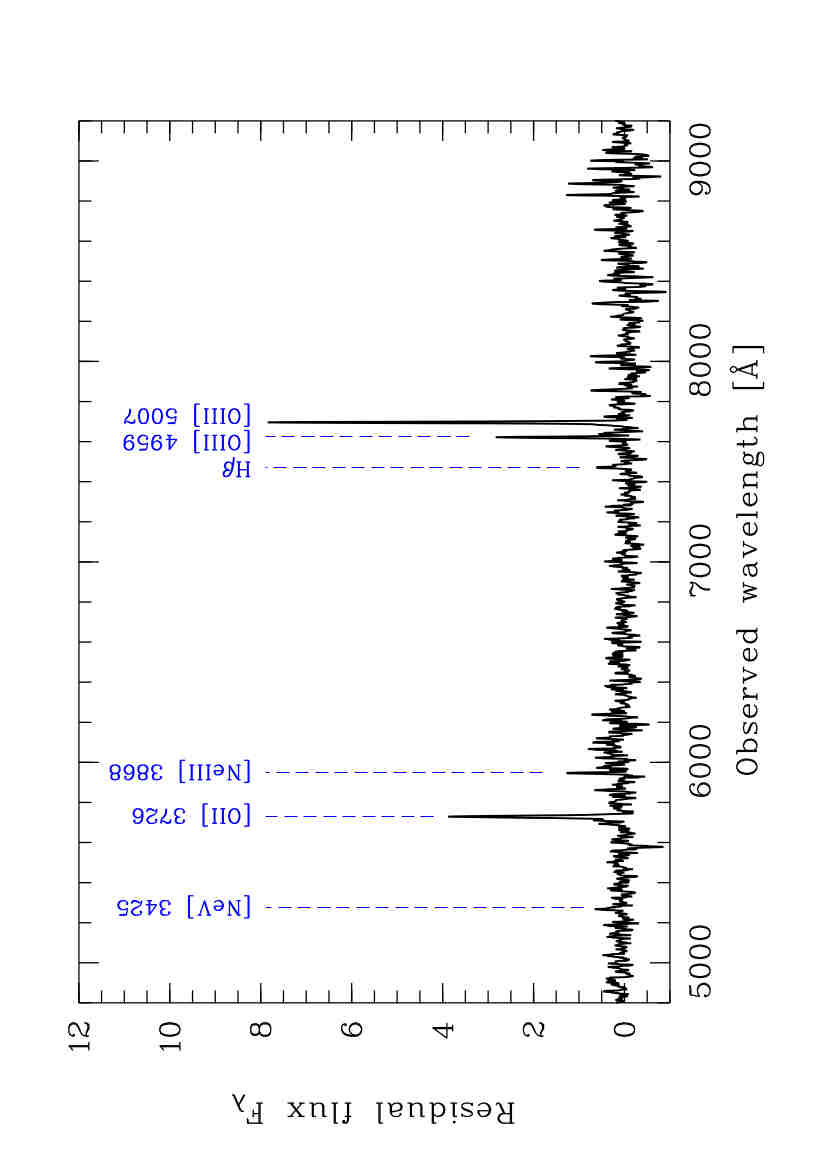}
\caption{
Residual spectrum after subtraction of the SDSS early-type galaxy template spectrum from the SDSS spectrum of  \object{SDSS J101714.23+390121.1}. 
The remaining emission lines are at $z = 0.537$.
}
\label{fig:spec_g1_residuum}
\end{figure}

\citet{Liao_2020} \citep[and also][]{Nascimento_2022} associated  the CSS source \object{4C\,+39.29} with the optical source \object{SDSS J101714.23+390121.1} at $z=0.211$. The SDSS spectrum shows the typical stellar continuum and absorption lines of an early-type galaxy at $z=0.211$, but it also contains a system of strong emission lines at higher redshift.  This means that the observed spectrum is the blend of two sources at different redshifts. We fitted the SDSS early-type galaxy template spectrum, redshifted to $z=0.211$, to the extinction-corrected galaxy spectrum. The difference between the observed SDSS spectrum and that of the matched template consists essentially of a system of strong emission lines at $z=0.5370 \pm 0.0005$  (Fig.\,\ref{fig:spec_g1_residuum}).  The high line ratio $\log\,[\ion{O}{iii}]/{\rm H\beta} = 1.48$ classifies the source as an AGN (Fig.\,\ref{fig:diagn_diagrams}).

\begin{figure}[htbp]
\begin{center}
\includegraphics[viewport=0 20 730 770,width=9.3cm,angle=0]{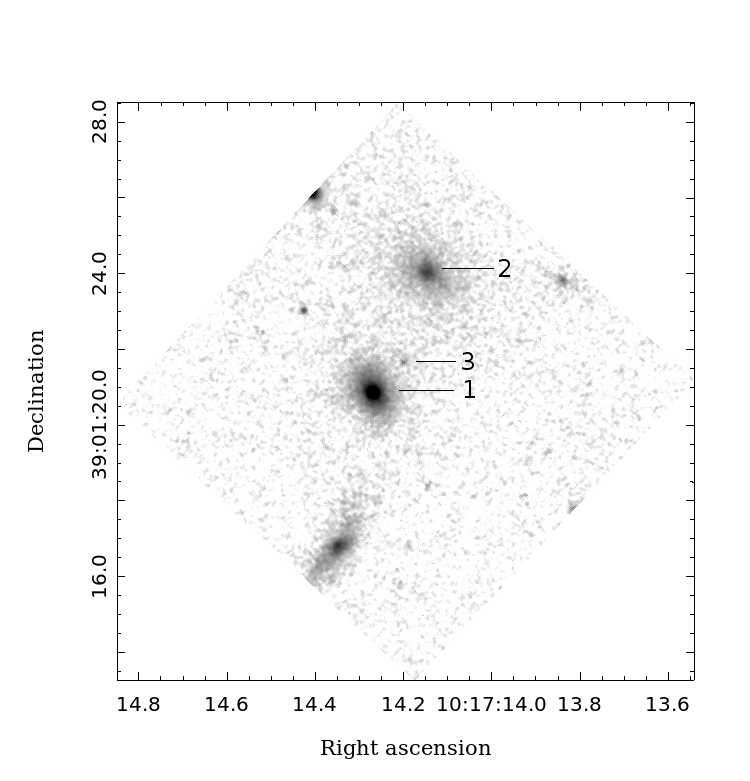}
\end{center}
\caption{
HST F845M image.  The numbers 1 to 3 mark the sources \object{SDSS J101714.23+390121.1},  \object{SDSS J101714.10+390123.8}, and \object{J101714.18+390121.8}, respectively.}
\label{fig:HST_SDSSJ1017}
\end{figure}

\begin{figure}[htbp]
\begin{center}
\includegraphics[viewport= 10 20 570 780,width=6.3cm,angle=270]{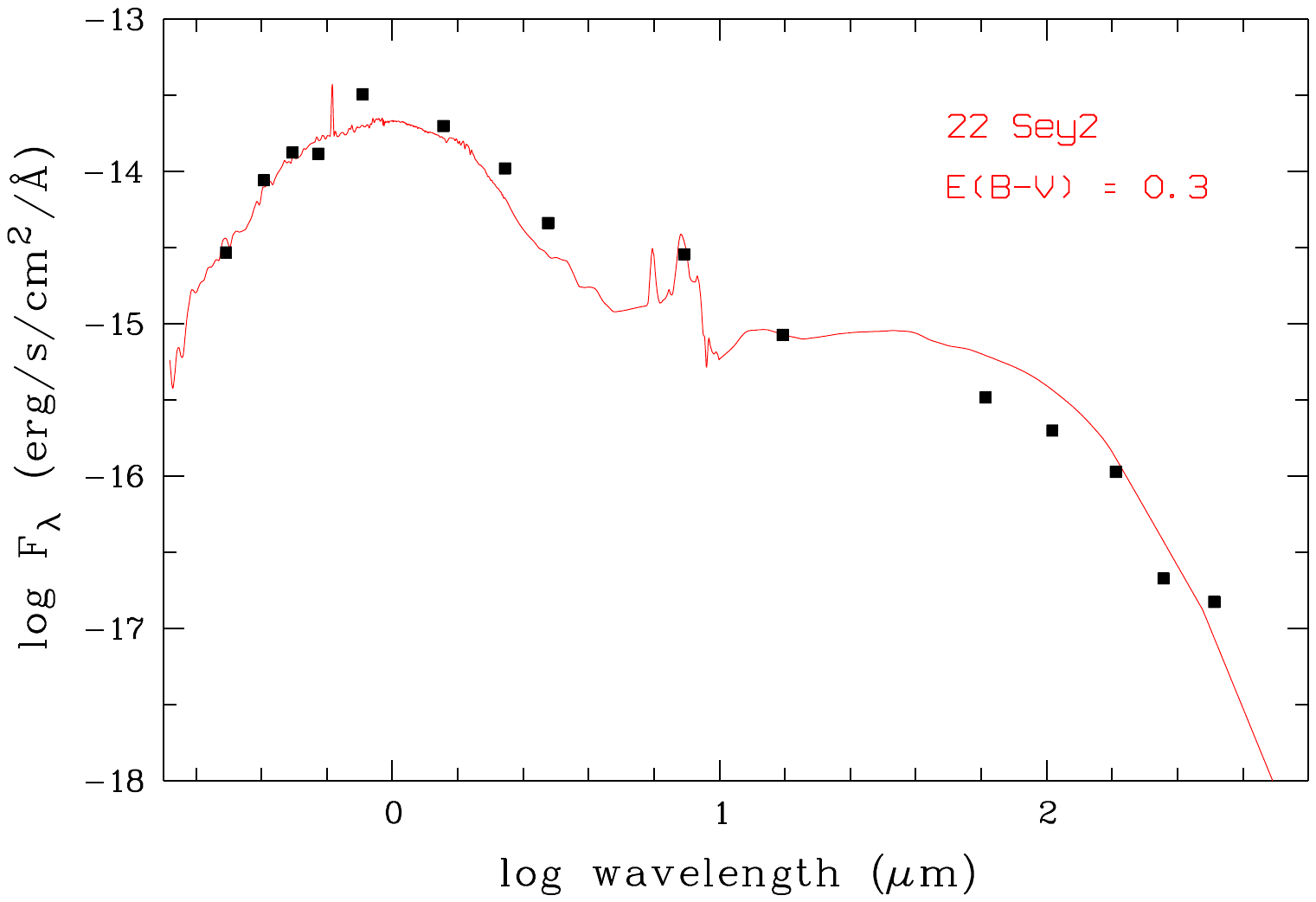}
\end{center}
\caption{
MBSED of \object{SDSS \,J101714.10+390123.8} (black squares) and best-fitting template (red). 
}
\label{fig:MBSED_J101714}
\end{figure}

We searched the Hubble Legacy Archive and found two images containing the field of 4C\,+39.29, both taken with the WFC3/UVIS camera through the F845M filter  (PropID 15245, PI: O'Dea).  Figure\,\ref{fig:HST_SDSSJ1017} shows the co-added two 700 sec exposures,  where we marked the two galaxies \object{SDSS J101714.23+390121.1} (galaxy 1) and \object{SDSS J101714.10+390123.8} (galaxy 2).   The only other optical source close to galaxy 1 is the faint, compact object  \object{J101714.20+390121.9} (source 3). It is seen clearly on both individual frames, so we can be sure that it is not an artefact.  The  distance of source 3  from  galaxy 1 is only $\sim 1\arcsec$, i.e. its emission  is captured by the aperture of the SDSS spectroscopic fibre centred on galaxy 1.  The SDSS spectrum of galaxy 1 can thus be easily understood if we assume that source 3 is a strong emitter of oxygen lines at  $z = 0.537$.

4C\,+39.29  was extensively discussed by  \citet{Gandhi_2006}. Based on a detailed analysis of observations at X-ray, optical, and radio wavelengths, the authors concluded that \object{SDSS J101714.10+390123.8}  (`source 1' in their notation,  galaxy 2 in Fig.\,\ref{fig:HST_SDSSJ1017}) is the most likely counterpart of \object{4C\,+39.29}.  They presented an optical spectrum of this galaxy, which shows strong oxygen emission lines at $z = 0.536$.  A long-slit spectrum presented by these authors reveals an extended emission line region (EELR) at the same redshift whose position is consistent with our source 3. \citet{Gandhi_2006} classified \object{SDSS J101714.10+390123.8} (galaxy 2) as a powerful type 2 quasar with a weak radio core and high core obscuration compared to other narrow-line radio quasars.  Here we note  in addition that it is positionally coincident with the {\it Herschel}-detected source \object{J101714.14+390124.4} which has been classified as a highly luminous type\,2 AGN \citep{Xu_2015}. We inspected the images from the IRAC camera at 3.6\,$\mu$m and 4.5\,$\mu$m and the MIPS camera at 24\,$\mu$m  aboard the Spitzer Space Telescope and confirmed that the bright mid-IR source is at the position of  galaxy 2.

The MBSED of galaxy 2 is shown in Fig.\,\ref{fig:MBSED_J101714}.  The set of observational data  consists of the infrared fluxes from \citet{Xu_2015} and the fluxes computed from the SDSS model magnitudes in the g, r, i, and z bands.  Among the SWIRE  template spectra (see Sect.\,\ref{sect:MBSED}), the one for Seyfert-2 galaxies provides by far the best agreement with the observed fluxes, in agreement with the classification by \citet{Xu_2015}.  To match the steep decline in the UV, we had to assume an additional intrinsic dust reddening with $\Delta A_V = 0.9$\,mag for the Milky Way reddening curve from \citet{Pei_1992}. \citet{Xu_2015} found a total infrared luminosity of $L_{\rm ir} = 1.9\cdot10^{11}\,L_\odot$, where the emission at $24\,\mu$m is dominated by the AGN. The stellar mass of  ${\mathcal M}_\ast = 2.8\cdot 10^{11}\,{\mathcal M}_\odot$ \citep{Xu_2015} is similar to  the masses of the PSS-CSO galaxies in our sample (Fig.\,\ref{fig:m_sfr}).

The VLA 8.4\,MHz radio map shown by \citet[][their Fig.\,1]{Gandhi_2006} reveals a weak core at $10^{\rm h}17^{\rm m}14\fs12 +39\degr 01\arcmin 24\farcs4$ (J2000), i.e. close to the optical core of galaxy 2. The radio core position is nearly in a line with two brighter separated radio components.  The linear extension from the core to the northern component is about 32 kpc and the core distance of the southern peak is about 15 kpc.  The northern component looks similar to a typical FR II lobe.  The brighter southern component appears more complex with the brightest part at  $10^{\rm h}17^{\rm m}14\fs2 +39\degr 01\arcmin 21\farcs8$ (J2000). This position coincides with the EELR and the optical source 3 in the HST image  (Fig\,\ref{fig:HST_SDSSJ1017}). \citet{Gandhi_2006} proposed that the morphology and emission-line properties of the EELR are consistent with an ongoing jet-cloud interaction.

\begin{figure}[htbp]
\begin{center}
\includegraphics[viewport=0 0 560 570,width=7.8cm,angle=0]{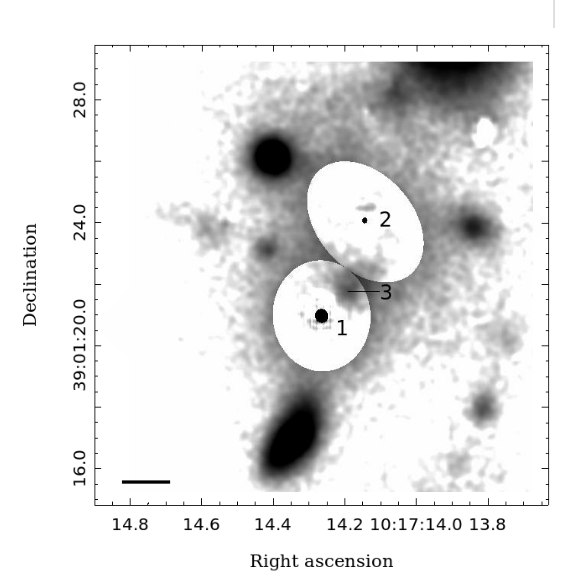} \\
\includegraphics[viewport=0 0 560 570,width=7.8cm,angle=0]{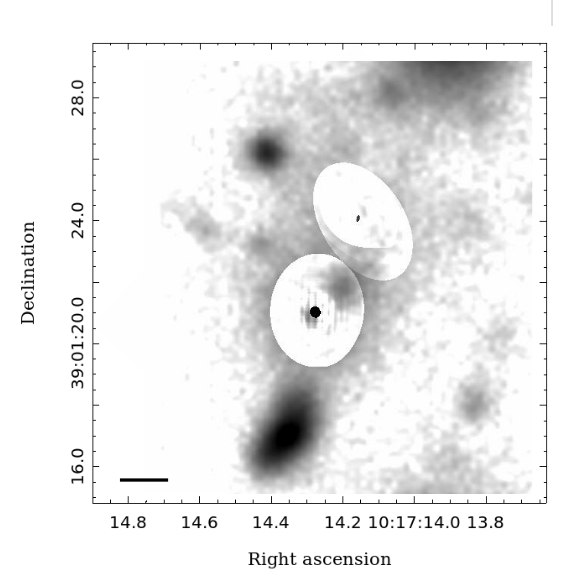}
\end{center}
\caption{
Suprime-Cam image of the field from Fig.\,\ref{fig:HST_SDSSJ1017} through the filters W-C-IC (left) and W-J-V (right). The brighter inner parts of the galaxies 1 and 2 were subtracted.  The bar in the lower left corner indicates a length of 10 kpc at $z = 0.537$. 
}
\label{fig:Subaru_SDSSJ1017}
\end{figure}

The field of \object{4C\,+39.29} is located on the outskirts of the galaxy cluster A\,963, which has been observed several times with the Subaru Prime Focus Camera  \citep[Suprime-Cam;][]{Miyazaki_2002}  mounted on the Subaru Telescope at Mauna Kea \citep{Baba_2002}.  We downloaded exposures through the  two broadband filters  W-J-V (centre at 5470\,\AA, FWHM = 970\,\AA) and W-C-IC (centre at 7970\,\AA, FWHM = 1400\,\AA) from the SMOKA science archive\footnote{https://smoka.nao.ac.jp/fssearch.jsp} and constructed co-added images with total exposure times of  3000 sec for  W-C-IC and  750 sec for W-J-V. The mean seeing was about $0\farcs7$, the pixel scale is 0.202 arcsec pixel$^{-1}$.  At the redshift of galaxy 2, the two filter bands contain the strong oxygen lines [\ion{O}{ii}]\,3727\AA\ (W-J-V) and [\ion{O}{ii}]\,5007\AA\ (W-C-IC), respectively. Both images show a faint extension of galaxy 1 towards the NW, at about the position of source 3, but the faint source is outshone by the two much brighter galaxies 1 and 2. We computed surface brightness models for the galaxies 1 and 2 based on their measured isophotes using the algorithm of \citet{Bender_1987} and subtracted the model images from the Subaru images. Source 3 is clearly visible in both two residual images, which trace the ionised gas  (Fig.\, \ref{fig:Subaru_SDSSJ1017}). In contrast, the filter band of the HST image (Fig.\,\ref{fig:HST_SDSSJ1017}) does not contain a strong emission line and thus traces the stellar component. Following \citet{Gandhi_2006}, we conclude that the radio jet of  \object{4C\,+39.29} most likely interacts with the interstellar medium of  a small, gas-rich satellite of galaxy 2, possibly part of an extended tidal structure.\footnote{For similar cases we refer e.g. to  3C\,133 \citep{Reddy_2021}, 3C\,346  \citep{Dey_1994, Worrall_2005}, and 3C\,277.3 = Coma\,A \citep{Tilak_2005, Worrall_2016}.}

The large galaxy  \object{SDSS J101713.86+390131.3} indicated at the upper edge of Fig.\,\ref{fig:Subaru_SDSSJ1017} has no  SDSS spectrum. We found a spectrum in the CfA Optical/Infrared Science Archive\footnote{https://oirsa.cfa.harvard.edu/} taken with the Hectospec Multi-Fiber Spectrograph  \citep{Fabricant_2005} (proposal ID 2008a-SAO-3, PI: K. Rines) for the Hectospec Cluster Survey \citep{Rines_2013}, which shows absorption lines at $z = 0.211$,  i.e. at the same redshift as \object{SDSS J101714.23+390121.1}  (galaxy 1). Both galaxies belong to the infall region of the foreground cluster A\,963 and are unrelated to \object{4C\,+39.29}. No redshift is available for the galaxy \object{SDSS J101714.32+390116.9} close to the bottom of Fig\,\ref{fig:Subaru_SDSSJ1017}.

%
\section{Images of morphologically peculiar host galaxies at $z < 0.3$}\label{sect:App_LS_images}
%

\begin{figure*}[ht]
\includegraphics[viewport=0 0 360 360,width=4.5cm,angle=0]{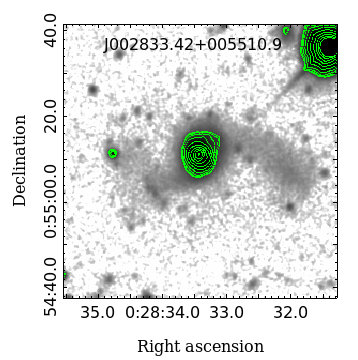}         
\includegraphics[viewport=0 -30 256 256,width=3.9cm,angle=0]{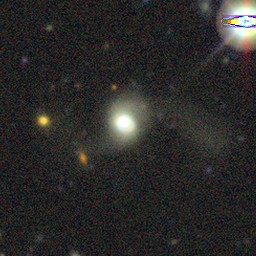}      
\hspace{0.5cm}
\includegraphics[viewport=0 0 360 360,width=4.5cm,angle=0]{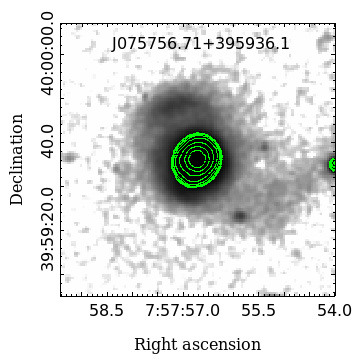}          
\includegraphics[viewport=0 -30 256 256,width=3.9cm,angle=0]{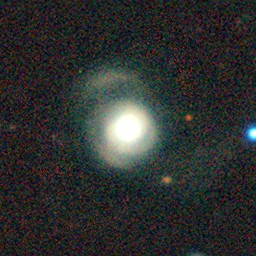} \\    
\vspace{0.3cm}
\includegraphics[viewport=0 0 360 360,width=4.5cm,angle=0]{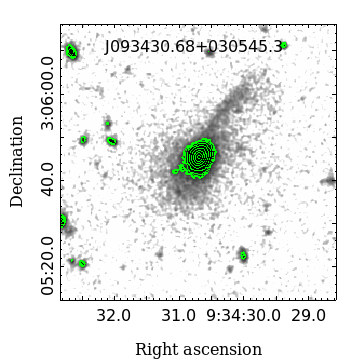}          
\includegraphics[viewport=0 -30 256 256,width=3.9cm,angle=0]{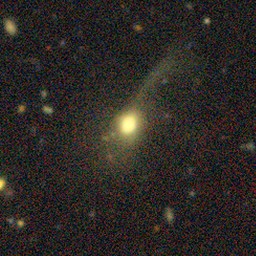}       
\hspace{0.5cm}
\includegraphics[viewport=0 0 360 360,width=4.5cm,angle=0]{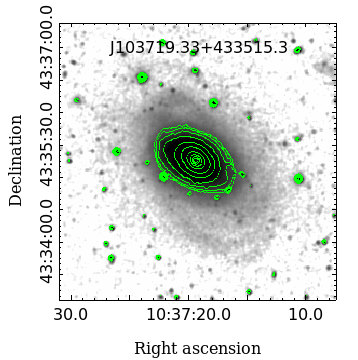}          
\includegraphics[viewport=0 -30 256 256,width=3.9cm,angle=0]{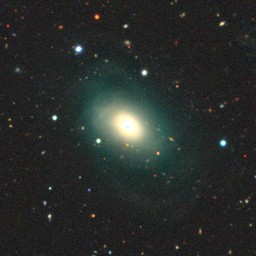} \\
\vspace{0.3cm}
\includegraphics[viewport=0 0 360 360,width=4.5cm,angle=0]{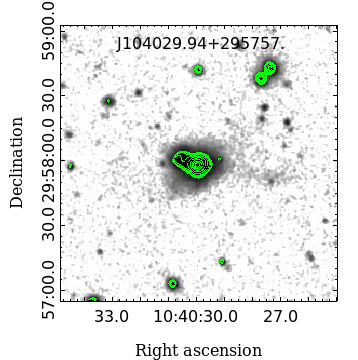}
\includegraphics[viewport=0 -30 256 256,width=3.9cm,angle=0]{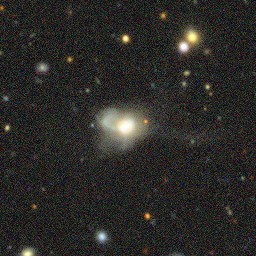}
\hspace{0.5cm}
\includegraphics[viewport=0 0 360 360,width=4.5cm,angle=0]{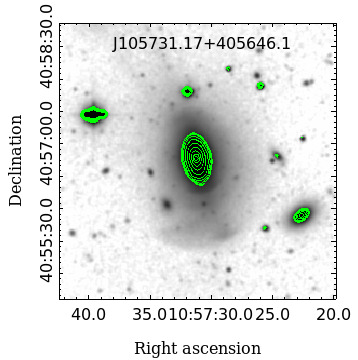}
\includegraphics[viewport=0 -30 256 256,width=3.9cm,angle=0]{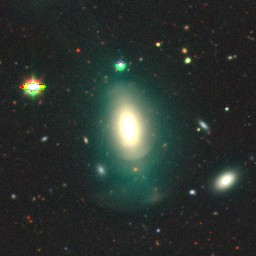} \\
\vspace{0.3cm}
\includegraphics[viewport=0 0 360 360,width=4.5cm,angle=0]{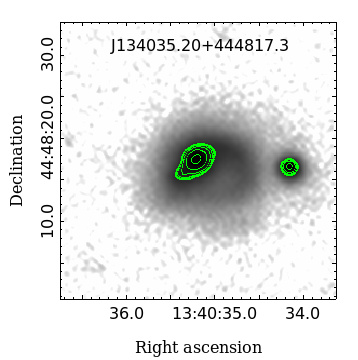}
\includegraphics[viewport=0 -30 256 256,width=3.9cm,angle=0]{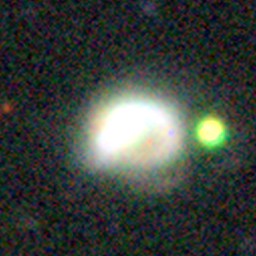}
\hspace{0.5cm}
\includegraphics[viewport=0 0 360 360,width=4.5cm,angle=0]{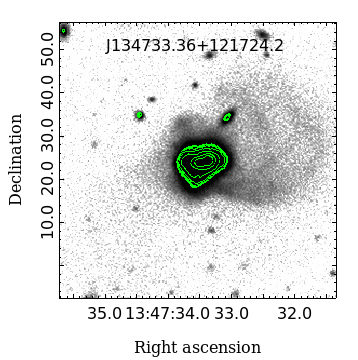}
\includegraphics[viewport=0 -30 256 256,width=3.9cm,angle=0]{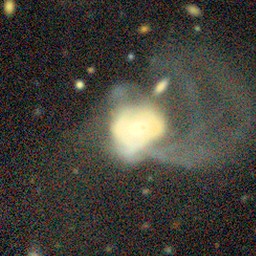} \\
\vspace{0.3cm}
\includegraphics[viewport=0 0 360 360,width=4.5cm,angle=0]{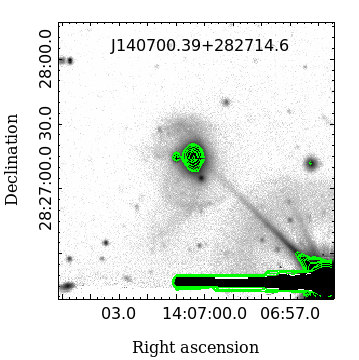}
\includegraphics[viewport=0 -30 256 256,width=3.9cm,angle=0]{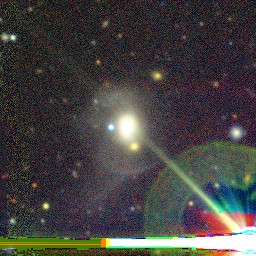}
\caption{
DESI Legacy Surveys images of the PSS-CSO host galaxies with $z < 0.3$ and $t_{\rm m} > 2.5$.  
A contrast-enhanced inverted greyscale image (left) and a RGB image (right) are shown for each galaxy.
}
\label{fig:PSS_merger}
\end{figure*}

\begin{figure*}[h]
\includegraphics[viewport=0 0 360 360,width=4.5cm,angle=0]{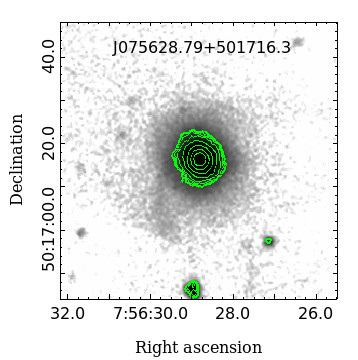}
\includegraphics[viewport=0 -30 256 256,width=3.9cm,angle=0]{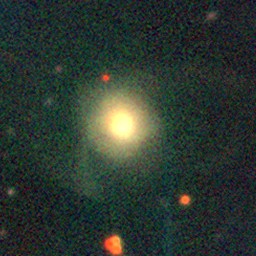}
\hspace{0.5cm}
\includegraphics[viewport=0 0 360 360,width=4.5cm,angle=0]{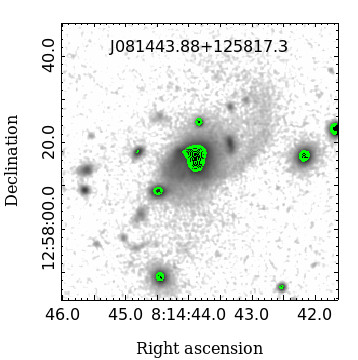}
\includegraphics[viewport=0 -30 256 256,width=3.9cm,angle=0]{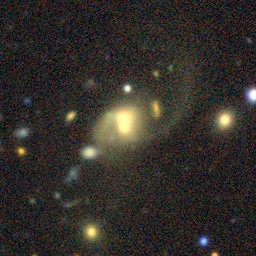} \\
\includegraphics[viewport=0 0 360 360,width=4.5cm,angle=0]{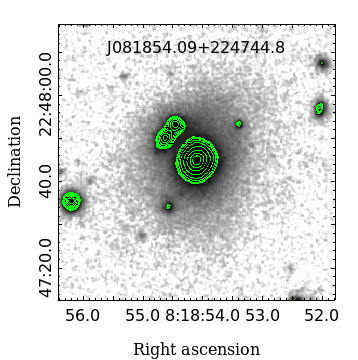}
\includegraphics[viewport=0 -30 256 256,width=3.9cm,angle=0]{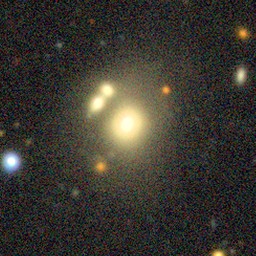}
\hspace{0.5cm}
\includegraphics[viewport=0 0 360 360,width=4.5cm,angle=0]{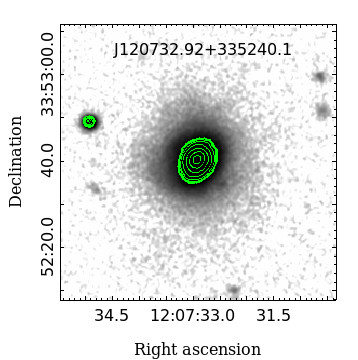}
\includegraphics[viewport=0 -30 256 256,width=3.9cm,angle=0]{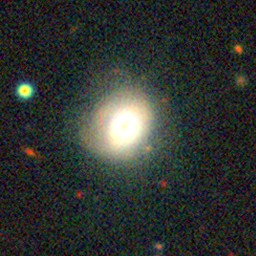} \\
\includegraphics[viewport=0 0 360 360,width=4.5cm,angle=0]{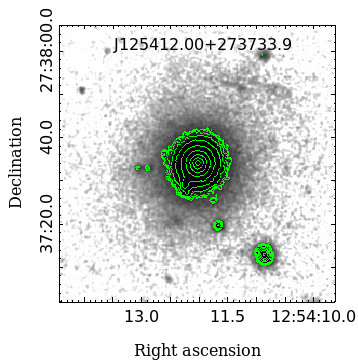}
\includegraphics[viewport=0 -30 256 256,width=3.9cm,angle=0]{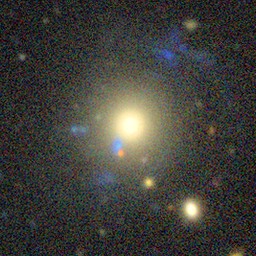} 
\hspace{0.5cm}
\includegraphics[viewport=0 0 360 360,width=4.5cm,angle=0]{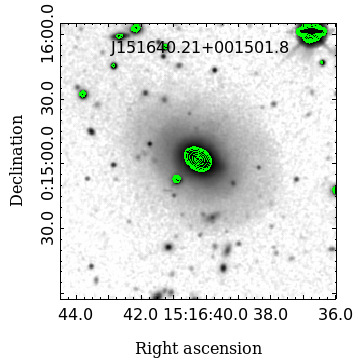}
\includegraphics[viewport=0 -30 256 256,width=3.9cm,angle=0]{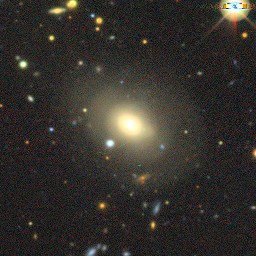}
\caption{
As in Fig.\,\ref{fig:PSS_merger}, but for the ECS host galaxies with $z<0.3$ and $t_{\rm m} \ge 2.0$. 
}
\label{fig:ECS_merger}
\end{figure*}

Figure\,\ref{fig:PSS_merger} shows the image cutouts from the  Legacy Surveys DR10 of the nine galaxies from the PSS-CSO  sample with $z \le 0.3$ and merger flag $t_{\rm m} > 2.5$.  For each galaxy, a RGB image is shown in combination with a contrast-enhanced inverted greyscale image. The latter was taken from the g, r, or i band, depending on which better reveals the peculiar structures. To show the structure in the blended bright areas, contour lines are plotted over them with lower and upper limits adjusted to the corresponding range of pixel values. A short description of the most remarkable peculiar features is given below.

(1) \object{SDSS J002833.42+005510.9} is  a relatively small galaxy with the lowest stellar mass  (log\,${\mathcal M}_\ast/{\mathcal M}_\odot = 10.58$) in our sample.  The contrast-enhanced r-band image reveals an extended structure of low surface brightness to the west of the main galaxy. This feature is barely visible on the image from the SDSS legacy survey,  but is similarly well seen on the r$_{\rm deep}$ co-adds from the IAC Stripe 82 Legacy Project \citep{Fliri_2016}. The structure  seems to be connected with the main body and extends up  to at least $\sim 45$\,kpc from the nucleus. A fainter and shorter structure is visible on the opposite side. The most likely interpretation of the extended structure is a low-surface brightness galaxy in tidal interaction with \object{SDSS J002833.42+005510.9}.

(2) \object{SDSS J075756.71+395936.1} shows a system of ring or shell-like structures around the main body in the RGB image similar to that in  \object{SDSS J105731.17+405646.1} (below).  In addition, an extended fan-like structure of low surface brightness stretches from the south of the main body to the west. Such structures are  most likely the result of gravitational distortion or a merger with a smaller galaxy.

(3) \object{SDSS J093430.68+030545.3} is the galaxy with the highest redshift ($z = 0.225$) in Figure\,\ref{fig:PSS_merger}. The morphology is best explained as an intermediate or  late stage of a merger, which produced a very long ($\sim 70$\,kpc)  tidal tail.  Such extended tidal structures are typically produced when at least one disk galaxy is involved  \citep{Toomre_1972, Springel_1999}.

(4) \object{SDSS J103719.33+433515.3}, along with SDSS J105731.17+405646.1 (below),  is the lowest redshift galaxy ($z = 0.025$) in the PSS-CSO sample.  It appears as a slightly lopsided but otherwise normal E/SO galaxy on the RGB image.  The contrast-enhanced r-band image shows a system of arc- or ring-shaped structures that are clearly asymmetric and cannot be interpreted as a spiral structure. SDSS J103719.33+433515.3 is a passive galaxy according to the WHAN diagram, the MBSED, the WISE colour-colour diagram, and the D4000n index.

(5) \object{SDSS J104029.94+295757.7} is an intermediate or late stage of a galaxy merger. The brightest tidal structure is clearly visible in the RGB image east of the core.  The contrast-enhanced greyscale image also shows an extended low-surface brightness tidal tail in the opposite direction. Such extended tidal structures indicate that at least one dynamically cold system is involved, which has been tidally disrupted (see also objects 4 above and 8 below). This is in agreement with the best-matching template of the MBSEB   (Table\,\ref{tab:low_z_PSS}).  The curved tail to the east of the core and the tilted structure of the core are  other remnants of this process.

(6) \object{SDSS J105731.17+405646.1} is the nearby E/SO galaxy NGC 3468. It clearly shows a system of shells and ripples, the most pronounced of which lie south of the main body. The most popular scenarios for the formation of such shell-like structures are based on gravitational distortions: fly-by interactions with  neighbouring galaxies \citep{Thomson_1991}  or  infall of one or possibly several dynamically cold galaxies into the potential of a giant E-type system  \citep{Schweizer_1980}.  Such shell galaxies often show also other signs of previous mergers, such as prominent central dust structures \citep{Richtler_2020} and unusually low central metallicity \citep{Carlsten_2017}.  The contrast-enhanced greyscale image in Fig.\,\ref{fig:PSS_merger} reveals in addition two faint streamers, extending from the main body 
towards the north-east, which support the infall scenario.

(7)  \object{SDSS J134035.20+444817.3} shows an elliptical ring-like structure with an offset core similar to the Cartwheel galaxy \citep{Fosbury_1977} or the ring galaxy  AM $2026-424$\footnote{https://esahubble.org/images/heic1919a/}.  Such ring structures are thought to be consistent with  the scenario of a  head-on collision of one or more companion galaxies with the target disk \citep{Elagali_2018}.

(8)  \object{SDSS J134733.36+121724.2}  is a well-studied merger host galaxy of a GPS source \citep{Evans_1999, Emonts_2016}.  Based on r and i band images from the Isaac Newton Telescope, \citet{Stanghellini_1993} suggested early that this object is a merger between a Seyfert galaxy and an elliptical radio galaxy. The deeper images from the Legacy Surveys clearly indicate an intermediate stage of a major merger with long tidal tails typically produced when at least one disk galaxy is involved.  There are two separate bright cores at a projected distance of $\sim 5$\,kpc.  The optical colours of both cores indicate strong reddening. The reddening is stronger for the western core, which is close to the position of the radio core.  The HST image (Fig.\,\ref{fig:HST}) reveals more details, such as dust absorption structures and bright knots close to the western core. The detection of the galaxy in both HI absorption, CO emission and as a bright IRAS source suggests the presence of large amounts of dust and cold gas in the nucleus of this GPS source  \citep[][and references therein]{Stanghellini_1993}. The SED is best matched by the spectrum of NGC\,6090, a star-forming galaxy with a similar merger morphology.

(9) \object{SDSS J140700.39+282714.6} is the host of a QSO. It displays a strange, complex morphology.  The image is slightly disturbed by the diffraction pattern and scattered light from a bright foreground star. South of the main body, and nearly concentric to it,  a relatively sharp fragment of a ring-like structure is seen, which, however, seems to bend towards the core in the south-west. On the other hand, there is an arch structure that extends from the south-east through the main body to the north and then bends to the north-east. If the latter forms a coherent structure, it is at least 100 kpc long.  The most likely explanation is a superposition of structures that result from a multiple merger.  A more exotic scenario would be an Einstein ring, with the lens being a merger galaxy. The grey-scale image is taken from the g band.

Figure\ref{fig:ECS_merger} shows  for comparison  the images of the six merger galaxies with  $z \le 0.3$ from the ECS sample.  The two galaxies at the top  were classified as $t_{\rm m} = 3$, the others as $t_{\rm m} = 2$.

%
\section{Object tables}\label{sect:comparison_sample}
%

For more details, readers can refer to Table\,\ref{tab:PSS}  and Table\,\ref{tab:ECS}.

\begin{table*}[htbp]
\caption{PSS-CSO sample. Only the first six rows of the table are shown here for illustration.  Uncertain data are indicated by a colon. The full list  is available in electronic form from the VizieR Service of the CDS Strasbourg.}
\begin{tabular}{ccrrccccccc} 
\hline\hline 
\noalign{\smallskip}
& SDSS name
& RA
& DEC
& $z$
& $R_{\rm i}$
& Type
& SDSS class
& $\log\,L_{\rm r}$
& log\,LLS 
& f\_L \\
& (1)
& (2)
& (3)
& (4) 
& (5)
& (6) 
& (7) 
& (8)
& (9)
& (10)\\
\hline
\noalign{\smallskip}
 1 &  J002225.42+001456.1 &    5.60592 &    0.24892 &  0.306 &  4.20  &  GPS &  GAL  &   26.9  &  $-0.57$  & - \\
 2 &  J002833.42+005510.9 &    7.13925 &    0.91969 &  0.104 &  2.52  &  CSS &  GAL  &   24.8  & \ \ $ 0.53$  & - \\
 3 &  J002914.24+345632.2 &    7.30933 &   34.94228 &  0.517 &  4.41  &  CSO &  GAL  &   27.2  &  $-0.70$  & - \\
 4 &  J005905.51+000651.6 &   14.77296 &    0.11433 &  0.719 &  3.84  &  CSS &  QSO  &   27.7  &     -     & - \\
 5 &  J014109.16+135328.3 &   25.28817 &   13.89119 &  0.621 &  4.54  &  CSS &  QSO  &   27.6  & \ \  $ 0.80$  & - \\
 6 &  J074417.47+375317.2 &  116.07279 &   37.88811 &  1.066 &  3.65  &  CSS &  QSO  &   27.8  & \ \ $ 1.25$  & - \\
\hline                 
\noalign{\smallskip}
\end{tabular}
\begin{tabular}{ccccccccccc} 
\hline\hline 
\noalign{\smallskip}
& $t_{\rm WHAN}$
& $t_{\rm SED}$
& $t_{\rm WISE}$
& D4000n
& e\_D4000n
& log\,SFR
& log\,$\mathcal{M}_\ast$
& f\_S
& $t_m$
& SDSS name of ECS \\
& (11)
& (12)
& (13)
& (14) 
& (15)
& (16)  
& (17)  
& (18) 
& (19) 
& (20) \\ 
\hline
\noalign{\smallskip}
 1  & wAGN & 3 & inter & 1.686 & 0.783  &     - &    - &  - &    - & J005727.88+020934.5 \\ 
 2  & sAGN & 2 & inter & 1.259 & 0.012  & 0.317 & 10.5 &  1 & 3.00 & J155700.17+413111.1 \\ 
 3  &  -   & 5 & inter & 1.518 & 1.418  &     - &    - &  - &    - & J113852.21+501602.3 \\ 
 4  &  -   & 6 & QSO   & -   - &     -  &     - &    - &  - &    - & J142456.92+200026.3 \\ 
 5  &  -   & 4 & QSO   & -   - &     -  &     - &    - &  - &    - & J121427.42+595814.8 \\ 
 6  &  -   & 6 & QSO   & -   - &     -  &     - &    - &  - &    - & J093033.53+360125.1 \\ 
\hline                 
\noalign{\smallskip}
\end{tabular}     
\tablefoot{
(1) SDSS name;
(2) - (3) right ascension and declination (degrees);
(4) redshift; 
(5) radio loudness parameter  (Sect.\,\ref{sect:Samples}); 
(6) radio type from \protect\citet{Liao_2020};
(7) SDSS spectral class; 
(8) 1.4 GHz radio luminosity (W\,Hz$^{-1}$);  
(9)-(10) largest linear size (kpc) and upper limit flag from \protect\citet{Liao_2020};
(11) type based on WHAN diagram;
(12) spectral type from MBSED (Table\,\ref{tab:templates});
(13) type based on WISE $W1-W2$ vs. $W2-W3$ diagram (Fig.\,\ref{fig:WISE_CCD}); 
(14) - (15)  4000\,\AA\ break index and uncertainty;
(16) - (18) SFR ($\mathcal{M_\odot}\,{\rm yr}^{-1}$), stellar mass ($\mathcal{M_\odot}$)
and uncertainty flag from \protect\citet{Chang_2015};
(19) merger index based on images from the DESI Legacy Surveys;                                                       
(20) counterpart in the ECS sample.\\                                                
}                                                                                    
\label{tab:PSS}
\end{table*}

\begin{table*}[h]
\caption{Comparison sample of extended radio sources (ECS). Only the first six rows of the table are shown here for illustration. 
Uncertain data are indicated by a colon.
The full list  is available in electronic form from the VizieR Service of the CDS Strasbourg.}
\begin{tabular}{ccrrccccrccc} 
\hline\hline 
\noalign{\smallskip}
& SDSS name
& RA
& DEC
& $z$
& $R_{\rm i}$
& SDSS class
& $\log\,L_{\rm r}$
& LLS  
& FR  
& $\alpha_{\rm r}$ 
& e\_$\alpha_{\rm r}$ \\
& (1)
& (2)
& (3)
& (4) 
& (5)
& (6) \ \ \ 
& (7) \ \ 
& (8)
& (9)
& (10)
& (11)\\
\hline
\noalign{\smallskip}
 1  &  J005727.88+020934.5  &  14.36620  &   2.15960  & 0.292  &  3.86  & GAL  &  25.9  &  182  & I    & $-0.80$  & 0.03   \\
 2  & J022507.93$-$003532.9 &  36.28307  & $-0.59248$ & 0.685  &  4.08  & QSO  &  27.3  &   97  & II   & $-0.69$  & 0.07   \\
 3  &  J073050.66+445600.8  & 112.71112  &  44.93359  & 0.072  &  1.33  & GAL  &  24.0  &   47  & II:  & $-0.78$  & 0.11   \\
 4  &  J074919.08+200753.7  & 117.32951  &  20.13160  & 0.368  &  3.00  & GAL  &  25.7  &  178  & II   & $-0.91$  & 0.11   \\
 5  &  J075529.95+520450.6  & 118.87473  &  52.08079  & 0.140  &  2.47  & GAL  &  24.7  &   61  & II   & $-0.97$  & 0.11   \\
 6  &  J075628.79+501716.3  & 119.11996  &  50.28787  & 0.134  &  1.79  & GAL  &  24.8  &  148  & II   & $-0.59$  & 0.09   \\
\hline                 
\noalign{\smallskip}
\end{tabular}
\begin{tabular}{ccccccccccc} 
\hline\hline 
\noalign{\smallskip}
& $t_{\rm WHAN}$
& $t_{\rm SED}$
& $t_{\rm WISE}$
& D4000n
& e\_D4000n
& log\,SFR
& log\,$\mathcal{M}_\ast$
& f\_S
& $t_m$
& SDSS name of PSS-CSO \\
& (12)
& (13)
& (14)
& (15) 
& (16)
& (17)  
& (18)  
& (19) 
& (20) 
& (21) \\ 
\hline
\noalign{\smallskip}
 1  &  -    & 2   & pass   & 1.86 & 0.61  &    -       &  -     & -  &  0.00    &  J002225.42+001456.1   \\
 2  &  -    & 3   & QSO    & -    &  -    &    -       &  -     & -  &  -    &  J154609.52+002624.6   \\
 3  & RG+P  & 1   & pass   & 2.00 & 0.03  &  $-3.67$   & 11.29  & 0.00  & 0.00  &  J083411.09+580321.4   \\
 4  &  -    & 2   & inter  & 1.88 & 0.97  &    -       &  -     & -  &  -    &  J084856.57+013647.8   \\
 5  & SF    & 2   & inter  & 1.43 & 0.04  &    0.24    & 10.57  & 1  & 0.00  &  J092405.30+141021.4   \\
 6  & RG+P  & 1   & pass   & 1.93 & 0.07  &  $-3.20$   & 11.77  & 1  & 3.00  &  J083139.79+460800.8   \\
\hline                 
\noalign{\smallskip}
\end{tabular}     
\tablefoot{
(1) SDSS name;
(2)-(3) right ascension and  declination (degrees);
(4) redshift; 
(5) radio loudness parameter  (Sect.\,\ref{sect:Samples}); 
(6) SDSS spectral class; 
(7) 1.4 GHz radio luminosity (W\,Hz$^{-1}$);  
(8) projected largest linear size (kpc);
(9) FR type;
(10) - (11) radio slope at $\nu > 1$\,GHz and uncertainty;
(12) type based on WHAN diagram;
(13) spectral type from MBSED (Table\,\ref{tab:templates});
(14) type based on WISE $W1-W2$ vs. $W2-W3$ diagram (Fig.\,\ref{fig:WISE_CCD}); 
(15) - (16) 4000\,\AA\ break and uncertainty;
(17) - (19) SFR ($\mathcal{M_\odot}\,{\rm yr}^{-1}$), stellar mass ($\mathcal{M_\odot}$)
and uncertainty flag from \protect\citet{Chang_2015};
(20) merger index based on images from the DESI Legacy Surveys;                                                       
(21) counterpart in the PSS-CSO sample.\\                                                
}                                                                                    
\label{tab:ECS}
\end{table*}

%
\section{VLASS-SDSS overlays of the sources from the ECS sample}\label{sect:FIRST_SDSS_all}
%

For more details, readers can refer to  Fig.\,\ref{fig:SDSS_VLASS_all}.

\begin{figure*}[htbp]
\begin{center}
\includegraphics[viewport= 0 0 360 360,width=4.45cm,angle=0]{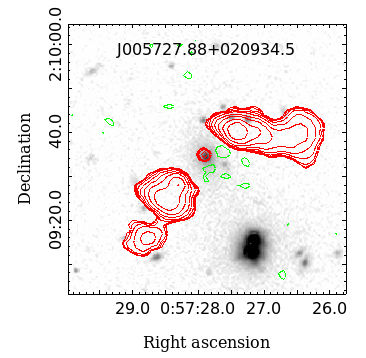}  
\includegraphics[viewport= 0 0 360 360,width=4.45cm,angle=0]{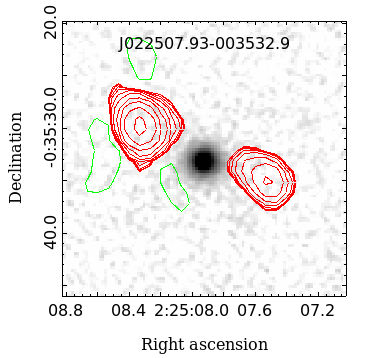}   
\includegraphics[viewport= 0 0 360 360,width=4.45cm,angle=0]{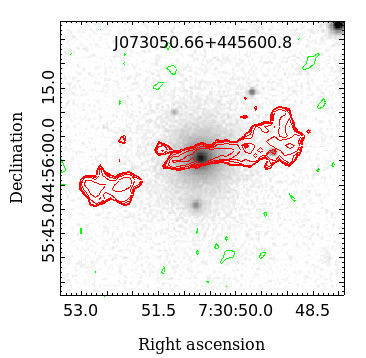}   
\includegraphics[viewport= 0 0 360 360,width=4.45cm,angle=0]{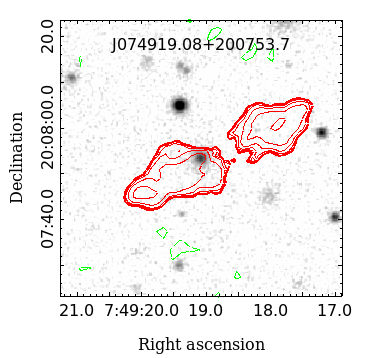}   \\    
\includegraphics[viewport= 0 0 360 360,width=4.45cm,angle=0]{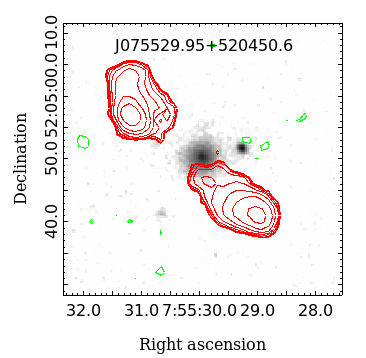}  
\includegraphics[viewport= 0 0 360 360,width=4.45cm,angle=0]{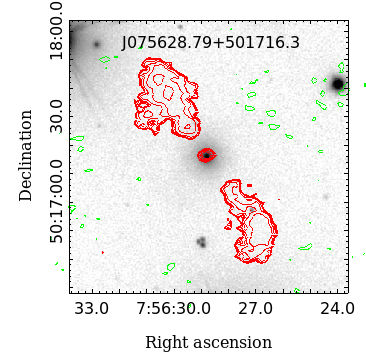}   
\includegraphics[viewport= 0 0 360 360,width=4.45cm,angle=0]{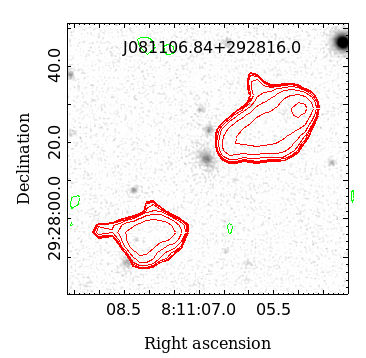}   
\includegraphics[viewport= 0 0 360 360,width=4.45cm,angle=0]{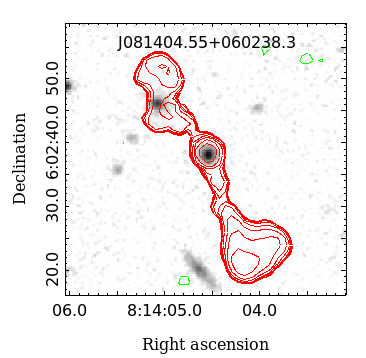}   \\  
\includegraphics[viewport= 0 0 360 360,width=4.45cm,angle=0]{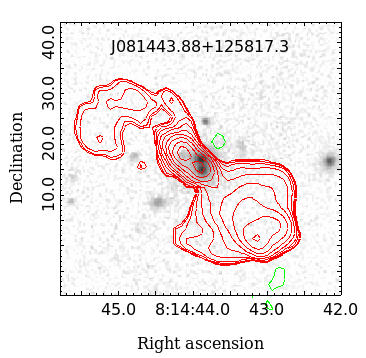}  
\includegraphics[viewport= 0 0 360 360,width=4.45cm,angle=0]{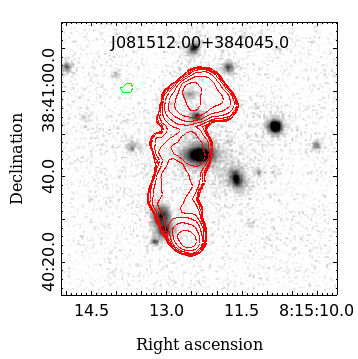}   
\includegraphics[viewport= 0 0 360 360,width=4.45cm,angle=0]{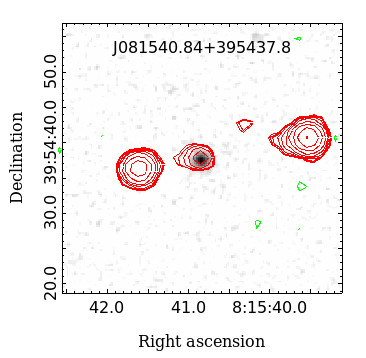}   
\includegraphics[viewport= 0 0 360 360,width=4.45cm,angle=0]{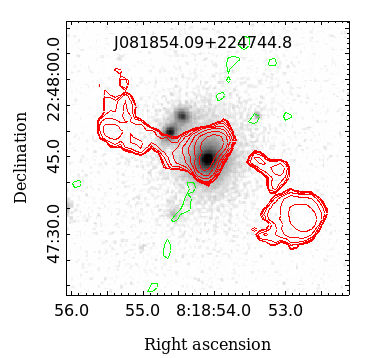}   \\  
\includegraphics[viewport= 0 0 360 360,width=4.45cm,angle=0]{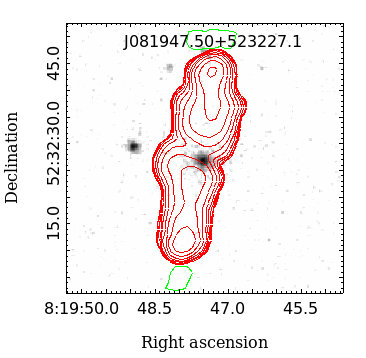}  
\includegraphics[viewport= 0 0 360 360,width=4.45cm,angle=0]{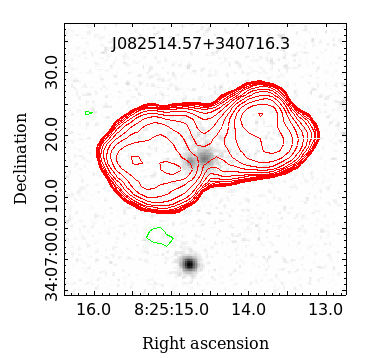}   
\includegraphics[viewport= 0 0 360 358,width=4.45cm,angle=0]{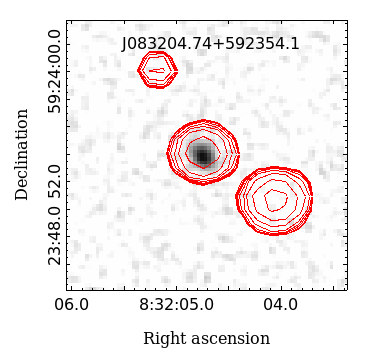}   
\includegraphics[viewport= -10 -10 360 359,width=4.5cm,angle=0]{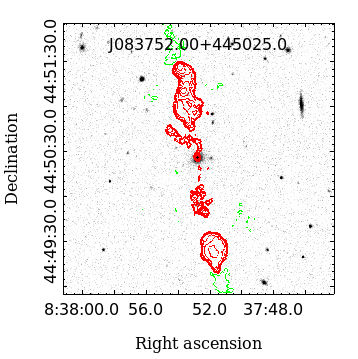}  \\
\includegraphics[viewport= 0 0 360 360,width=4.45cm,angle=0]{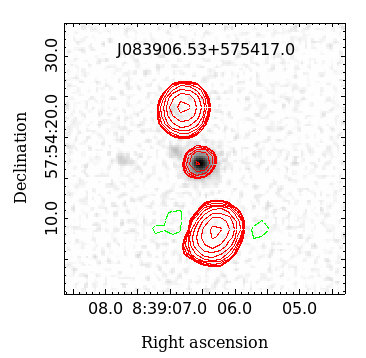} 
\includegraphics[viewport= 0 0 360 360,width=4.45cm,angle=0]{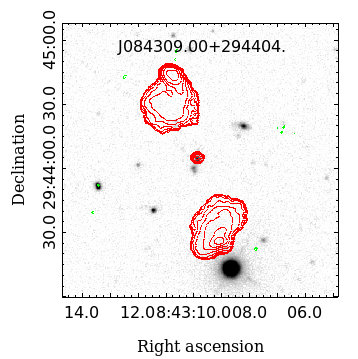}
\includegraphics[viewport= 0 0 360 360,width=4.45cm,angle=0]{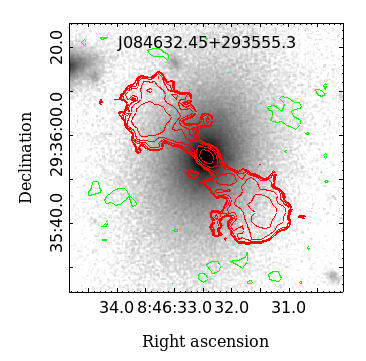}  
\includegraphics[viewport= 0 0 360 360,width=4.45cm,angle=0]{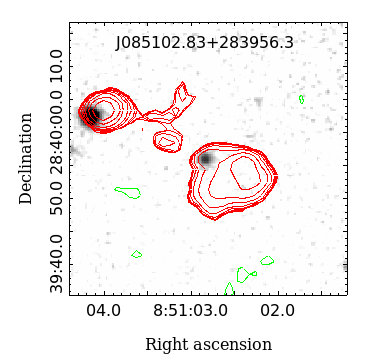} \\
\end{center}
\caption{
VLASS contours of the sources from the ECS sample.   The ten red contour lines are logarithmically spaced between $5\sigma$ of the local noise and the maximum flux of the source.  The green dashed lines are negative contours at $-5\sigma$ to $-3\sigma$. The background images are from the SDSS i band.              
}
\label{fig:SDSS_VLASS_all}
\end{figure*}

\setcounter{figure}{0}
\begin{figure*}[htbp]
\begin{center}
\includegraphics[viewport= 0 0 360 360,width=4.45cm,angle=0]{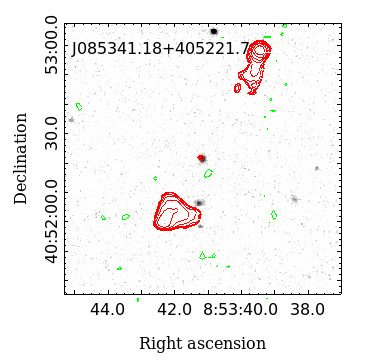}
\includegraphics[viewport= 0 0 360 360,width=4.45cm,angle=0]{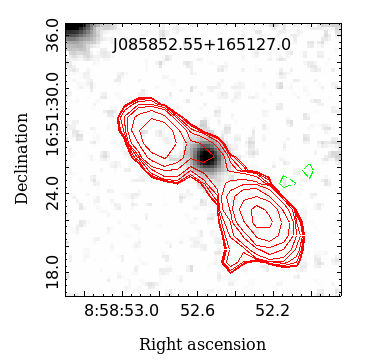}
\includegraphics[viewport= 0 0 360 360,width=4.45cm,angle=0]{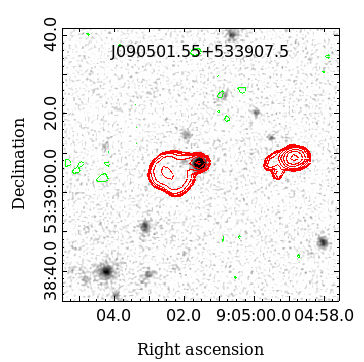}  
\includegraphics[viewport= 0 0 360 360,width=4.45cm,angle=0]{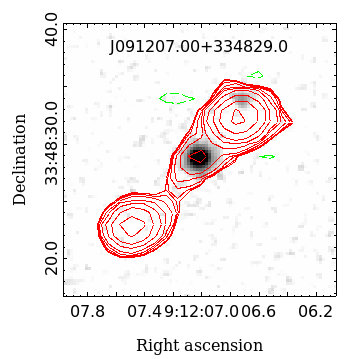} \\   
\includegraphics[viewport= 0 0 360 360,width=4.5cm,angle=0]{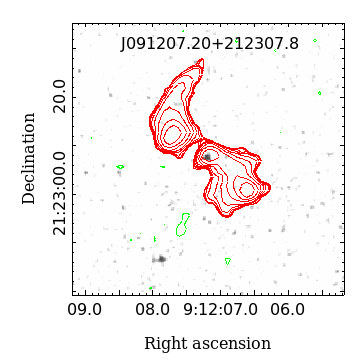}   
\includegraphics[viewport= 0 0 360 360,width=4.45cm,angle=0]{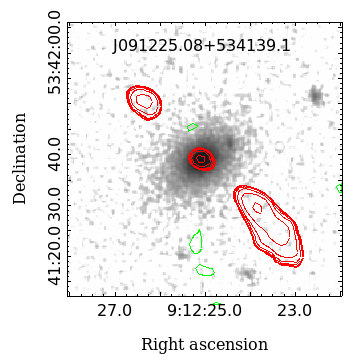}       
\includegraphics[viewport= 0 0 360 360,width=4.45cm,angle=0]{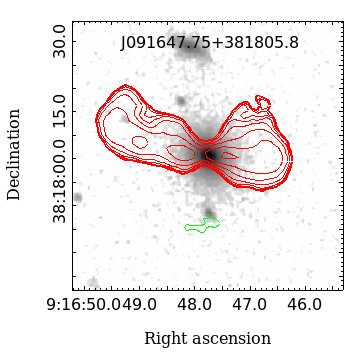}  
\includegraphics[viewport= 0 0 360 360,width=4.45cm,angle=0]{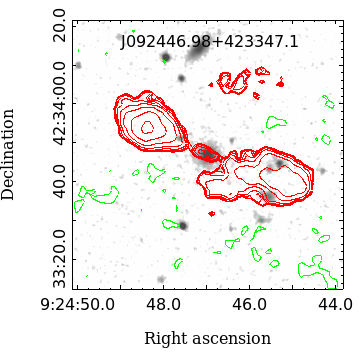}   \\
\includegraphics[viewport= -20 -10 358 358,width=4.7cm,angle=0]{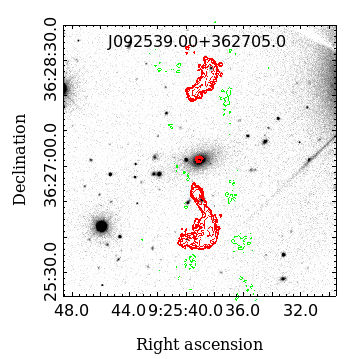}   
\includegraphics[viewport= 0 0 358 358,width=4.45cm,angle=0]{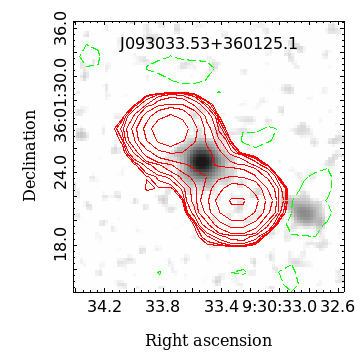}
\includegraphics[viewport= 0 0 358 358,width=4.45cm,angle=0]{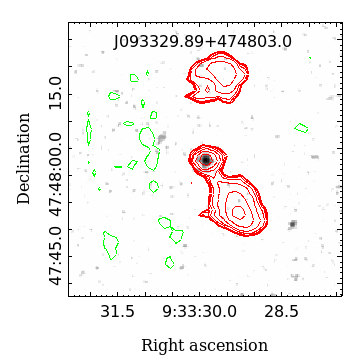}
\includegraphics[viewport= -10 -10 358 358,width=4.5cm,angle=0]{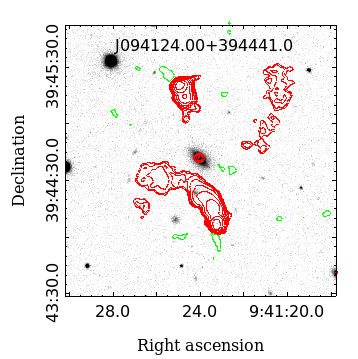} \\
\includegraphics[viewport= 0 0 360 360,width=4.45cm,angle=0]{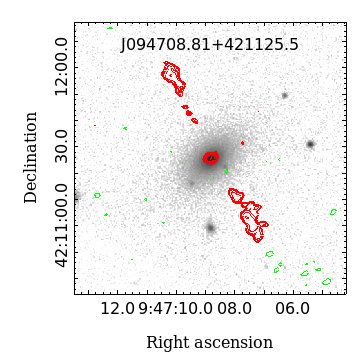}  
\includegraphics[viewport= 0 0 360 360,width=4.45cm,angle=0]{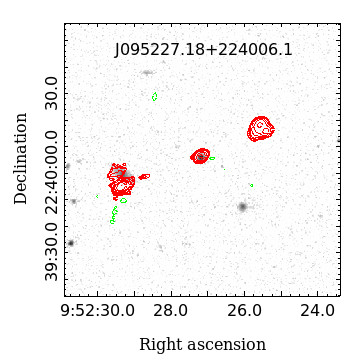} 
\includegraphics[viewport= -10 -10 358 358,width=4.5cm,angle=0]{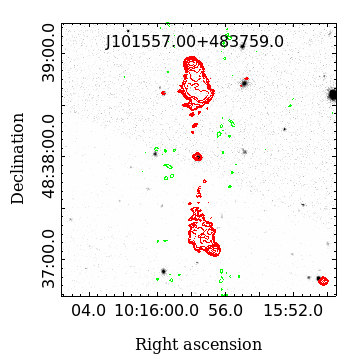} 
\includegraphics[viewport= 0 0 360 360,width=4.45cm,angle=0]{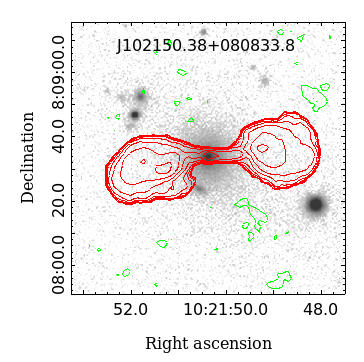} \\
\includegraphics[viewport= 0 0 360 360,width=4.45cm,angle=0]{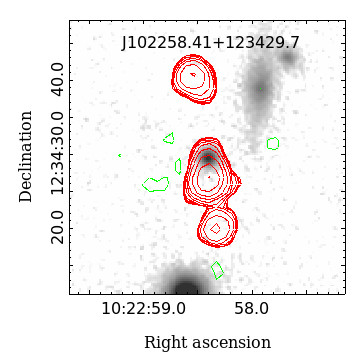}
\includegraphics[viewport= 0 0 360 360,width=4.45cm,angle=0]{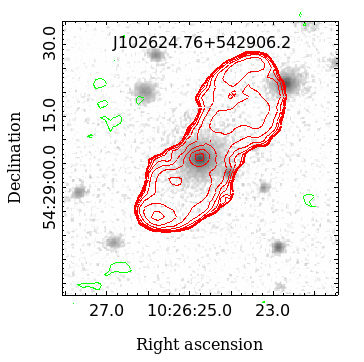}
\includegraphics[viewport= 0 0 360 360,width=4.45cm,angle=0]{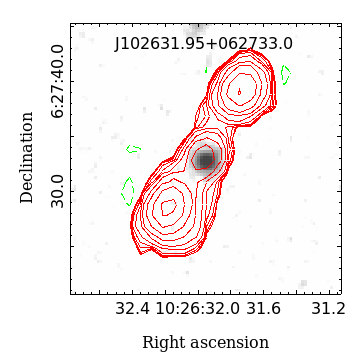}
\includegraphics[viewport= 0 0 360 360,width=4.45cm,angle=0]{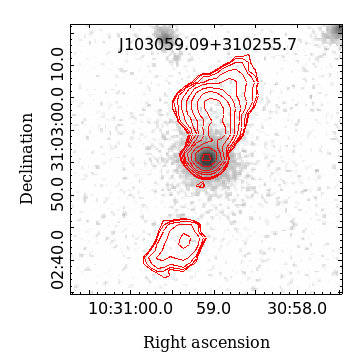}
\end{center}
\caption{continued.
}
\end{figure*}

\setcounter{figure}{0}
\begin{figure*}[htbp]
\begin{center}
\includegraphics[viewport= 0 0 360 360,width=4.45cm,angle=0]{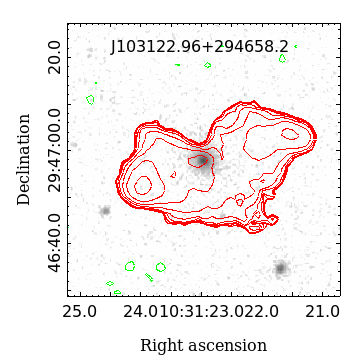}
\includegraphics[viewport= 0 0 360 360,width=4.45cm,angle=0]{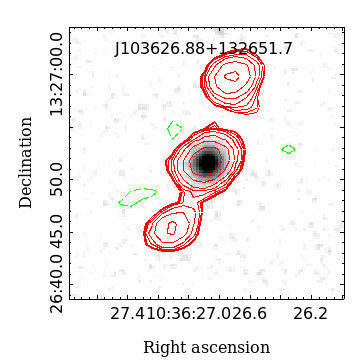}
\includegraphics[viewport= 0 0 360 360,width=4.45cm,angle=0]{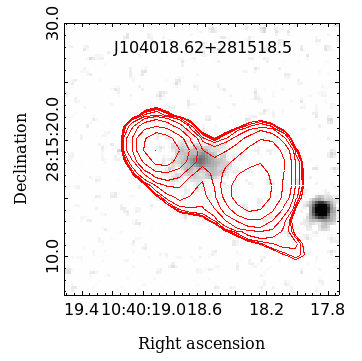}  
\includegraphics[viewport= 0 0 360 360,width=4.45cm,angle=0]{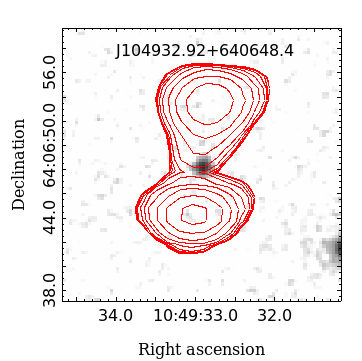}  \\
\includegraphics[viewport= 0 0 360 360,width=4.45cm,angle=0]{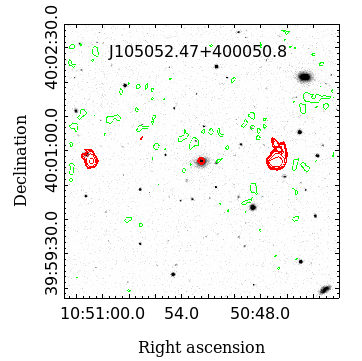}
\includegraphics[viewport= 0 0 360 360,width=4.45cm,angle=0]{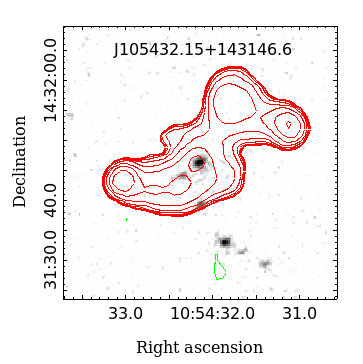}
\includegraphics[viewport= 0 0 360 360,width=4.45cm,angle=0]{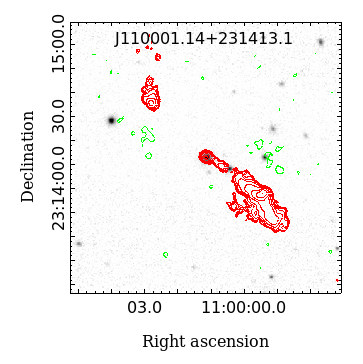}
\includegraphics[viewport= -10 -10 360 360,width=4.55cm,angle=0]{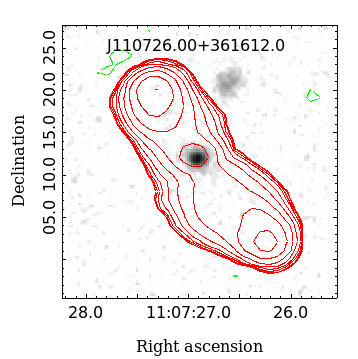} \\
\includegraphics[viewport= 0 0 360 360,width=4.45cm,angle=0]{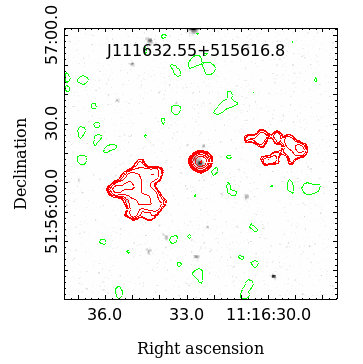}
\includegraphics[viewport= 0 0 360 360,width=4.45cm,angle=0]{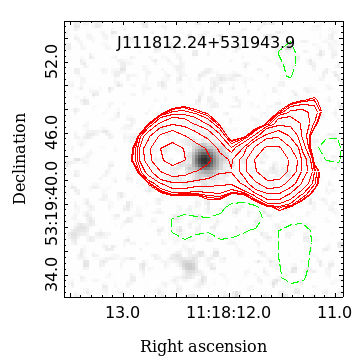}
\includegraphics[viewport= 0 0 360 360,width=4.45cm,angle=0]{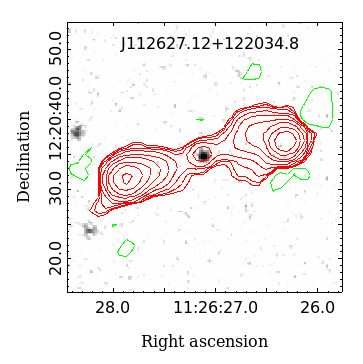}
\includegraphics[viewport= 0 0 360 360,width=4.45cm,angle=0]{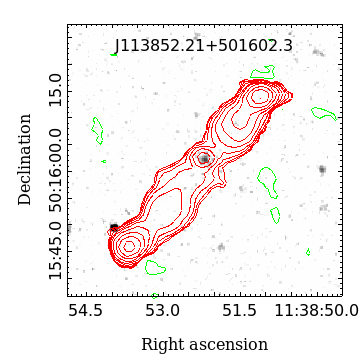} \\
\includegraphics[viewport= 0 0 360 360,width=4.45cm,angle=0]{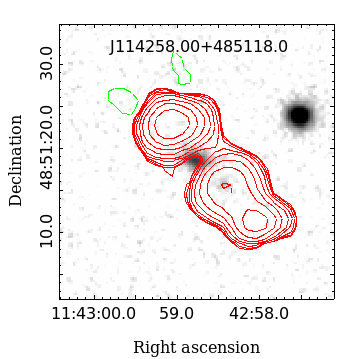}
\includegraphics[viewport= 0 0 360 360,width=4.45cm,angle=0]{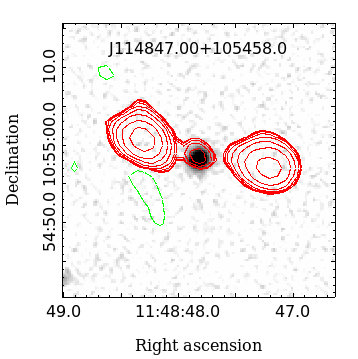}
\includegraphics[viewport= 0 0 360 360,width=4.45cm,angle=0]{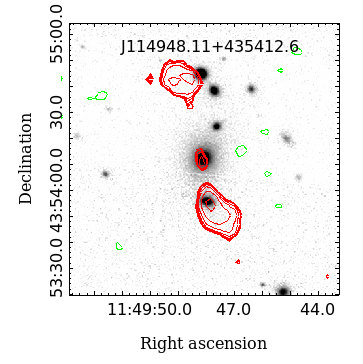}
\includegraphics[viewport= 0 0 360 360,width=4.45cm,angle=0]{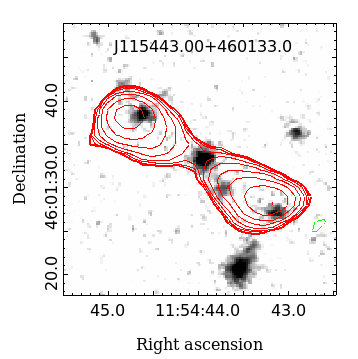} \\
\includegraphics[viewport= 0 0 360 360,width=4.45cm,angle=0]{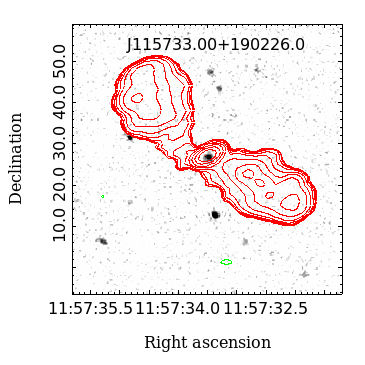}
\includegraphics[viewport= 0 0 360 360,width=4.45cm,angle=0]{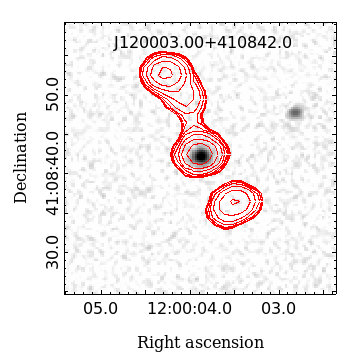}
\includegraphics[viewport= 0 0 360 360,width=4.45cm,angle=0]{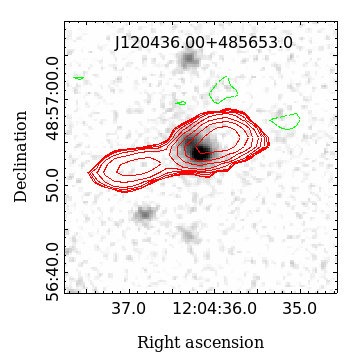}
\includegraphics[viewport= 0 0 360 360,width=4.45cm,angle=0]{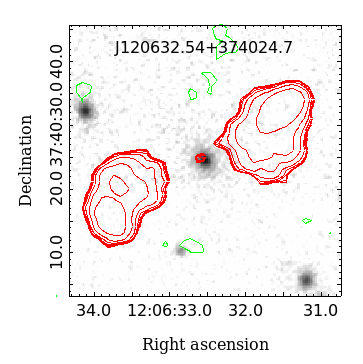}
\end{center}
\caption{continued.
}
\end{figure*}

\setcounter{figure}{0}
\begin{figure*}[htbp]
\begin{center}
\includegraphics[viewport= 0 0 360 360,width=4.50cm,angle=0]{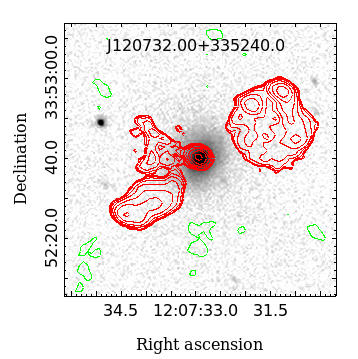}
\includegraphics[viewport= 0 0 360 360,width=4.45cm,angle=0]{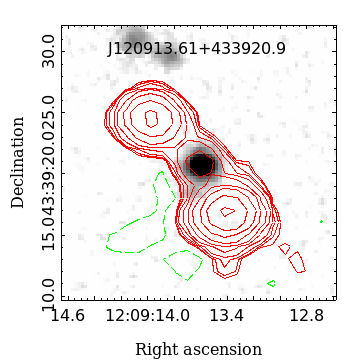}
\includegraphics[viewport= 0 0 360 360,width=4.45cm,angle=0]{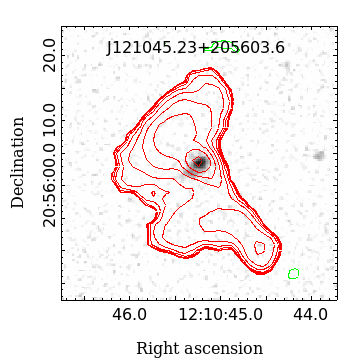}  
\includegraphics[viewport= 15 0 360 360,width=4.25cm,angle=0]{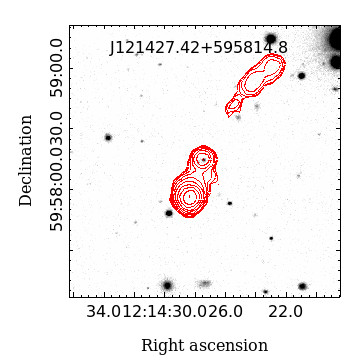}  \\
\includegraphics[viewport= 0 0 360 360,width=4.45cm,angle=0]{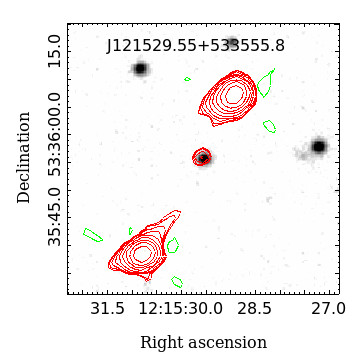}
\includegraphics[viewport= 0 0 360 360,width=4.45cm,angle=0]{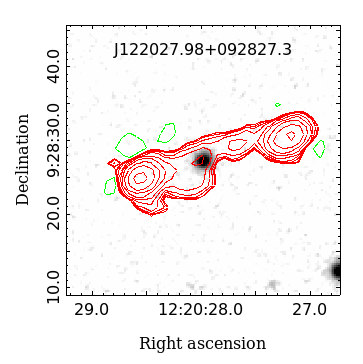}
\includegraphics[viewport= 0 0 360 360,width=4.45cm,angle=0]{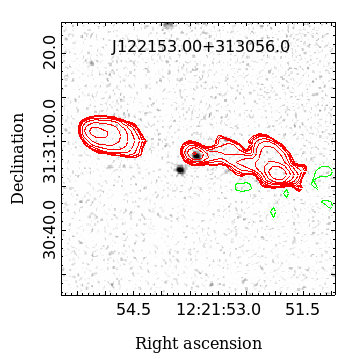}
\includegraphics[viewport= 0 0 360 360,width=4.45cm,angle=0]{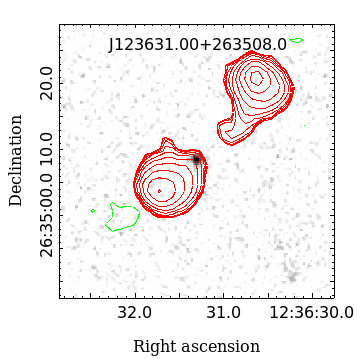} \\
\includegraphics[viewport= 0 0 360 360,width=4.45cm,angle=0]{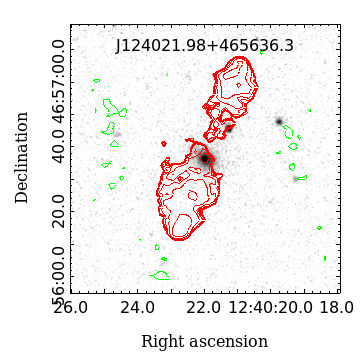}
\includegraphics[viewport= 0 0 360 360,width=4.45cm,angle=0]{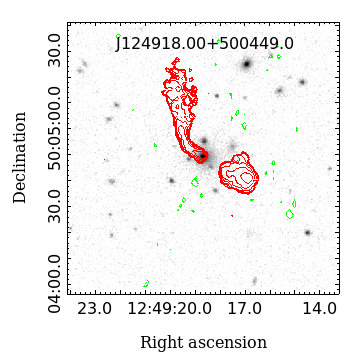}
\includegraphics[viewport= 0 0 360 360,width=4.45cm,angle=0]{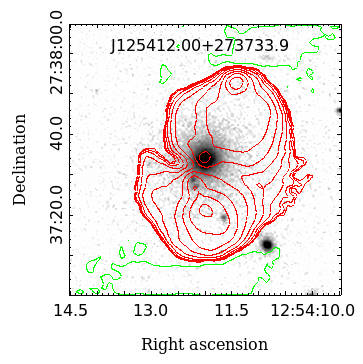}
\includegraphics[viewport= 0 0 360 360,width=4.45cm,angle=0]{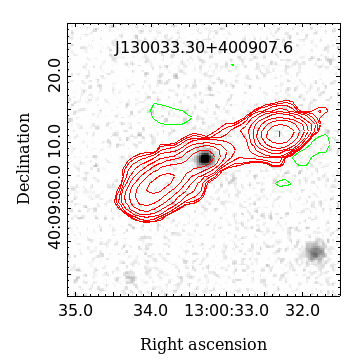} \\
\includegraphics[viewport= 0 0 360 360,width=4.45cm,angle=0]{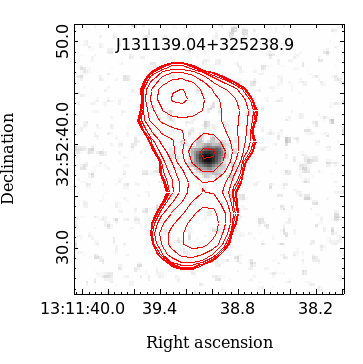}
\includegraphics[viewport= 0 0 360 360,width=4.45cm,angle=0]{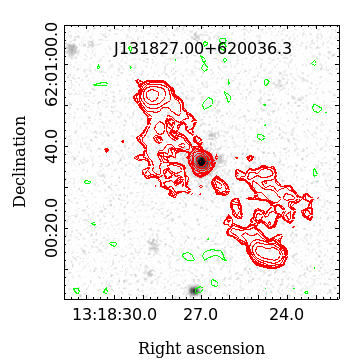}
\includegraphics[viewport= 0 0 360 360,width=4.45cm,angle=0]{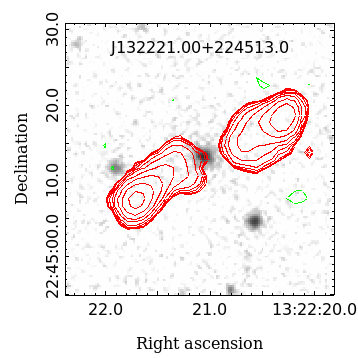}
\includegraphics[viewport= 0 0 360 360,width=4.45cm,angle=0]{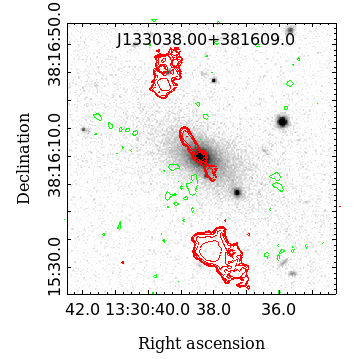} \\
\includegraphics[viewport= 0 0 360 360,width=4.45cm,angle=0]{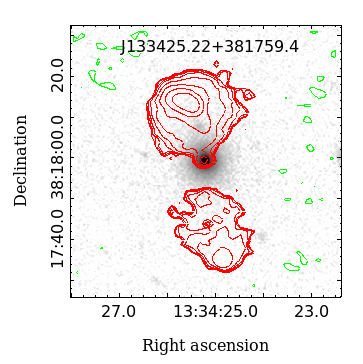}
\includegraphics[viewport= 0 0 360 360,width=4.45cm,angle=0]{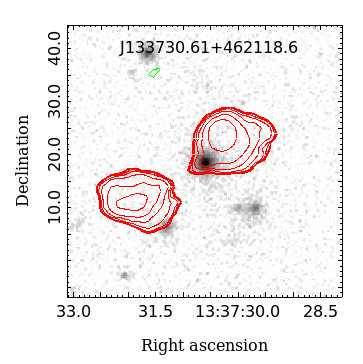}
\includegraphics[viewport= 0 0 360 360,width=4.45cm,angle=0]{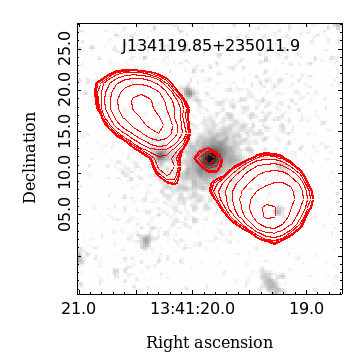}
\includegraphics[viewport= 0 0 360 360,width=4.45cm,angle=0]{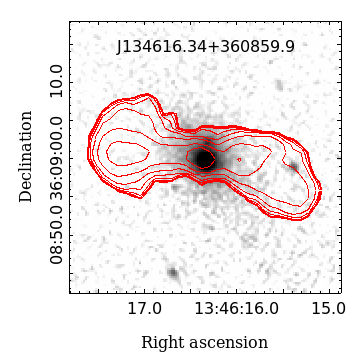}
\end{center}
\caption{continued.
}
\end{figure*}

\setcounter{figure}{0}
\begin{figure*}[htbp]
\begin{center}
\includegraphics[viewport= 0 0 360 360,width=4.45cm,angle=0]{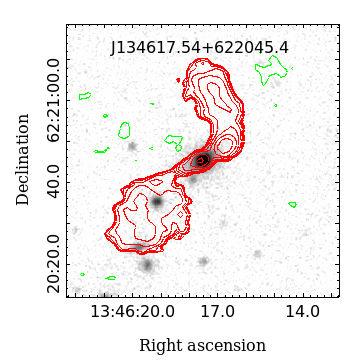}
\includegraphics[viewport= 0 0 360 360,width=4.45cm,angle=0]{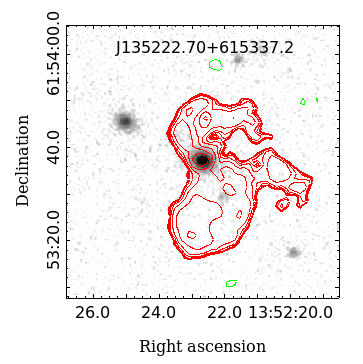}
\includegraphics[viewport= 0 0 360 360,width=4.45cm,angle=0]{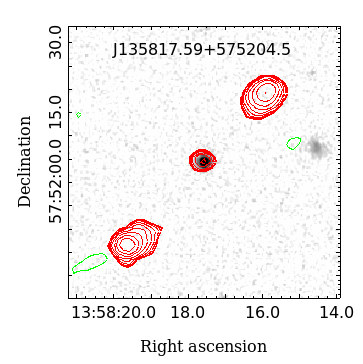}
\includegraphics[viewport= 0 0 360 360,width=4.45cm,angle=0]{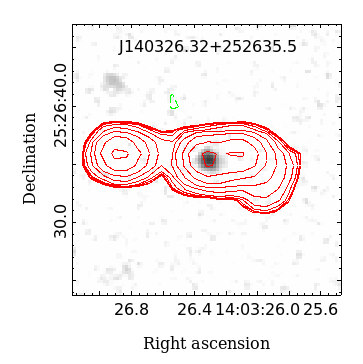} \\
\includegraphics[viewport= 0 0 360 360,width=4.45cm,angle=0]{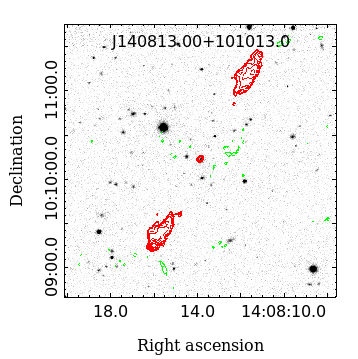}
\includegraphics[viewport= 0 0 360 360,width=4.45cm,angle=0]{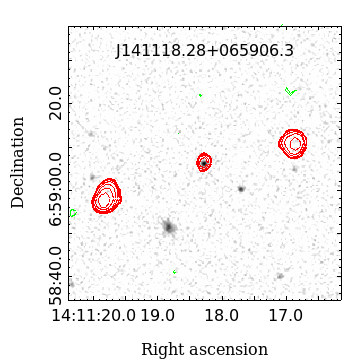}
\includegraphics[viewport= 0 0 360 360,width=4.45cm,angle=0]{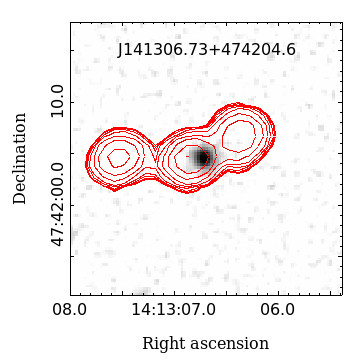}
\includegraphics[viewport= 0 0 360 360,width=4.45cm,angle=0]{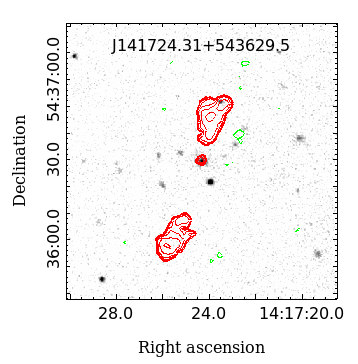} \\
\includegraphics[viewport= 0 0 360 360,width=4.45cm,angle=0]{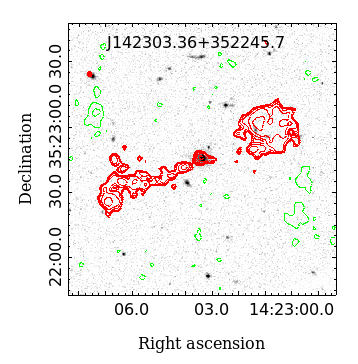}
\includegraphics[viewport= 0 0 360 360,width=4.45cm,angle=0]{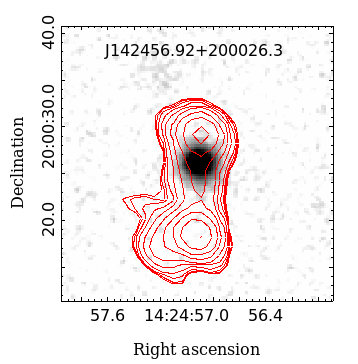}
\includegraphics[viewport= 0 0 360 360,width=4.45cm,angle=0]{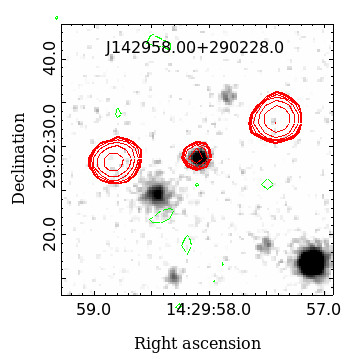}
\includegraphics[viewport= 0 0 360 360,width=4.45cm,angle=0]{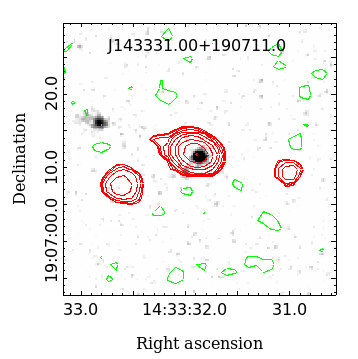} \\
\includegraphics[viewport= 0 0 360 360,width=4.45cm,angle=0]{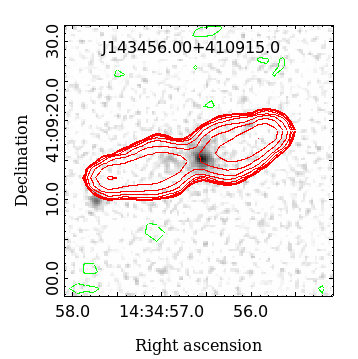}
\includegraphics[viewport= 0 0 360 360,width=4.45cm,angle=0]{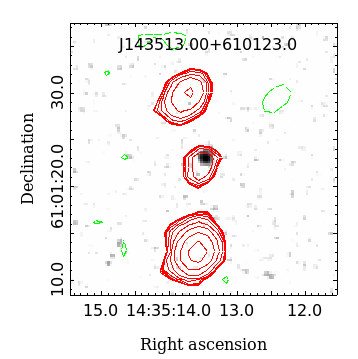}
\includegraphics[viewport= 0 0 360 360,width=4.45cm,angle=0]{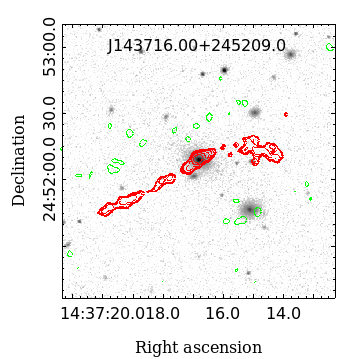}
\includegraphics[viewport= 0 0 360 360,width=4.45cm,angle=0]{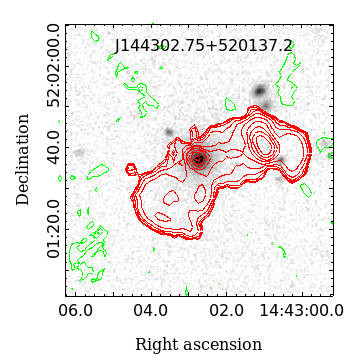} \\
\includegraphics[viewport= 0 0 360 360,width=4.45cm,angle=0]{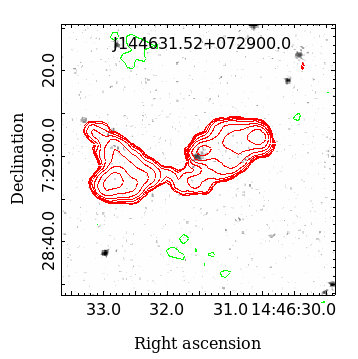}
\includegraphics[viewport= 0 0 360 360,width=4.45cm,angle=0]{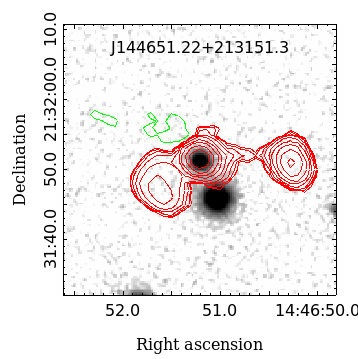}
\includegraphics[viewport= 0 0 360 360,width=4.45cm,angle=0]{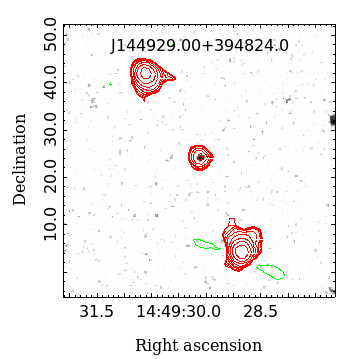}
\includegraphics[viewport= 0 0 360 360,width=4.45cm,angle=0]{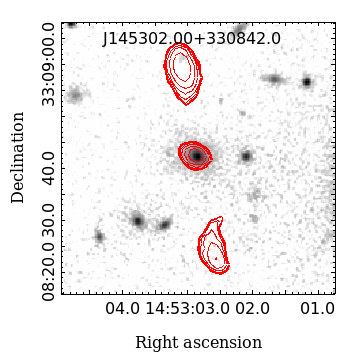}
\end{center}
\caption{continued.
}
\end{figure*}

\setcounter{figure}{0}
\begin{figure*}[htbp]
\begin{center}
\includegraphics[viewport= 0 0 360 360,width=4.45cm,angle=0]{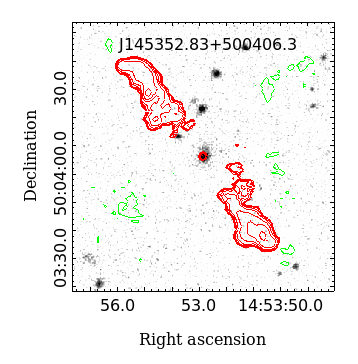} 
\includegraphics[viewport= 0 -10 360 360,width=4.40cm,angle=0]{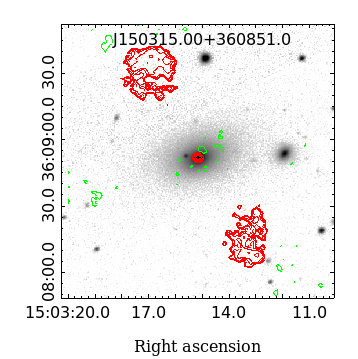}
\includegraphics[viewport= 0 0 360 360,width=4.45cm,angle=0]{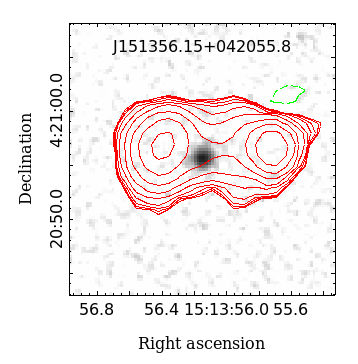}
\includegraphics[viewport= 0 0 360 360,width=4.45cm,angle=0]{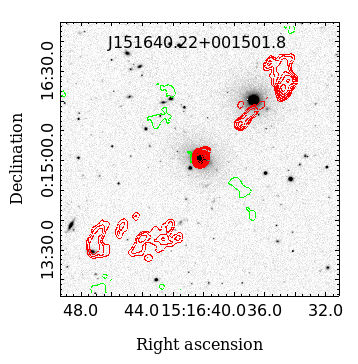} \\
\includegraphics[viewport= 0 0 360 360,width=4.45cm,angle=0]{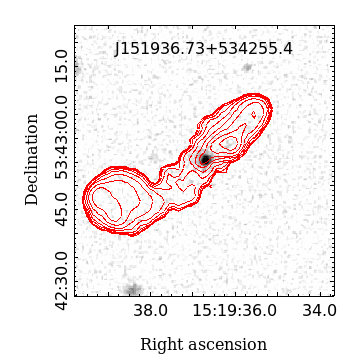} 
\includegraphics[viewport= 0 0 360 360,width=4.45cm,angle=0]{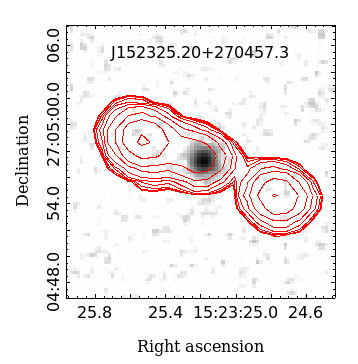}
\includegraphics[viewport= 0 0 360 360,width=4.45cm,angle=0]{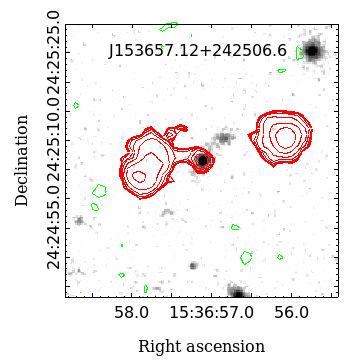}
\includegraphics[viewport= 0 0 360 360,width=4.45cm,angle=0]{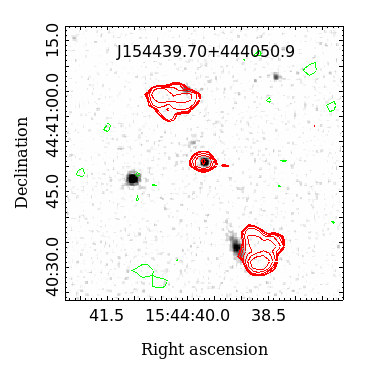} \\
\includegraphics[viewport= 0 0 360 360,width=4.45cm,angle=0]{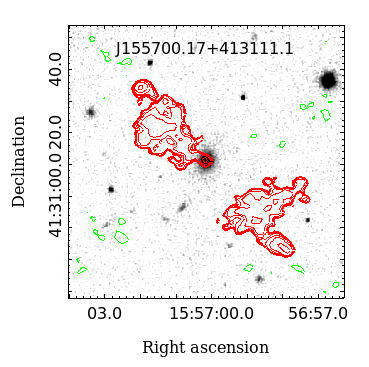} 
\includegraphics[viewport= 0 0 360 360,width=4.45cm,angle=0]{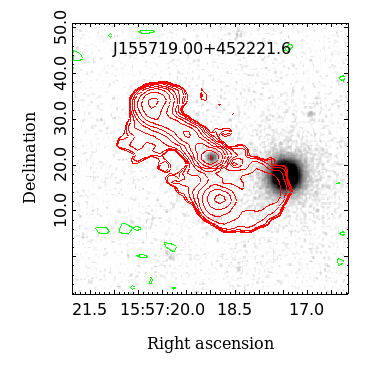}
\includegraphics[viewport= 0 0 360 360,width=4.45cm,angle=0]{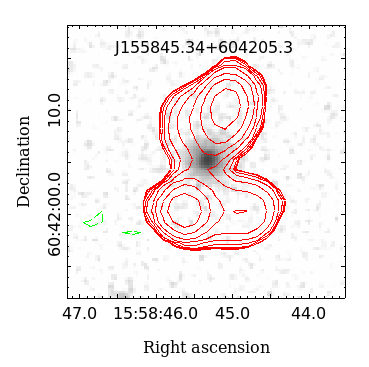}
\includegraphics[viewport= 0 0 360 360,width=4.45cm,angle=0]{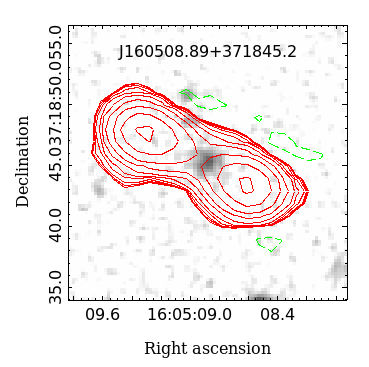} \\
\includegraphics[viewport= 0 0 360 360,width=4.45cm,angle=0]{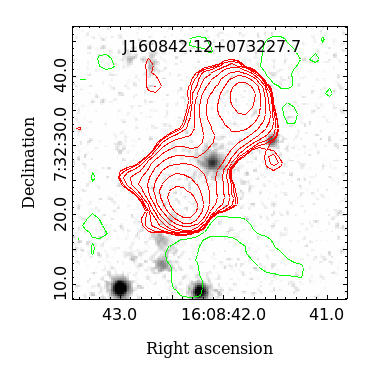} 
\includegraphics[viewport= 0 0 360 360,width=4.45cm,angle=0]{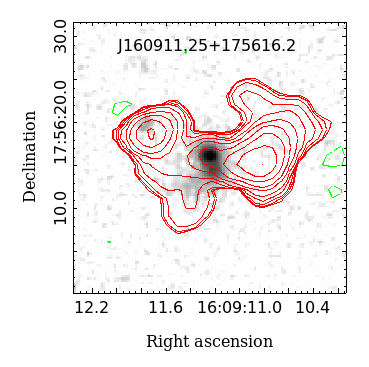}
\includegraphics[viewport= 0 0 360 360,width=4.45cm,angle=0]{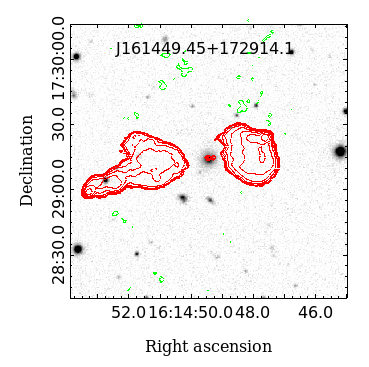}
\includegraphics[viewport= 0 0 360 360,width=4.45cm,angle=0]{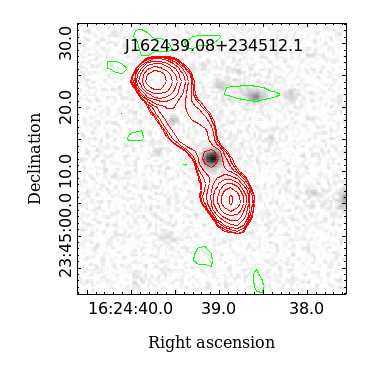} \\
\includegraphics[viewport= 0 0 360 360,width=4.45cm,angle=0]{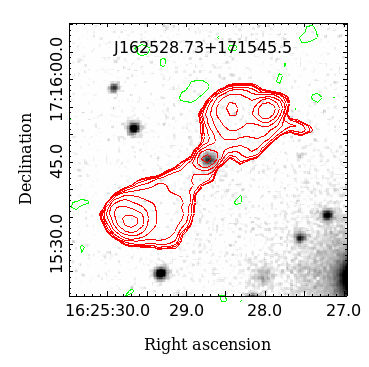} 
\includegraphics[viewport= 0 0 360 360,width=4.45cm,angle=0]{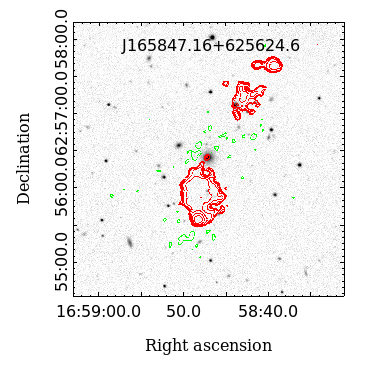}
\includegraphics[viewport= 0 0 360 360,width=4.45cm,angle=0]{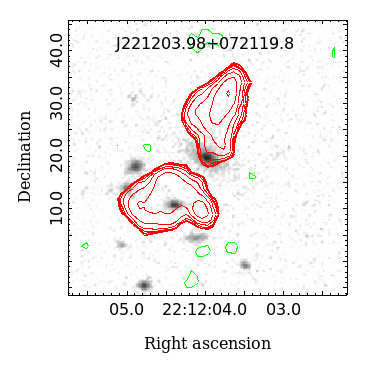}
\includegraphics[viewport= 0 0 360 360,width=4.45cm,angle=0]{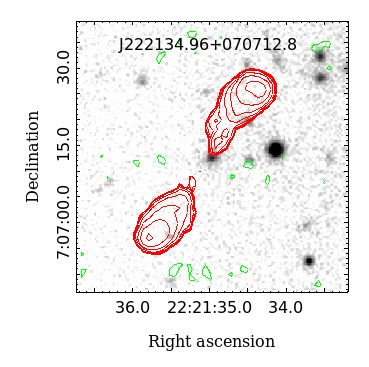} \\
\end{center}
\caption{continued.
}
\end{figure*}

\setcounter{figure}{0}
\begin{figure}[htbp]
\includegraphics[viewport= 0 0 360 360,width=4.45cm,angle=0]{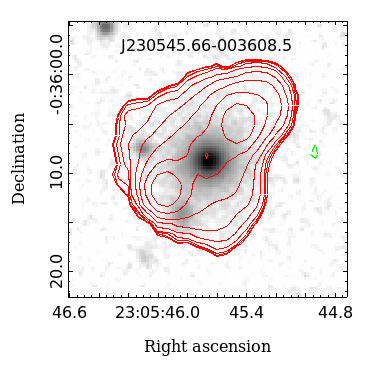} 
\caption{continued.
}
\end{figure}

\end{appendix}

\end{document}